**DEVELOPMENT OF GAMIFICATION MODEL FOR**

**PERSONALIZED E-LEARNING**

**BY**

**IBISU AFVENSU ENOCH**

**TP18/19/H/1824**

**B.Sc. (Information Technology), Lokoja**

**A THESIS WRITTEN IN THE DEPARTMENT OF COMPUTER SCIENCE AND ENGINEERING, FACULTY OF TECHNOLOGY AND SUBMITTED TO THE POSTGRADUATE COLLEGE, OBAFEMI AWOLOWO UNIVERSITY, ILE-IFE, NIGERIA, IN PARTIAL FULFILMENT OF THE REQUIREMENTS FOR THE AWARD OF THE DEGREE OF MASTER OF SCIENCE (M.Sc.) IN INFORMATION SYSTEMS.**

2023

# ANTI-PLAGIARISM DECLARATION

I, Ibisu, Afvensu Enoch, with Registration number TP18/19/H/1824 of the Department of Computer Science and Engineering, Faculty of Technology understand the interpretation of plagiarism and I am well aware of the University's policy in this regard. I submit that this thesis is my own original work and other people's work utilized, (either from a printed source, the internet, or any other source) has been properly acknowledged and referenced in accordance with University's policy.

04/09/2023

**Signature**                                                                                      **Date**

**OBAFEMI AWOLOWO UNIVERSITY, ILE-IFE, NIGERIA**

**HEZEKIAH OLUWASANMI LIBRARY**

**POSTGRADUATE THESIS**

**AUTHORIZATION TO COPY**

**AUTHOR:**   IBISU Afvensu Enoch

**TITLE:**   Development of a gamification model for personalized E-learning

**DEGREE:**   M.Sc. Information Systems

**YEAR:**   2023

I, Ibisu, Afvensu Enoch, hereby authorize Hezekiah Oluwasanmi Library to copy my thesis in whole or in part in response to request from individual researchers and organizations for the purpose of private study or research.

                                                                    **04/09/2023**
**Signature**                                                          **Date**

# CERTIFICATION

This is to certify that this thesis was written by IBISU Afvensu Enoch with the Registration Number TP18/19/H/1824 as part of the requirements for the award of Master of Science in Information Systems at the Department of Computer Science and Engineering, Faculty of Technology, Obafemi Awolowo University, Ile-Ife, Nigeria.

07/09/2023

Dr. R. N. Ikono                                    Date

Supervisor

07/9/2023

Prof. A. O. Oluwatope                              Date

Head of Department

15/11/2023

                                                   Date

Provost, Postgraduate College

iv

# ACKNOWLEDGEMENTS

I am deeply grateful to acknowledge the following individuals and groups who have played a significant role in making this research a reality: First and foremost, my sincere appreciation goes to my supervisor, Dr. (Mrs.) R.N. Ikono, for her invaluable guidance, unwavering support, and encouragement throughout the entire research process. I extend special thanks to the Health Information Systems research group for their constructive inputs, which have ensured that this research maintains its relevance and rigor. I am also indebted to the staff and students of the Department of Computer Science and Engineering, Obafemi Awolowo University, Ile-Ife, for creating a conducive environment and making my academic program truly worthwhile.

To my beloved parents, Mr. & Mrs. B.M. Ibisu, and my dear siblings Ikpesu, Nyaga, Eminya, Eding, and Buga, I am forever grateful for your unwavering support and boundless love.

I am thankful to Dr M. Garba, Dr A. Baba, and Dr S. Aminu of Taraba State University for providing their expert opinions on game elements and their suitability for different MBTI cognitive pairs. To the El-spice Media/Skillcom Valley team, Mr. V.P. Dickson, Maxwell, and St. Charles, I extend my deepest gratitude for generously providing access to your learning management system, which was instrumental in conducting this research.

I express my heartfelt appreciation to the families of Mr. & Mrs. George and Mr. & Dr Mrs. Lakan for their warm hospitality and making my research stay in Jos a homely experience. Lastly, I would like to thank my dear friends, Bouqui, Chidimma, Alex, Timothy, Alhassan, Maria, Joy, Victor, Emmanuel, Tolu, Babalola, Daniel, Zenret, Emeka, and Plangwer, in no particular order, for their constant support and being a sounding board for my ideas. Your presence has made this journey all the more enjoyable.

Once again, thank you all for being part of this remarkable endeavour. Your contributions have made a lasting impact on this research, and I am truly grateful for each and every one of you.

# TABLE OF CONTENTS

# LIST OF FIGURES

# LIST OF TABLES

**ABSTRACT**

This study designed a personality-based gamification model for E-learning systems. It also implemented the model and evaluated the performance of the gamification model implemented. These were with a view to developing a model for gamifying personalization of e-learning systems.

Personalization requirements for motivational tendencies based on the Myers-Briggs Type Indicator (MBTI) and gamification elements were elicited for e-learning from existing literature and from education experts using interview and questionnaire. The gamification model for personalized e-learning was designed by mapping motivational tendencies to corresponding gamification elements using set theory and rendered using Unified modelling language (UML) tools. The model was implemented using Hypertext Markup Language for the front end, Hypertext Preprocessor (PHP) for the backend and Structured Query Language (SQL) for database on WordPress. The model was evaluated using appeal, emotion, user-centricity as well as satisfaction as engagement criteria, and clarity, error correction as well as feedback for educational usability.

The results collected from the implemented system database and questionnaires administered to learners showed an average appeal rating of 4.3, an emotion rating of 4.5, a user-centricity rating of 4.4, and a satisfaction rating of 4.4 in terms of engagement on a 5.0 scale. The results also showed that clarity, error correction and feedback received an average rating of 3.9, 4.7, and 4.8 respectively on a 5.0 scale concerning educational usability. In addition, when comparing educational usability (4.5) to engagement (4.4), educational usability received slightly higher ratings.

The study concluded that the gamification model for personalized e-learning was suitable for increasing learner motivation and engagement within the personalized e-learning environment.

<div align="center">

**CHAPTER ONE**

**INTRODUCTION**

</div>

1.1. **Background to the Study**

E-learning involves the utilization of digital platforms to enhance the learning process, delivering instructional content to learners within or outside a physical classroom environment. This emerged from advancements in web-based technologies which support creation and management of electronic contents accessible to multiple users simultaneously. E-learning systems are not only capable of rapidly providing contents but also supporting content management, interaction among large number of users (both faculty and students) and administrative functions (Fawareh, 2013).

E-learning has become an important tool for learning and teaching, enabling flexibility to learners and instructors alike. It provides an avenue for learners to engage in learning from diverse locations simultaneously or and at different times (Crawford, Butler-Henderson, Rudolph, Malkawi, Glowatz, Burton, Magni, Lam, 2020). Platforms like Udemy, Duolingo, uLesson and Coursera can provide courses and engage with students as a group or privately, enhancing communication and interaction between teachers and students. The availability of interactive contents, multimedia materials, and downloadable resources on these platforms can motivate students and spark interest in the learning materials. Additionally, the collaborative learning environment fostered by these platforms encourages social interactions facilitated by technology.

Instructional systems design is another way to further increase engagement and motivation in students. This approach systematically understands the variables related to a learning process and designs learning activities accordingly. Instructional design has become increasingly popular in creating effective and engaging learning materials (Lee & Jang, 2014; Suartama, Setyosari, Ulfa, 2019). To achieve this, instructional designers often draw inspiration from prominent motivational theories proposed by influential psychologists like Maslow, Herzberg, and McClelland.

Maslow's hierarchy of needs, suggests that individuals have a hierarchy of needs ranging from basic physiological needs to higher-order needs like self-esteem and self-actualization (Maslow, 1943). When addressing these needs within the learning environment, instructors can foster a sense of fulfillment and motivation in students, thereby enhancing their desire to engage actively in the learning process. Incorporating Herzberg's two-factor theory is another effective approach to instructional design. This theory posits that certain factors, such as achievement and recognition, act as motivators, while other factors, like working conditions, act as hygiene factors that can lead to dissatisfaction if not adequately addressed (Herzberg, Mausner, & Syderman, 1993). By leveraging this theory, instructors can design learning activities that incorporate motivating factors, encouraging student enthusiasm and active participation. Furthermore, McClelland's theory of needs highlights three primary needs: the need for achievement, affiliation, and power. Understanding and catering to these needs allow instructors to personalize learning experiences and tailor them to individual students' motivational preferences, thereby creating a more effective and engaging learning environment (McClelland, 1961).

In addition to these theories, instructional design benefits from insights provided by expectancy theory (Wigfield & Eccles, 2000), self-efficacy theory (Bandura, 1977), goal setting theory (Locke & Latham, 2006), and self-determination theory (Deci & Ryan, 2010). These theories explain concepts that facilitate intrinsic motivation leading to engagement in activities to fulfil the need for competence and autonomy, as well as extrinsic motivation such as getting rewarded for desired behaviours (Shillingford & Karlin, 2013).

In the literature, intrinsic motivational drive has consistently been suggested to be more sustainable than extrinsic motivation in learning (Shillingford & Karlin, 2013; Freitas, 2015; Dahlstrøm, 2017; Pink, 2009; Zichermann & Cunningham, 2011) as individuals intrinsically motivated find personal fulfilment in the learning process itself, making them less reliant on external rewards. This holds especially true in e-learning environments, where learners can derive motivation from rewards gained on completed tasks or through interactions with peers in discussions, collaboration, or competition— common features reminiscent of video games. The documented ability of games to promote high levels of motivation and engage players in immersive challenges has been well-established (Yee, 2006). Players often invest significant time and effort in overcoming these challenges, showcasing their commitment and deep engagement. Leveraging these gaming features, the gamification of e-learning emerges as a promising approach to harmonize learning and pleasure. By incorporating gaming elements into e-learning systems, educators can capitalize on learners' intrinsic motivation while still acknowledging the value of extrinsic motivators.

Gamification in e-learning fosters both intrinsic and extrinsic motivation among learners. Incorporating gaming elements provides satisfaction, fulfilment, and a sense of accomplishment, promoting intrinsic motivation (Chou, 2015). Learners are also motivated by the rewards they receive, contributing to extrinsic motivation (Papp, 2017). Research reveals common game elements utilized in e-learning gamification to include points, challenges, badges, leaderboards, and levels, which significantly motivate learners (Koivista & Hamari, 2019), although with varying extent impact and effectiveness (Zaric, 2018). A reason for this variation could be similar to what has been observed with in gamers. While numerous games are available to gamers, they don't all evoke the same level of motivation, and some may not fully engage players (Roy & Zaman, 2017). Similar effects are observed in gamified e-learning systems. Although gamification offers extensive opportunities to enhance the learning experience (Aljabali, Ahmad, Yusof, Miskon, Ali, Musa, 2020), not all gamified systems yield desired results, often due to the limitations of a "one size fits all" approach (Böckle, Micheel, Bick, & Novak, 2018). Researchers have found that personality significantly influences behaviours, including academic performance, indicating that personalized learning systems are more effective than generic ones (Ghaban & Hendley, 2019; Hassan, Habiba, Majeed, & Shoaib, 2019; Kang & Kusuma, 2020). Consequently, it becomes essential to investigate and develop a gamification model which uses learners' personality to personalize the learning experience for improved engagement and academic achievements.

**1.2. Statement of the Research Problem**

The increasing popularity of e-learning platforms has undoubtedly brought a surge of learners into the digital educational landscape. However, despite its accessibility and convenience, a persistent challenge looms large: low motivation and engagement among these learners (Hassan, *et al.,* 2019). One critical factor contributing to this issue is the similarity in learning experiences, where all learners are offered the same content and activities, irrespective of their diverse learning styles and individual preferences (Böckle *et. al.,* 2018).

While gamification has emerged as a promising strategy to enhance learner motivation and engagement in e-learning, research has unveiled a complexity of effects. Despite its potential to foster both intrinsic and extrinsic motivation (Chou, 2015; Papp, 2017), the impact and effectiveness of gamification appear to vary across different contexts and learner profiles (Zaric, 2018). Such discrepancies mirror those observed in the gaming community, where gamers' motivation levels towards specific games diverge significantly (Roy & Zaman, 2017). In response to the challenge of low motivation and engagement in e-learning, personalized learning systems have surfaced as a compelling solution, exhibiting superior outcomes compared to generic approaches (Ghaban & Hendley, 2019; Kang & Kusuma, 2020). This growing body of evidence points to the potential of incorporating personality-based gamification in e-learning. By understanding and leveraging learners' unique personality traits, a personalized and more immersive learning experience can be crafted, effectively addressing the issue of disengagement and lack of motivation.

The key objective of this research endeavour is, therefore, to investigate and develop a gamification model that harnesses learners' personalities to personalize the e-learning experience. By doing so, the study aims to foster enhanced motivation, engagement, and academic achievements in e-learning platforms, leading to a more satisfying and rewarding educational journey for learners.

### 1.3. Scope of the Research

This research is centred around exploring the implementation of gamification in personalized e-learning environments within open asynchronous settings. The focus is on allowing learners to enrol without prerequisites or study time restrictions, promoting accessibility and flexibility in the learning process. Specifically, the study delves into the incorporation of gamification elements customized to learners' cognitive pairs, as identified through the Myers-Briggs Type Indicator (MBTI) personality model. The objective is to examine how these personalized gamification strategies impact learner motivation, engagement, and academic achievements.

### 1.4. Aim and Objectives of the Study

This study aimed at developing a gamification model for personalized e-learning.

The specific objectives of the research were to;

i.      design a gamification model for personalized e-learning;

ii.     implement the model in (i); and

iii.    evaluate the performance of the system in (ii).

### 1.5. Methodology

The following methodologies were employed to achieve the aforementioned objectives:

i.      Requirements were elicited on motivational tendencies based on the Myers-Briggs Type Indicator and gamification elements for E-learning from existing literature and educational experts using interview and questionnaire.

ii.     A personalized E-learning model was designed by mapping motivational tendencies to corresponding gamification elements using Set Theory and rendered using UML tools.

iii.    The model was implemented using HTML for the front end, PHP for the backend and SQL for database.

iv.    The system was evaluated with appeal, emotion, user-centricity and satisfaction for engagement and clarity, error correction as well as feedback for educational usability using data collected from the system database and questionnaires.

## 1.6. Justification of the Study

Researchers have reported that despite the potential of gamification to spur motivation and engagement, the use of numerous game elements do not necessitate the desired outcome (Aljabali, et al., 2020; Aldemir, Celik, & Kaplan, 2017). This research will provide a gamification model that provides useful engagement to learners geared towards greater satisfaction with the learning process and achieving learning objectives.

## 1.7. Operational Definition of Terms

This section provides clear definitions of key terms or concepts that will be used throughout the thesis. It ensures that there is a common understanding of the terminology used in the research study.

- **Cognitive Core:** The central aspects of an individual's cognitive functioning that influence their decision-making, problem-solving, and interaction with their environment. In the context of the study, it refers to the key cognitive patterns identified within the Myers-Briggs Type Indicator (MBTI) framework.

- **Educational Usability:** The effectiveness, ease of use, and appropriateness of educational technology in facilitating learning and achieving learning goals.

- **E-learning:** Learning that happens online using digital tools and resources, making education accessible from anywhere with an internet connection.

- **Engagement:** The level of interest, attention, and active involvement that learners have while studying, which makes learning more effective and enjoyable.

- **Gamification:** A strategic approach aimed at enhancing user engagement and motivation by integrating game-like elements, such as rewards, challenges, and competition, into non-game contexts, such as education or business.

- **Human-Computer Interaction (HCI):** The study of how people interact with computers and software, encompassing usability, user interface design, and user experience.

- **Instructional Design:** The systematic process of designing effective and engaging learning experiences by applying principles of pedagogy, technology, and assessment. It involves planning, developing, and evaluating educational content and activities to meet specific learning objectives.

- **Learning Management Systems (LMS):** Software platforms that facilitate the management, delivery, and tracking of educational content and activities, typically used in online and blended learning environments.

- **Model:** A simplified representation of a complex concept, system, or process that helps to understand, analyse, and predict its behaviour. In the context of the study, it refers to a conceptual framework used to structure and guide the development of the personalized gamified e-learning system.

- **Motivation:** The inner drive or enthusiasm that pushes individuals to start learning, stay focused, and persist in their educational efforts to achieve their goals.

- **Pedagogy:** The art and science of teaching and learning, encompassing educational theories, strategies, methods, and practices aimed at facilitating effective and meaningful learning experiences for learners.

- **Personalization:** Tailoring learning experiences, content, and activities to meet the individual needs, preferences, and characteristics of each learner, enhancing engagement and learning outcomes.

- **Taxonomy:** A hierarchical classification system that categorizes concepts, objects, or information based on their characteristics and relationships.

- **Unified Modelling Language (UML):** A standardized visual modelling language used for specifying, visualizing, constructing, and documenting software systems.

- **User Experience:** The overall experience and perceptions of users when interacting with a product, service, or system. It includes aspects of usability, satisfaction, emotions, and effectiveness in achieving user goals. In the context of the study, it relates to learners' perceptions and engagement within the gamified e-learning environment.

- **User Interface (UI):** The visual and interactive elements through which users interact with a computer system, application, or software.

# CHAPTER TWO

# LITERATURE REVIEW

## 2.1. Introduction

This chapter elaborates on the concepts that form and support the basis of this research thesis. It discusses constructs in information systems, learning management systems, instructional design, learning psychology and game design as well as other related works.

## 2.2. Information Systems

Since the last two decades, the relevance of information systems in organizations has enjoyed continued growth. This growth was majorly influenced by the rise of personal computers (PC) in the 1980s and the creation of the internet in the 1990s (Bourgeois, 2014). Information systems were no longer manual but either fully computer-based or hybrid of manual and computer-based systems. Information System (IS) refers to any organized combination of resources that are useful in storage, retrieval, transformation, and dissemination of information within an organization (Al-Mamary, Shamsuddin, & Aziati, 2014). These systems can either stand alone or form part of a larger enterprise system.

According to Paul (2007), an Information System is the outcome of the utilization of the Information Technology (IT) delivery system by the users. This usage is made up of the following two parts:

i. The formal processes, determined by the choice of technology employed.

ii. The informal processes, comprising of people who use the information technology, and the type of outputs produced by the technology in order to ensure that useful work is accomplished.

Evidently, information systems, like any other system comprise of a number of components. They include the following:

i. Human Resource (People) / Procedures

    ii.    Tools

        a.   Computer Hardware

        b.   Computer Software

  iii.   Data / Information

  iv.   Network / Communication System

These components interact to provide organizations and businesses with benefits such as increased speed in accessing, processing and retrieval of information; maintaining accuracy, reduction in overall costs, and increased functionality of various departments.

Computer-based information systems can be broadly categorized into three (3) types, Management Support System, Knowledge Support System and Operational Support System (Zwass, 2017) which is as illustrated in Figure 2.1.

### 2.2.1. Management support systems

This category of information systems assists in management of an organization by utilizing and harnessing information from other systems within and outside the organization. Management support systems provide support in various levels of management in form of prediction, decision support judgement and analysis as well as recommendation. Management support systems include systems that provide executives with easy access to relevant information and data (executive information systems), assist decision-makers in identifying and evaluating options (decision support systems), collect and analyse data to help inform business decisions (business intelligence systems) and use spatial data to inform decision making (geographic information systems).

### 2.2.2. Knowledge support systems

Knowledge support systems are designed to improve information collection and retrieval, collaboration / teamwork among peers and professional, knowledge transfer. The focus of these systems is flow of volumes information in multiple formats rather than actual processing of information. Knowledge support systems may include professional support systems, collaboration systems, and knowledge management systems.

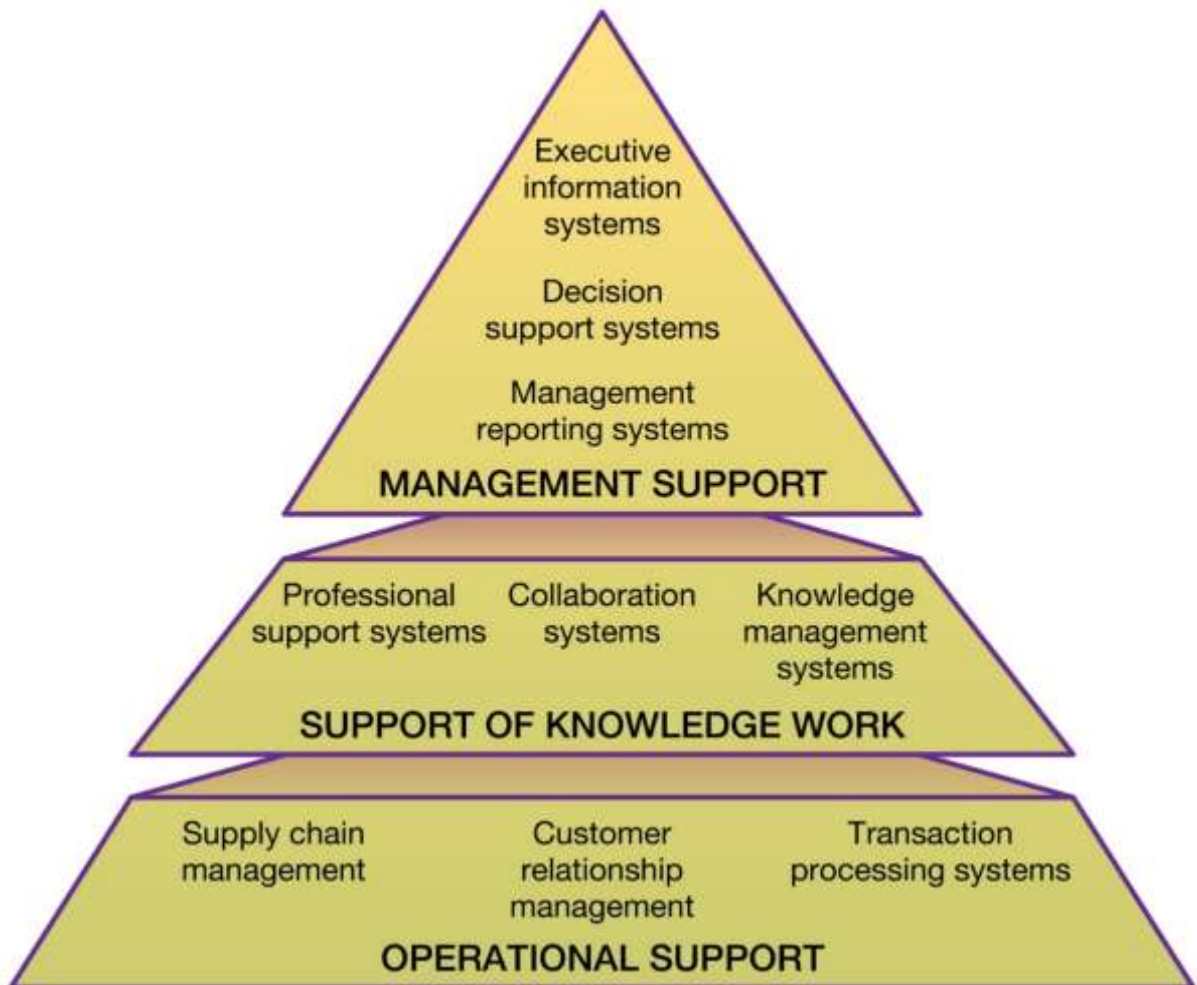

**Figure 2.1** Types of information systems (Zwass 2017)

### 2.2.3. Operational support systems

These systems provide support for routine processes and operations such as inventory management, data entry and retrieval, marketing, distribution and communication. Operational support systems include transaction processing systems, human resource management systems, customer relationship management systems and supply chain management systems. Computer-based information systems such as knowledge support systems are popular in academic institutions where the primary goal is efficient dissemination of educational contents and supporting administrative functions.

### 2.2.4. Learning management systems

Learning management systems (LMS) have emerged as pivotal tools for knowledge management and online learning processes (AbuShawar & Al-Sadi, 2010). LMS, a web-based information system, integrates interactive learning environments with administrative functions (Ifenthaler, 2012). The prevalence of e-learning strategies, adopted by over 40% of global Fortune 500 firms, emphasizes the significance of LMS for employee training and development (Pappas, 2014). The term "Learning Management System" encompasses a virtual learning environment where computers and internet technology facilitate knowledge delivery and communication (AbuShawar & Al-Sadi, 2010). Core to LMS is its role in managing various aspects of courses, including module arrangement, student registration, interaction, task assignment, submissions, and digital result computation (Epignosis LLC, 2014). This technology-assisted teaching backbone plays an indispensable role in modern education.

Simanullang and Rajagukguk (2020) highlight Moodle as a popular LMS application, supporting various online learning activities. From videos to discussion forums, chat, materials, and quizzes, Moodle facilitates interactive engagement. Their study investigates the impact of Moodle-based learning on student activities, emphasizing its suitability for online education. In the context of improving educational systems, Akhmadaliyev (2023) underscores the importance of efficient teaching, particularly in informatics and information technologies. The integration of information provision, education standards, and pedagogical improvements highlights the multifaceted approach taken by governments and institutions. Rabiman, Nurtanto, and Kholifah (2020) emphasize the transformational impact of digital content and Information and Communication Technology (ICT) in education. Their research and development aim at creating an LMS-based E-Learning system that enhances learning experiences. Their findings reveal positive outcomes, showcasing improved learning quality and satisfaction through LMS-based education. The evaluation and selection of Open-Source Software in Learning Management Systems (OSS-LMS) are explored by Abdullateef, Elias, Mohamed, Zaidan, and Zaidan, (2016). Their study presents insights into evaluating and selecting appropriate OSS-LMS packages. They emphasize criteria identification, gap analysis, and multi-criteria decision-making techniques as crucial steps for effective selection.

## 2.3. Instructional Systems Design

Instructional Systems Design is a systematic approach used extensively in the fields of education and training to craft effective learning experiences (Tomei, 2010). Within this framework, various key aspects come into play. Pedagogy guides the overall teaching approach, while educational psychology delves into the intricacies of learning and motivation theories, as well as the diverse spectrum of learning styles. Additionally, instructional systems design embraces a range of instructional design models, providing structured roadmaps for the development of impactful instructional materials and processes. This comprehensive approach ensures that learners' needs are analysed, instructional strategies are well-designed, materials are effectively developed, and instruction is thoughtfully implemented and evaluated for optimal outcomes.

### 2.3.1. Pedagogy

Pedagogy refers to the examination of instructional techniques, encompassing educational objectives and strategies to attain these objectives (Peel, 2017). This area draws extensively from educational psychology, which delves into scientific principles of learning, and to a certain degree, from the philosophy of education, which contemplates the purposes and significance of education from a philosophical stance (Peel, 2017). E-pedagogy is the practice of teaching with technology for active learning that creates participatory, connected and reflective classrooms (Jimoyiannis, Tsiotakis, & Roussinos, 2011). According to Dempster (2012), a technology-assisted learning design includes integration of educational qualities, teaching efficacy, values, learning, and assessment activities. Nine aspects of pedagogy are illustrated in Figure 2.2 and are elucidated as follows:

i.  **Interactive lectures:** intellectually engaging students as active participants in a lecture-based class. Techniques such as role playing, demonstration and think-pair-share are employed to foster engagement and enhance the value of each lecture segments.

ii.  **Assessment:** this provides better insight on students learning experience with prior knowledge.

iii.  **Investigative case:** students explore the science in realistic situations thereby expanding their learning capacity.

iv.  **Strong industry-institute interaction:** students can gain hands-on experience with live cases and experiences in the industry.

v.  **Teaching with the case method:** by engaging in the work of the discipline, students are given the opportunity to go beyond just watching or reading about it. Through this active preparation, they can gain a better understanding of the concepts, techniques, and methods of the discipline and be able to apply them more effectively. Furthermore, this practice will also help them to increase their knowledge and understanding of the vocabulary, theories, and methods discussed in the course.

vi.  **Using media to enhance teaching and learning:** this involves the use of multimedia (videos, documentaries, popular and contemporary music, film and television shows) to drive deep, memorable and meaningful learning experience.

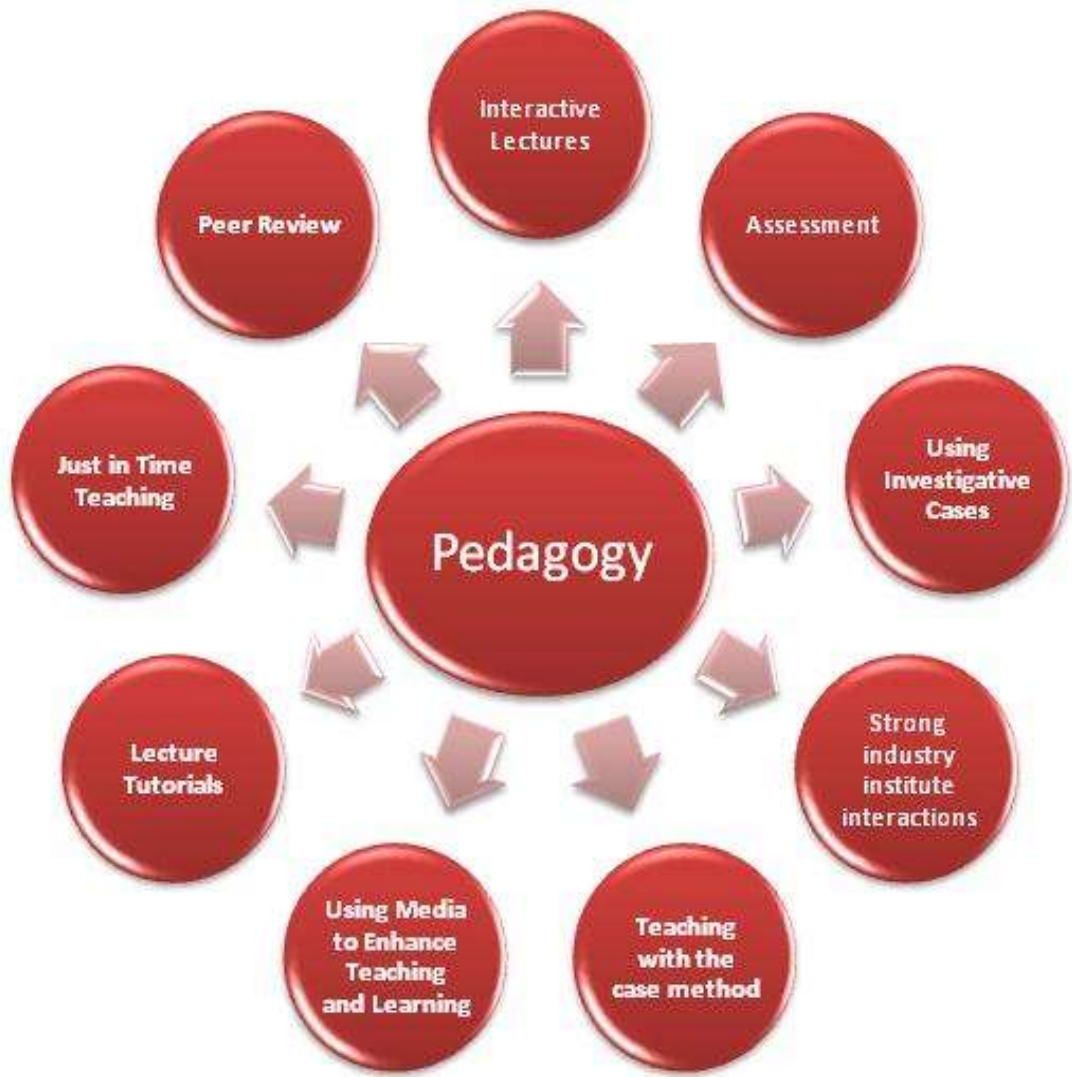

**Figure 2.2** Aspects of pedagogy (SIMCA, 2016)

vii. **Lecture tutorials:** these are short-term worksheets, tasks or activities which students perform and complete in class to improve interactivity of the class session and address common misconceptions and difficulties students encounter in the course.

viii. **Just in time teaching:** this refers to the process of stimulating active and collaborative learning within and outside the classroom. It is achieved by using web-based platforms to deliver questions before a class meeting. The instructor is able to gain insight to the level of students understanding of the course and tailors the class meeting towards meeting students actual learning needs.

ix. **Assessment / peer review:** students are reviewed again by instructor and by other students during interactions, discussions or group activities based on recently acquired knowledge.

## 2.3.2. Educational psychology

Psychology is a field of scientific enquiry that investigates mental states, processes, and behaviours in humans and animals. Educational psychology is a "branch of modern psychology focused on the learning process and psychological problems associated with teaching and training students" (Encyclopaedia Britannica, 2019).

### a. Learning and motivation theories

Within educational psychology, learning theories play a crucial role in understanding how students learn and how teaching strategies can be enhanced. These theories offer theoretical frameworks that shed light on effective approaches to education. Alongside these, motivation stands out as a pivotal element influencing the learning process. Complementing this, motivation emerges as a central factor influencing the learning process, encompassing the mental and emotional drive directing individuals toward their desired objectives (Kim, Song, Lockee, & Burton, 2018). It is the force that fuels engagement and persistence in activities, propelling learners to acquire new skills and knowledge. The significance of motivation in learning is undeniable, serving as the catalyst for proactive participation in the learning journey (Cenić, Petrović, & Cenić, 2019). Conversely, a lack of motivation leads to disengagement from educational opportunities, hindering the learning experience (Park, 2018). This section provides a concise overview of some prominent learning and motivation theories, highlighting their contributions to understanding the intricate relationship between motivation and learning.

i.　　**Behaviourism and reinforcement theory**

Behaviourism, a learning paradigm deeply rooted in the influence of external stimuli, places paramount importance on the interplay between stimuli and responses as the crux of learning (Becker, 2017; Informational Literacy Group, 2018). This perspective asserts that learning is essentially driven by conditioning, a process that links stimuli to corresponding responses. Within the educational context, behaviourists argue that educators act as the primary source of stimuli, and learners associate instructional cues with desired behaviours. Consequently, behaviourism promotes methods such as direct instruction and lecture-based approaches, creating an orchestrated learning environment with specific targeted outcomes.

At the heart of behaviourism lies the concept of reinforcement, a principle seamlessly interwoven with the reinforcement theory (Barnett, 2020). This theory elucidates the art of shaping behaviour by manipulating the consequences that follow it. Analogous to the principle of cause and effect, the reinforcement theory employs rewards or punishments to amplify desired behaviours or deter undesirable ones (Skinner, 1965). Its strategies encompass positive reinforcement, where behaviour repetition is encouraged through favourable outcomes, and negative reinforcement, which alleviates unwanted consequences. Furthermore, punishment curtails undesirable behaviour, while extinction withdraws the favourable outcomes of a behaviour to diminish its occurrence.

ii.　　**Constructivism and self-efficacy theory**

Embedded within the constructivist paradigm is the foundational notion that learners actively construct knowledge rather than passively assimilate information, inherently shaping their own mental frameworks and contributing to the collective pool of knowledge (Center for Educational Innovation, 2020). This pedagogical approach centers on positioning the learner as the focal point of the educational journey, utilizing techniques like project-based tasks and inquiry-driven learning to cultivate deep engagement in the learning process (Center for Educational Innovation, 2020). The malleable nature of this approach accommodates diverse learning settings, ranging from traditional classrooms to outdoor environments, where students find inspiration in the abstract or symbolic intersections with their prior knowledge (Duffy & Raymer, 2010).

Within this constructivist paradigm, Bandura's self-efficacy theory (1977) weaves a crucial thread by underscoring the profound impact of an individual's belief in their own capability to achieve performance goals. Rooted in four key sources—performance accomplishments, vicarious experiences, verbal persuasion, and physiological states—self-efficacy emerges as a potent determinant of one's endeavour. The potency of these sources directly influences the depth of one's self-efficacy, demonstrating the intricate interplay between personal perception and motivation (Carey, 2009). As learners actively engage in constructing their understanding, their confidence in their capabilities enhances the quality of their educational pursuits, elevating their journey towards effective self-directed learning (Bandura, 1977; Carey, 2009).

### iii.      **Social constructivism and self-determination theory**

Social Constructivism views learning as a dynamic and collaborative process in which teachers and learners coalesce to create an environment rich with social interactions and guidance. Teachers thoughtfully design tasks that bridge learners' existing knowledge with novel information, fostering shared experiences that catalyse learning and cognitive maturation (Becker, 2017). This approach, renowned for its emphasis on communal learning, underscores the intricate interplay between educators and learners, contributing to the tapestry of knowledge acquisition (Becker, 2017). Complementing this perspective is the Self-Determination Theory (SDT), a motivational framework that illuminates the intricacies between individuals' volition and the influence of their surroundings. SDT illuminates how environmental factors, encompassing social and cultural elements, intertwine with personal attributes like personality development and self-regulation. A central tenet of SDT is the recognition of individuals' universal psychological needs, which encompass autonomy, competence, and relatedness (Deci & Ryan, 2000; 2008; Kim *et al.,* 2018). As individuals set and pursue goals, the fulfilment of these basic psychological needs becomes a driving force. Autonomy grants control over behaviours and their consequences, competence offers the assurance of adequate abilities, and relatedness fosters a sense of community connection.

iv.    **Liberationism/humanism and expectancy-value theory**

In a Liberationism/Humanism educational framework, the central focus shifts towards students, placing them at the heart of the classroom dynamic. This approach fosters a democratic atmosphere where teachers assume the role of both educators and learners, engaging in mutual discovery with their students. This philosophy employs unconventional examples and creative illustrations to enliven the subject matter, whether through students crafting speeches or partaking in dance performances. This approach emanates from the affirmative belief that learners possess an innate drive for personal growth, development, and active citizenship. Consequently, educators are tasked with affording students opportunities to flourish and evolve (Becker, 2017).

Correspondingly, the Expectancy-Value Theory posits that an individual's behaviour in relation to their achievements is most proximately influenced by their expectations of success and the value they ascribe to a given task (Wigfield & Eccles, 2000). This theoretical framework asserts that individuals make decisions regarding their achievements based on their anticipation of successful outcomes and their subjective assessment of the task's value in a specific domain (Leaper, 2011). As delineated by Studer and Knecht (2016), the expectancy-value theory can be partitioned into four discrete dimensions: attainment value (the significance attributed to achieving high performance), intrinsic value (the gratification derived from task engagement), utility value (the task's relevance for future objectives), and relative cost (the invested resources). By integrating Liberationism/Humanism with the Expectancy-Value Theory, educators can design learning environments that empower students through a student-centred approach while considering individual motivations and perceived task values as integral drivers of academic engagement and success.

Other motivation theories and models include the following:

v.      **Intrinsic & extrinsic motivation**

Intrinsic motivation is a type of motivation driven by an individual's own pleasure, interest or curiosity (Kim, *et. al.,* 2018). The individual performs an action or engage in an activity because of the fun or joy derived rather than some other external reward or benefit. In terms of academic achievement, it has been found that intrinsic motivation is more significant than extrinsic motivation (Deci & Ryan, 2000). Extrinsic motivation is driven by external, environmental factors like punishment, reward and pressure, which are unrelated to the individual. Scanlon, Anderson, and Sweeney (2016) suggest that students can be both intrinsically and extrinsically motivated, while Kim, *et. al.* (2018) and one form of motivation may be more suited for certain scenarios. Considering both motivation types simultaneously could prove to be more beneficial than dealing with only one type of motivation (Cerasoli, Nicklin, & Ford, 2014).

vi.      **Achievement goal theory**

The Achievement goal theory (Elliott & Dweck, 1988) suggests that people can get motivated by their belief or desire to meet specific objectives. It consists of dominant goal types

i.    Mastery / task involved goals which are desires to acquire certain abilities that are needed to perform a task or understand a concept. This leads to self-improvement and competence development.

ii.    Performance / ego involved goals which are desires to show higher accomplishments and achievements than other people. This is exhibited in social comparison (Seifert, 2004).

vii.      **Goal setting theory**

This theory focuses on setting specific and measurable goals. It assumes that an individual is more inclined and motivated to execute a task or embark on an activity if specified measurable targets or goals to be achieved are clearly stated (Locke & Latham, 2006; Young, 2017). The goal setting theory presents five (5) principles for effective goal setting:

i.    Clarity - The aim of a goal should be to be clearly defined and unambiguous.

ii.    Challenging – Goals should be such that they are neither overly simple nor overly hard to achieve.

iii.    Commitment – Stakeholders should have a clear understanding of the goals and be involved in their realization from an early stage.

iv.    Feedback - Provide regular feedback throughout the process to maintain focus on the goal.

v.    Complexity of task – The task should be achievable within a reasonable time frame, divided into smaller goals and checked on regularly.

These motivation theories are key to this research as they help explain how games and game design principles can be used to stir up motivation in learning and other activities.

viii.    **ARCS model**

The ARCS model proposes ways by which student's attention could be gained and sustained. It shows how to emphasize the relevance within the instruction to increase motivation, design challenges that build confidence and satisfaction from achievements. There are four steps in the ARCS instructional design process. These steps are Attention, Relevance, Confidence, and Satisfaction (ARCS), as proposed by John Keller in 1988 (Clark, 2010).

**b. Learning styles**

Ho (2019) described a learning style model as a framework that defines the mechanism by which new skills or information could be learnt. The following is a brief overview of some popular learning styles.

i.    **Ocean model (the big five personality traits)**

The OCEAN model presents five main components of personality: Openness, Consciousness, Extraversion, Agreeableness and Neuroticism (Tremolada, 2015; Belyh, 2019).

i. **Openness** describes people who are insightful and imaginative. Individuals with high openness trait are prone to be more dynamic with a wide range of interest. They can be very adaptive, curious and innovative.

ii. **Consciousness** is a trait that is exhibited by individuals keen on planning, organizing and following laid procedures.

iii. **Extraversion** trait is seen in individuals who are often opportunistic, loving to take advantage of voids and be in the forefront of events.

iv. **Agreeableness** is the personality trait that enables an individual flow easily with most people they encounter. Agreeable individuals are often perceived to be caring, full of empathy and great at paying attention.

v. **Neuroticism** is often exhibited by anxiety, fear, worry loneliness and the likes. Individuals who exhibit this trait hardly deal well with uncertainty and will often need to verify and crosscheck everything for more reassurance.

**ii. Kolb's learning style inventory (KLSI)**

Also known as the "experiential learning theory," the Kolb learning model describes the learning cycle in four stages: Concrete Experience, Reflective Observation, Abstract Conceptualization and Active Experimentation. At the start, the learner is exposed to something either new or familiar. By reflecting upon the experience, they develop a personal understanding of what has transpired. Subsequently, they enter the abstract conceptualization stage, where new ideas are created or current concepts are altered. Finally, the learner puts what they have learned to the test in reality. The outcome of these experiments then restarts the cycle. (Ho, 2019).

Based on this cycle, the current version of Kolb's learning style inventory, the LSI4.0 presents nine types of learners (Kolb & Kolb, 2013)

i. The **Initiating style** is characterized by having the capability to take action to tackle experiences and circumstances.

ii.   The **Experiencing style** is identified by the capability to derive significance from thorough involvement in experience.

iii.   The **Creating style** is identified by the skill to produce significance by observing and reflecting on experiences.

iv.   The **Reflecting style** is identified by the capacity to link experience and concepts through sustained contemplation.

v.   The **Analysing style** is identified by the ability to consolidate and systematise thoughts through contemplation.

vi.   The **Thinking style** is identified by the capacity for structured engagement in abstract reasoning, mathematics, and logic.

vii.   The **Deciding style** is identified by the capability to employ theories and models to resolve problems and pick courses of action.

viii.   The **Acting style** is identified by a strong drive for goal-oriented behaviour that joins people and tasks.

ix.   The **Balancing style** is identified by the ability to adjust adaptively by weighing the pros and cons of acting vs. reflecting and experiencing vs. thinking. (Kolb & Kolb, 2013).

**iii.   VARK learning style model**

VARK, an acronym for Visual, Auditory, Reading/Writing, and Kinaesthetic, is a learning styles which suggests that each learner experiences learning through any one of the VARK processes (Fleming, 1995).

i.   **Visual learners** remember information they see better than information they hear. This could be information could be presented using charts, graphs, diagrams, arrows, illustrations and other visualization including photographs and videos.

ii.   **Auditory learners** excel when information presented to them comes vocally. Such learners would usually rather listen and participate in class discussions than take notes which to them may be distracting.

iii. The **reading / writing learners** demonstrate an intense desire to learn through the written word. This includes being presented with written content during class, such as handouts, notes, or slides and the chance to apply course material in writing assignments.

iv. **Kinaesthetic Learners** excel in hands-on, active and participatory learning where they play an active physical role in the learning process. They might struggle with theoretical courses but excel easily in practical and experimental classes.

The VARK model suggests that learners who have a single mode preference are slow learners while the multimodal learners who can switch between multiple learning styles as the need arises are faster learners (VARK, 2011).

### iv. Herrmann brain dominance instrument

Based on this model, it is possible to determine an individual's learning preferences. The model classifies individuals as theorists (thinkers), organizers, humanitarians, or innovators (Ho, 2019; Boer, Toit, Scheepers, & Bothma, 2013).

i. **Thinkers** prefer to be precise and validate information. They search through resources to identify logical and theoretic rationales.

ii. **Organizers** are better able to process and retain new knowledge if it is organized into a logical structure. They appreciate a systematic, structured way of learning, along with keeping temporal order and having clear instructions and practicality.

iii. **Humanitarians** excel in interpersonal settings, giving them the opportunity to learn from and express emotions, feelings, and ideas. Group activities are often an integral part of their learning experience.

iv. **Innovators** focus on utilizing existing knowledge to broaden their imagination. Their unique style is characterized by spontaneity and originality, emphasizing activities that require problem-solving skills and critical thinking.

**v.    Felder-Silverman learning style model**

The basis of this model rests on the idea that people have their own ways of taking in new information. Some people may have multiple ways they prefer, some may switch between different ways, and some may only have one. The model contains four dichotomies: Sensing-Intuitive (which determines how information is perceived and absorbed), Visual-Verbal (which determines preference of information presentation), Active-Reflective (which determines preferred method of processing information), and Sequential-Global (which determines preferred method of organizing and understanding information) (Felder & Silverman, 1988).

i.     Students who are Sensing are characterized by being concrete, practical, and focusing on facts/procedures.

ii.    Intuitive Students are those who are conceptual, creative and seek out theories and meanings.

iii.   Visual Students prefer to use visuals to aid in their understanding of information e.g. pictures, diagrams and flow charts.

iv.    Verbal Students favour written/spoken explanations over visuals.

v.     Active Students learn best by carrying out activities and working with others.

vi.    Reflective Students learn best through analysing/thinking and prefer to work alone or with familiar partners.

vii.   Sequential Students have an ordered approach to learning, progressing with small incremental steps.

viii.  Global Students take a holistic view of their learning and make large leaps in knowledge. individuals who are concrete, practical and orientated toward facts and procedures (Boer, *et. al.,* 2013; Felder & Silverman, 1988).

**vi.   Myers-Briggs type indicator (MBTI)**

MBTI is a psychometric instrument based on Carl Jung's theory from 1923 that people perceive and make decisions using four psychological functions – sensing, intuiting, feeling, and thinking (Faust, 2019). This is represented by the four dichotomies forming 16 different personality types (The Myers

& Briggs Foundation, 2014b). The MBTI measures the psychological preferences of individuals in how they make decisions and view the world as described below:

i. **Energy** - Extraversion vs. Introversion (E/I): This dimension is about how people direct their energy. Extraverts are more outgoing and interested in seeking out social interaction, while introverts are more inwardly focused and prefer working alone.

ii. **Perception** - Sensing vs. Intuition (S/N): This dimension focuses on how people take in and process information. Sensors rely on facts and details, while intuitives tend to be more abstract thinkers, making connections and seeing patterns that others might not.

iii. **Judgement** - Thinking vs. Feeling (T/F): This dimension is about how people make decisions. Those who favour thinking are more likely to make decisions based on logic, while those who favour feeling make decisions based on their values and emotions.

iv. **Orientation -** Judging vs. Perceiving (J/P): This dimension is about how people approach their outer world. Judgers are more organized and prefer structure and order, while perceivers are more flexible and prefer to keep their options open.

Upon completing a self-administered questionnaire, individuals are assigned a four-letter code that describes their MBTI personality type. Through understanding this code, individuals can gain insight into their preferences, strengths, and weaknesses, and use this information to develop strategies to improve their lives, relationships, and work performance. The MBTI is widely used in educational, business, and counselling settings. Figure 2.3 illustrates the 16 personality types of the MBTI.

The four basic psychological/cognitive functions – sensing (**S**), intuiting (**N**), feeling (**F**), and thinking (**T**) which coordinates information processing and decision making can be paired into what is referred to as function pairs or cognitive core (The Myers & Briggs Foundation, 2014a; Pearman & Albritton, 1994) is as follows:

i. Individuals who demonstrate the Sensing plus Thinking (ST) characteristics are typically objective, analytical, and focused on reality and practical applications. Furthermore, they are often disinterested in nurturing and supporting the growth and development of others. Individuals with the ST personality types are ESTP, ESTJ, ISTP, and ISTJ.

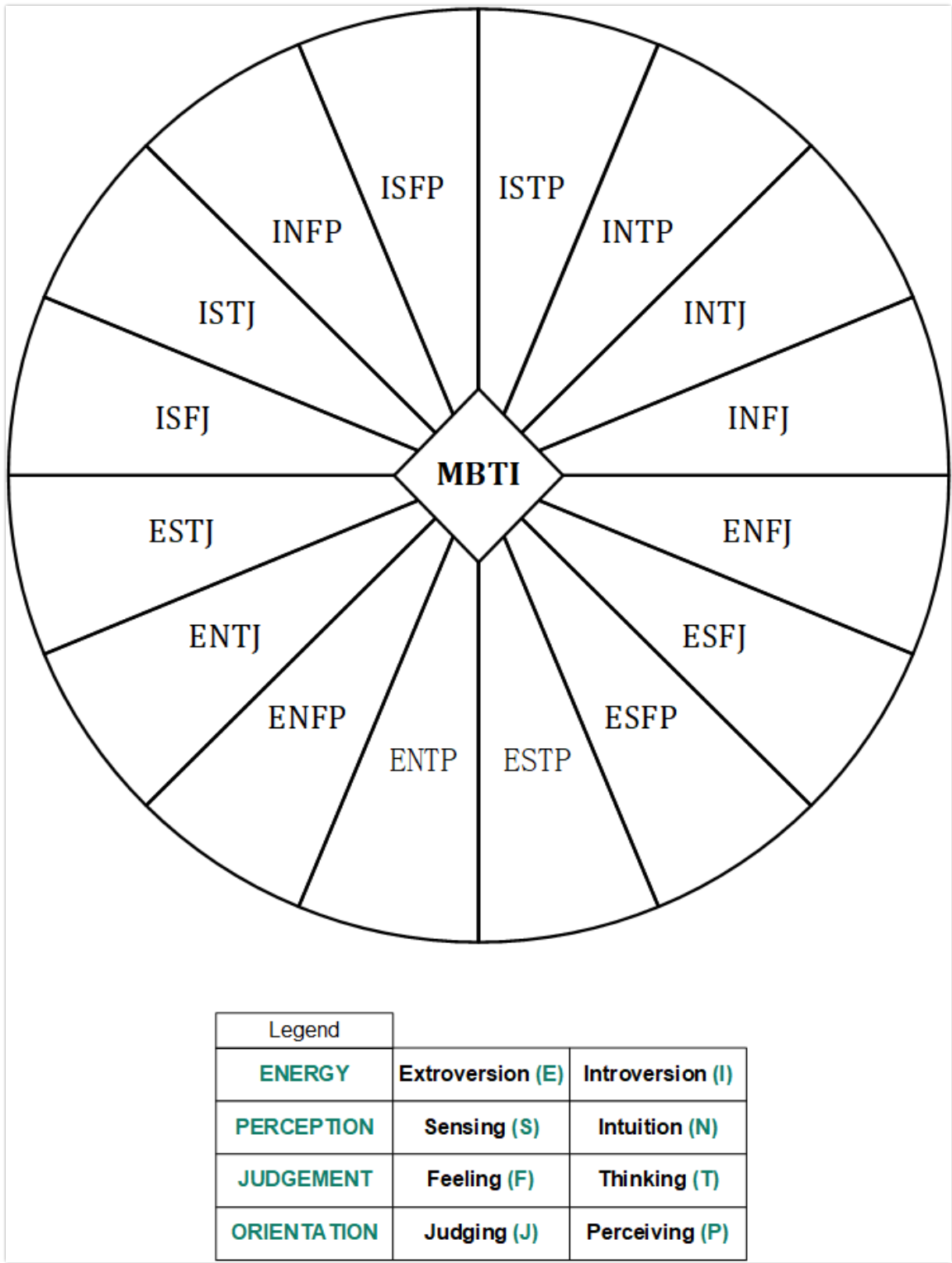

**Figure 2.3** The 16 MBTI Personalities

ii. Those with the Sensing plus Feeling (SF) personality type tend to be warm, people-oriented, and sympathetic. They focus more on the given realities and caring for those around them than analytical approaches or detached ideas. ESFP, ESFJ, ISFP, and ISFJ are all examples of SF personalities.

iii. Those with the NF personality type attribute themselves with warmth and enthusiasm, have good communication abilities, and favour ideas and possibilities over factual data. People of this personality type, which includes ENFP, ENFJ, INFP, and INFJ, often have an understanding of others, a good focus on the abstract, and a distaste for a technical approach to things.

iv. Individuals with Intuition plus Thinking (NT) personality characteristics are often logical and objective thinkers with a knack for being ingenious and focused on technical possibilities. NT personalities tend to be somewhat impersonal and have an analytical attitude towards ideas, information and even people. They typically do not desire warmth or sympathy and favour logical problem-solving over a hands-on approach to helping people. Types of NT personalities include ENTP, ENTJ, INTP, and INTJ.

### c. Learning goals / objectives

In the design of instructional contents, the first step is to always identify specific expected outcomes of the learning process. These outcomes are linked to the learning objectives which are benchmarking statements that are used to assess if a learning has taken place. Learning objectives are what learners are expected to know upon completing a learning process. They are used to create syllabi and course material and assess how effective the learning process was (Gogus, 2002).

### 2.3.3. Instructional design

Instructional design involves analysing learning needs, systematically developing learning experiences, and combining the art of creating engaging interactions with the science of brain function (Kearsley & Culatta, 2018). This process is used to enhance instruction or teaching and is defined by Goethe (2019) as the steps required to create and implement instructional content. According to Dick, Carey and Carey (2015), the instructor, learners, materials, instructional activities, delivery system, and learning environments interconnect and collaborate to achieve desired student learning outcomes. Hence, an adjustment in one element can send ripples through others, influencing the final learning results. Consequently, neglecting any component's conditions can jeopardize the entire instructional process, hence, the need for instructional design models.

### a. Instructional design models

In order to craft an efficient learning experience, a well-defined instructional design model or process should be followed (May, 2018). A typical instructional design process at its most basic form, will have three components: instructional analysis, instructional strategy and evaluation (Smith & Ragan, 2005). There are several instructional design models and processes, some of which are discussed as follows:

### i. ADDIE model

This is general-purpose process has been adopted traditionally by instructional designers (Kearsley & Culatta, 2018). The five phases of ADDIE are Analysis, Design, Development, Implementation and Evaluation – represent a dynamic framework for the formation effective teaching and training. According to Branch (2009), the application of ADDIE follows an educational philosophy that encourages student-centred learning, encourages the use of creative and original approaches, requires authentic activities, and inspires learners.

**ii.  Systems approach model**

The Dick and Carey instructional design model, also known as the systems approach model, involves a systematic nine-step process to create effective instructional materials that promote learning outcomes (Dick, Carey, & Carey, 2015). The steps within the systems approach model are briefly highlighted as follows:

1.  Identify Instructional Goal(s): Define the new skills and knowledge learners should acquire through the instruction. Goals can be derived from various sources such as needs assessments, performance analyses, or practical experience.

2.  Conduct Instructional Analysis: Break down the instructional goal into specific steps, subskills, and entry skills. Determine what learners are expected to do and identify the skills, knowledge, and attitudes (entry skills) required for successful learning.

3.  Analyse Learners and Contexts: Understand the learners' characteristics, preferences, and current skills. Also, analyse the learning environment and the context in which the skills will be applied.

4.  Write Performance Objectives: Create specific, measurable statements outlining what learners will achieve upon completing the instruction. Objectives define the skills, conditions, and criteria for successful performance.

5.  Develop Assessment Instruments: Design assessments aligned with the performance objectives. These assessments measure learners' ability to perform the skills outlined in the objectives and can include various types of assessments such as tests, performances, and portfolios.

6.  Develop Instructional Strategy: Design a theoretical framework for instruction based on current learning theories. Include components like pre-instructional activities, content presentation, active participation, practice with feedback, and follow-through activities to assess learning and real-world application.

7.      Develop and Select Instructional Materials: Create or select instructional materials that align with the instructional strategy. These materials could include guides, multimedia resources, presentations, and other learning resources.

8.      Design and Conduct Formative Evaluation: Test the instruction through evaluations that help identify issues and opportunities for improvement. Different types of evaluations, such as one-to-one, small-group, and field trial evaluations, provide insights for refining the instruction.

9.      Revise Instruction: Use data from formative evaluations to revise and improve the instruction. Continuously reassess assumptions, decisions, objectives, and instructional strategy based on feedback and information gathered throughout the process.

Summative evaluation concludes the Dick and Carey Model, assessing overall instruction effectiveness. It ensures alignment with goals, knowledge transfer, and learner impact. Independent evaluation addresses evolving challenges, like cross-context efficacy and e-learning. An integral step in the dynamic instructional landscape.

### iii.   4C/ID model

The 4C/ID model proposed by Jeron J. G. van Merriënboer in 1997 (van Merriënboer, 2020) stands for Four Component Instructional Design. It is a design tool instructional designers employed in development of educational programs aimed at professional competencies and teaching complex skills (van Merriënboer, 2019). The 4C/ID is based on the assumption that learning can be described in terms of four basic components which are ordered sequentially according to difficulty. The four components of the model are: learning tasks, supportive information, procedural information, and part-task practice.

### iv.   5E learning circle

This model, according to Duram and Duram (2004), is based on constructivist learning theory, cognitive psychology and best practices in science education. It focuses on supporting comprehension of concepts over time by learners through five phases. These phases include engage, explore, explain, elaborate and lastly, evaluate. Through this model, learners are able to redefine, reorganize, elaborate and transform the initial concept through self-reflection and interacting with the environment and peers (Duran & Duran, 2004).

**v.    Instructional design for teachers (ID4T)**

ID4T involves creating classroom instruction using a systematic process: setting goals, defining objectives, analysing students, designing tests, selecting materials, creating activities, choosing media, implementing, and revising lessons (Carr-Chellman, 2016). It empowers teachers to create effective and engaging learning experiences.

**vi.   Gagné's 9 events of instruction**

This model consists of nine (9) step processes (events) of instruction. Gagne's 9 events of instruction provide a systematic framework to developing effective strategies and creating activities for instructional classes. Each event in the model addresses a form of communication in favour of the learning process, promoting engagement of students and comprehension of contents (The Peak Performance Center, 2020). The nine events are as follows: Receiving attention (reception), informing students of the goal (expectancy), triggering recollection of prior knowledge (retrieval), displaying the stimulus (selective perception), supplying learning support (semantic encoding), inciting action (responding), giving feedback (reinforcement), evaluating performance (retrieval), and enhancing both retention and application (generalization).

**vii.   ASSURE instructional model**

The ASSURE Instructional Model is a well-established framework designed to effectively integrate multimedia technology into the learning environment (Bajracharya, 2019). This model consists of six key steps, each represented by a letter in the acronym "ASSURE." The steps are as follows:

i.    Analyse Learners: In this initial phase, instructors carefully examine the characteristics and needs of the learners. This analysis helps tailor the instructional approach to suit the specific audience.

ii.   State Objectives: Clear and measurable objectives are established to define what the learners should be able to accomplish by the end of the instruction. These objectives guide the design and delivery of the learning experience.

iii.    Select Methods, Media, and Materials: The appropriate instructional methods, multimedia tools, and materials are chosen to support the learning objectives. This step involves identifying the most suitable resources that align with the desired outcomes.

iv.    Utilize Methods, Media, and Materials: In this stage, the selected methods, media, and materials are implemented in the instructional process. Educators effectively utilize the chosen resources to facilitate the learning experience and engage the learners.

v.    Require Learner Participation: Learner engagement and active participation are essential components of the ASSURE model. Instructors create opportunities for learners to actively engage with the content, encouraging their involvement and enhancing their understanding.

Evaluate and Revise: Evaluation is a crucial step in the ASSURE model. Educators assess the effectiveness of the instruction and gather feedback from learners. Based on the evaluation results, revisions are made to improve the instructional design and ensure continuous improvement.

**b.    Technology mix in instructional design**

The use of information technology in implementing instructional designs could take several dimensions, some of which are highlighted below.

i.    **Simulated learning**: involves students exploring new concepts in a controlled, yet realistic environment, wherein they practice particular behaviours and experience the effects of their choices (Schmucker, 1999). These simulations are generally rooted in underlying models of the environment, phenomena or experience, with an element of uncertainty or unpredictability (Goethe, 2019).

ii.    **Mobile learning**: the increased usage of wireless technology, such as smartphones, has revolutionized the way people obtain information and acquire knowledge (Kim, Rueckert, Kim, & Seo, 2013). The emergence of mobile learning has extended the traditional model of learning to an era of engaging with mobile devices in the learning process (Su & Cheng, 2013).

iii. **Game-based learning**: is an educational tool that utilizes play in order to improve student engagement in the learning process. Digital games are often developed to cover a particular academic subject matter and may be used as a pedagogical tool for blended learning systems (Bychkov, Zabrodina, Netesova, & Mapelli, 2018).

iv. **Gamified learning**: is a concept that applies game design principles and elements to the entire learning process, rather than the game itself being used solely for learning (Becker, 2017). A similar yet distinct concept is that of a serious game. These are games created to offer users stimulating and enjoyable experiences, interactive learning environments and group learning activities as opposed to simply providing amusement (Anastasiadis, Lampropoulos, & Siakas, 2018).

## 2.4. Human Computer Interaction (HCI)

The fields of information systems and human-computer interaction (HCI) are interconnected by their shared focus on effective system design. This encompasses recognizing both the human and technical aspects of crafting and assessing systems for work and leisure purposes (Balbo, Bentley, & Collings, 2006). The designer adopting an HCI perspective places emphasis on optimizing the system's ease of use, enabling users to seamlessly carry out tasks and accomplish objectives within the system. The overarching goal is to establish systems that possess an innate level of intuitiveness, allowing users, even those with limited external guidance, to navigate and engage (Culatta, 2020). HCI is relevant in implementing e-learning systems because it ensures users are at the receiving end of the best experience the system can afford. An efficient pedagogy implemented together with a memorable user experience goes a long way in improving learner motivation.

### 2.4.1. User experience

User experience as a field is concerned with the creation and shaping of experience through technology (Interaction Design Foundation, 2014). Borne out of HCI, user experience has continually evolved, keeping up with new trends in the field of HCI design such as voice, touch (Nichols & Chesnut, 2014) and game-like interaction. User experience design is built on a wide array of disciplinary, methodological, and conceptual points of view, making it valuable in systems development (Wright, Blythe, & McCarthy, 2005).

### 2.4.2. Usability and user interface design

McCain, Ntuen, and Park (1996) described usability is a cornerstone of effective human-computer interaction. It involves designing interfaces that are intuitive, efficient, and user-friendly, enabling learners to navigate e-learning platforms with ease. Central to usability, user interface design prioritizes the development of visually captivating arrangements and interaction models that amplify engagement and comprehension (Oppermann, 2002). Through the prioritization of user-centricity in the design of e-learning systems, educators can cultivate an enriching and gratifying learning encounter.

### 2.4.3. Emotional design and aesthetics

Emotional design recognizes that user experiences are deeply influenced by emotional responses (Wang & Yuan, 2018). Aesthetics, including visual and interactive design elements, play a pivotal role in shaping users' emotional engagement with e-learning platforms (Faisal, Fernandez-Lanvin, Andrés, & Gonzalez-Rodriguez, 2020). Incorporating emotionally resonant design elements can positively impact motivation, satisfaction, and overall learning outcomes (Wong & Adesope, 2021).

### 2.4.4. User behaviour modelling

User behaviour modelling involves the analysis and prediction of how individuals interact with systems, such as e-learning platforms, to anticipate their actions, preferences, and responses (Han, Hao, & Liu, 2012; Kompan, Kassák, & Bieliková, 2017). By examining historical data and patterns, system designers can gain insights into learners' tendencies, enabling the design of tailored experiences that cater to specific needs (Shen, Liang, Law, Hemberg, & O'Reilly, 2020). This approach enables proactive adaptation of content, interactions, and interventions, fostering an environment conducive to optimized learning outcomes.

## 2.5. Gamification

Gamification has garnered several definitions from various scholars. For example, Zichermann and Cunningham (2011) described it as the process of using game-thinking and game mechanics to engage users and solve problems. Witt, Scheiner, and Robra-Bissantz (2011) defined gamification as the integration of principles and mechanics of games, like points, levels, and leader boards, in a serious context while Liu, Alexandrova and Nakajima, (2011) asserts that non-gaming systems are "gamified" when game mechanics are added to their main functions. In this thesis, gamification is considered to be a strategy for enhancing user experience by introducing carefully chosen game attributes.

Gamification seeks to incentivize non-gaming systems and foster a sense of Gameful behaviour in users by incorporating game elements such as badges, points, and the freedom to fail and retry in order to solve specific problems. While it is not on the same scale as serious games, it is still a series of relevant activities and procedures that are used for a particular purpose, which outweighs merely using game elements without any goal (Marache-Francisco & Brangier, 2014). There are some essential concepts to consider in gamifying systems. These concepts are considered in the following sections:

### 2.5.1. Engagement and flow

Engagement refers to the heightened simultaneous experience of immersion, interest and enjoyment (Kim, *et. al.,* 2018) in an activity or a connection (Zichermann & Cunningham, 2011) between a user and a system. Flow on the other hand is a psychological state of optimum experience described by being fully concentrated and engaged in an activity (Hamari, 2014; Nakamura & Csikszentmihalyi, 2014). The flow and engagement a student would experience is determined by the student's current skill level and the difficulty of the challenge the student is exposed to. As illustrated in Figure 2.4, different skill levels and challenge / difficulty levels result in different psychological states. In order to attain flow, the learners' skill level and difficulty of task must be within the student's ability.

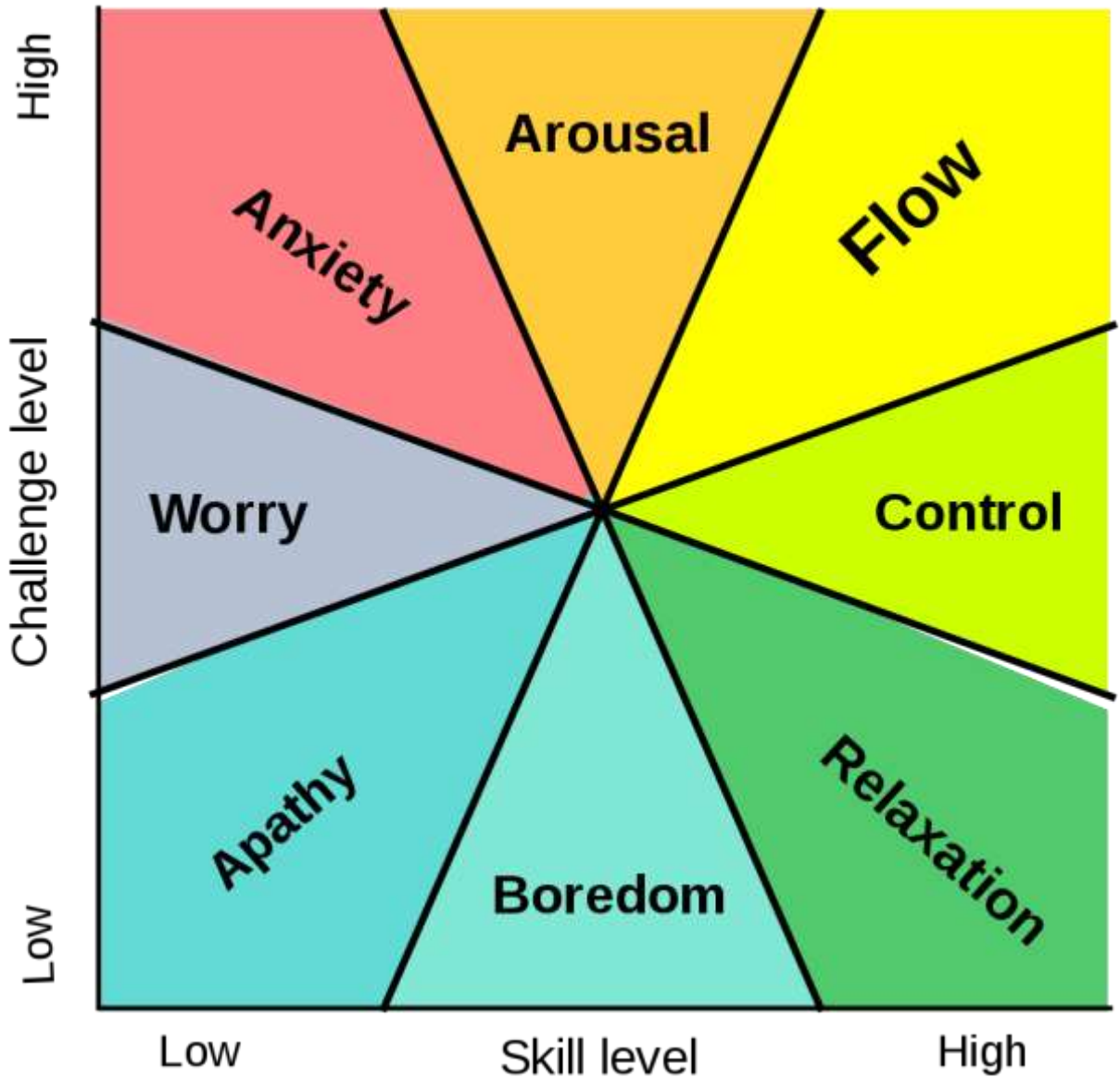

**Figure 2.4** Flow state (Nakamura & Csikszentmihalyi, 2014)

### 2.5.2. Game thinking

A game is an activity embarked upon for fun which is often guided by well-defined rules for actions and repercussion. As a teaching and learning tool, gaming usually acts as a learning trigger inducing lively discussion on learning concepts among users during or after the gameplay as they enliven topics and are peculiarly effective for dealing with problem-solving and key concepts (Goethe, 2019).

The concept of game thinking is a user experience design process by which game design methods are adopted in the planning, designing and implementation of digital educational materials and systems to introduce a game-like experience. By breaking down and reengineering the learning process and experience with game artifacts, learners are encouraged to problem solve and sustain their interest from armature to expert level while tackling sizable challenges in manageable steps. Additionally, diverse interests and skillsets are supported and a confident, optimistic attitude is cultivated (Werbach & Hunter, 2012).

### 2.5.3. Game elements / artifacts

These are the various individual gaming artifacts that can be integrated into a process or system to make it have game properties. Marczewski (2014) proposed a hierarchical framework for categorizing game element as illustrated in Figure 2.5.

1. **Mechanics**: A set of rules administered by the designer that define what can be done within the system.

2. **Schedules**: sets of rules that dictate when and how certain activities take place within the system. These rules may include the number of points or scores that must be earned to move up to the next level, when awards, badges, or trophies will be given, or what combinations of factors must come together to reach a desired outcome.

3. **Dynamics**: How the mechanics and the user interact in real-time. They often can lead to unexpected / emergent outcomes.

4. **Feedback**: Results of events occurring in the system are often represented by feedback artifacts

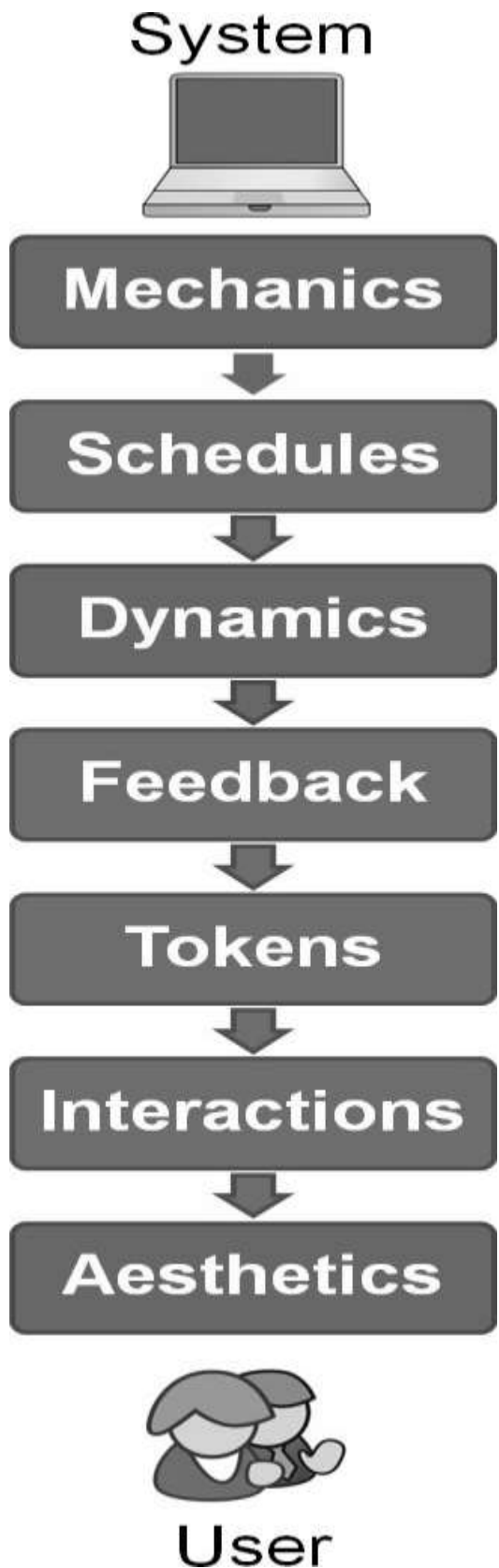

**Figure 2.5** Framework of game elements (Marczewski, 2014)

such as points, badges, progress bars and messages.

5. **Tokens**: These can include virtual items such as points, rewards, and collectables such as Easter eggs, certificates, and coupons.

6. **Interactions**: The user can interact with the system through clicks of a mouse, pressing of keys, gestures, or commands spoken aloud.

7. **Aesthetics**: How the user emotionally responds to the system, such as joy, fear, and frustration.

i. **Taxonomy of game elements**

Fundamentally, all games comprise of a player or players, objectives, a system of rules, and feedback when viewed from a high level. However, when broken down to smaller units, these have manifested in a number of game elements which according to the gamification taxonomy presented by Toda *et. al*., (2019), have been classified under the following categories illustrated in figure 2.6:

a. **Ecological:** These are elements that relate to the environment where the gamification is being implemented. The elements in this dimension include Chance (randomness, luck, fortune or probability), Alternate paths / Imposed Choice, Economy (transactions, market, exchange), Rarity (limited items, collectibles, exclusivity) and Time Pressure (timed events and clock counts).

b. **Social**: These elements are used to foster interactions among the members of the e-learning system. The elements in this dimension include Competition (conflict, leaderboards, scoreboards, player vs player contests), Cooperation (teamwork, co-op, groups), Reputation (classification or status), and Social Pressure (peer pressure or guild missions).

c. **Personal**: These elements provide meaning to learners within learning environment by providing new information, changes to the system and challenges. The elements in this dimension includes Sensation (haptic, visual or audio stimulation), Objectives (missions, side-quests, milestones), Puzzle (challenges, cognitive tasks, actual puzzles), Novelty (updates and changes), and Renovation (boosts, extra life, renewal).

d. **Fictional**: These elements provide meaning and context to events and activities within the

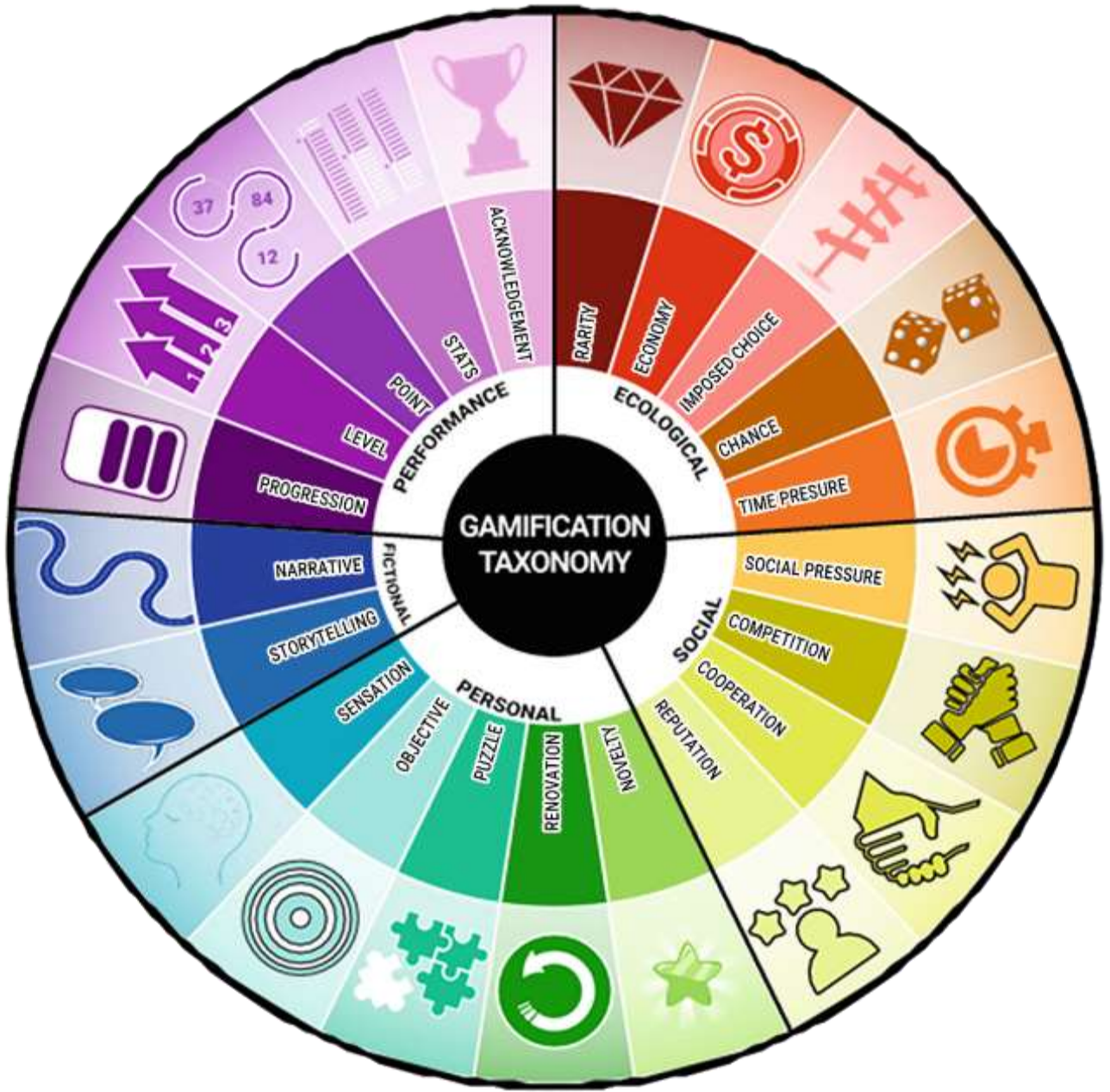

**Figure 2.6** Gamification taxonomy (Todo *et. al*., 2019)

environment. They include storylines (where learners follow a predefined script of events and activities) and narratives (where learners actions and decisions have an impact on future events).

e. **Performance**: These are elements provide feedback to learners in response to events and activities going on within the environment. The elements in this dimension include Point (scores, experience points, skill points), Progression (progress bars, steps, maps), Levels, Performance Statistics, and Acknowledgement (badges, medals, trophies and achievement cards).

## ii. Benefits of gamification

Gamification should not be used as a way to take advantage of people, and instead be utilized to create genuine satisfaction and help them reach their goals. Werbach and Hunter (2012) suggest this strategy as a way to inspire people to take part in tasks they may not have wanted to partake in before. Kim, *et. al.,* (2018) highlited some benefits of gamification in a learning environment. They include:

1. Stimulated student engagement and enthusiasm.
2. Enhanced learning performance and academic success.
3. Boosted recollection and sustained memory.
4. Instantaneous feedback on students' development and endeavour.
5. Incentivized behavioural transformation.
6. Allowed students to observe their advancement.
7. Increase in collaborative capacities.

## 2.6. Unified Modelling Language (UML)

UML is a visual language used by software engineers to design systems, with the goal of providing a standardized way to communicate the design of complex systems. UML is used to model the structure and behaviour of a system by using a collection of diagrams. The diagrams are used to represent the different elements of a system, such as the classes, objects, relationships, activities, components, and interactions between different elements. UML provides a visual language for modelling and documenting software systems. UML diagrams can be used to model any type of system, from a simple system to a complex distributed system (Wazlawick, 2014). Use cases are useful for outlining the general goals of the system. Activity diagrams can be used to visualize the steps a user will take when interacting with the system. Class diagrams can be used to illustrate the different classes that make up the system, their relationships, and the tasks each class must be able to perform. Together, these diagrams can create a comprehensive model of a gamified personalized e-learning system.

### 2.7. Set Theory

A set is a collection of all elements that satisfy a certain given property (Jech, 1971). To mildly state, everything that can be categorized can be described in terms of set.

The intersection of two sets A and B, denoted by A∩B, is a set with all of their common elements. A union of A and B, denoted by A∪B, is a set comprising the elements of both A and B. The complement of A, denoted by $A^C$, is the set of elements that are in the universal set U but not in A. The null set, denoted by ∅, is a set with no elements at all.

Other symbolic representations of sets include:

    i.      Subsets A ⊆ B

Subset has few or all elements equal to the set as in {1, 2} ⊆ {1, 2, 3, 4}

    ii.     Not subset A ⊄ B.

Left set is not a subset of right set as in {1, 2} ⊄ B'

### Functions

F is a mapping from a set X to a set Y that assigns a unique element f(x) ∈ Y to each x ∈ X. This relationship is also referred to as a function, map, or transformation. The domain of f is the set X, and the codomain, which is the set Y in which f(x) ∈, is the set Y. To illustrate the mapping of x to f(x), we can write f : x → f(x) (Hunter, 2014).

A function is a "map" that takes an element from the domain and assigns it to a corresponding element in the codomain. This can be seen graphically as an arrow pointing from the domain to the codomain (Figure 2.7). This mapping can be used to show the complex relationships such as personalities and game elements, enabling system designers to determine which elements of the game will best suit different types of users. For example, designers can consider difficulty level, customization options, and rewards structure when creating a gamification strategy, in order to make an experience that appeals to a variety of personalities.

## 2.8. Related Work

This section contains an overview of some relevant literature that delves into the field of personalized gamification in the context of e-learning. It explores approaches and methodologies, as well as highlighting the benefits, drawbacks, and key considerations associated with the utilization of personalized gamification in the field of e-learning.

In the landscape of e-learning gamification research, various methodologies have been employed to investigate the effectiveness of personalized strategies, game design elements, learning style preferences, and user-centered approaches. Each methodology offers distinct insights into the intricate relationship between gamification and e-learning outcomes, shedding light on the most effective ways to engage learners and enhance their educational experiences.

Experimental designs have played a pivotal role in assessing the impact of gamification on e-learning outcomes. Abbasi, Montazer, Ghrobani, and Alipour (2021) used a pretest-posttest experimental design to evaluate the effects of personalized gamification on learners' motivation and personality traits. This methodological choice allowed for a controlled comparison of gamification variants, showcasing the potential of personalized approaches to enhance e-learning effectiveness. A similar experimental approach was utilized by Aljabali, et. al., (2020) where personalized gamified learning based on learning styles led to improved student scores and perceived usefulness. Personalized strategies have also gained attention, with researchers like El Bachari, Abelwahed, and El Adnani (2011) emphasizing personalization based on dynamic learners' preferences, using the dominant MBTI cognitive functions to design distinct learning strategies and media formats. Adewale, Agbonifo & Osajiuba (2019) further incorporated the MBTI to tailor e-learning experiences to individual learners' cognitive functions. This approach acknowledged the fluidity of individual differences in learning preferences and aimed to improve user-friendly experiences. Meanwhile, Amirhosseini & Kazemian (2020) leveraged machine learning algorithms to predict personality types from social media data, illustrating the potential of tailoring gamification through data-driven insights.

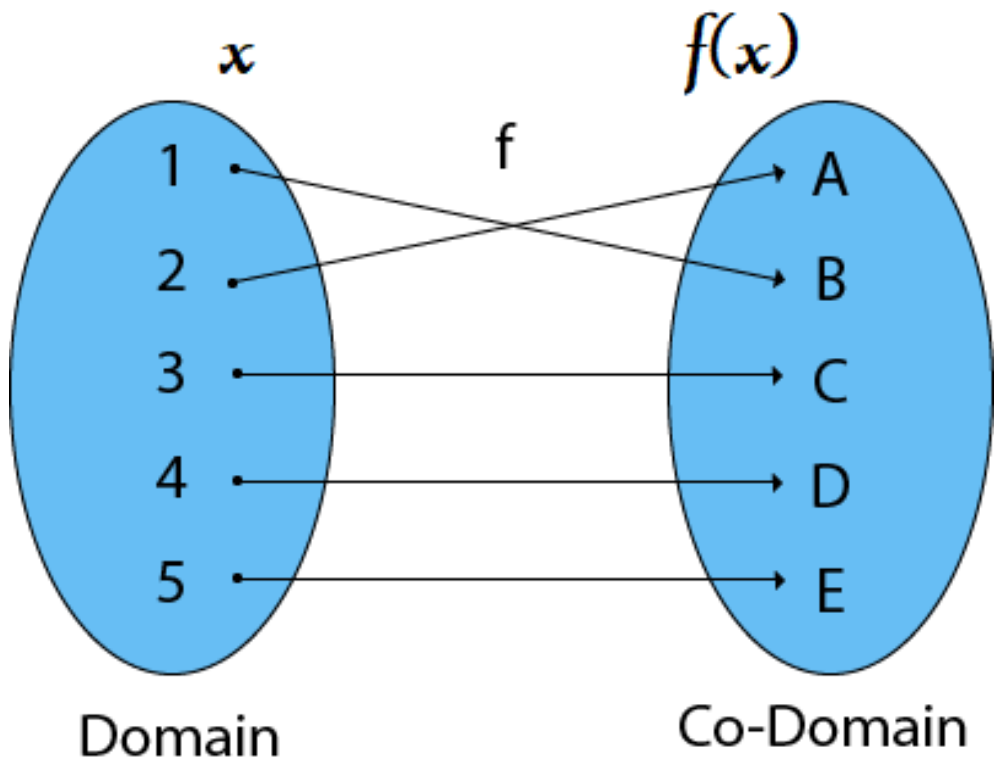

Figure 2.7 Mapping function

Conceptual frameworks have been instrumental in understanding the relationship between game design elements, user experience, and learning outcomes. Alexiou & Schippers (2018) undertook a comprehensive literature review to construct a framework that delineates the connection between these elements. This methodology offered a holistic perspective on how game design shapes user engagement and learning outcomes, guiding the development of more effective gamified educational environments. User-centered approaches, such as those employed by Santoso, Puntra & Hendra (2021), emphasize the importance of designing e-learning modules that align with learners' preferences. Through usability tests and surveys, this methodology assesses the effectiveness of the module in meeting learners' visual and global learning style preferences. This user-centric approach reflects a deeper understanding of learners' needs and preferences, ultimately enhancing engagement and satisfaction.

Iterative development models, as showcased by Kamunya, Mirirti, Oboko, & Maina, (2020) underscore the dynamic nature of gamification strategy development. Through Design Science Research Methodology (DSRM), the researchers developed and tested an adaptive gamification model in iterative cycles. This methodology highlights the need for continuous refinement and adjustment to ensure the effectiveness of gamified interventions over time. These methodological choices collectively contribute to the comprehensive understanding of the interplay between gamification and e-learning outcomes. Experimental designs offer empirical evidence, personality-based approaches cater to individual differences, conceptual frameworks provide theoretical insights, user-centered methodologies prioritize learner preferences, and iterative development models ensure continuous improvement. By employing these diverse methodologies, the researchers enrich the discourse on gamification's potential to transform the landscape of e-learning.

Comparing the outcomes of various studies have light on the multifaceted impact of gamification on e-learning, revealing both its potential and limitations across different methodologies. Several studies emphasize the potency of personalized gamification strategies. Abbasi et al. (2021) and Aljabali et al. (2020) both show improved learning outcomes and increased engagement through personalized approaches. The former highlights the significance of considering learners' motivation and personality, while the latter focuses on learning style personalization. These findings underscore the efficacy of aligning gamification with individual learner attributes. Addressing user experience and usability as in the research by Adewale et. al., (2019) highlights the importance of user-friendliness and ease of navigation in e-learning systems. Santoso et. al., (2021) contribute by demonstrating that catering to visual and global learning preferences enhances learner satisfaction and usability. The study by Hurpur & de Villiers (2015) adds a framework for evaluating mobile learning environments, emphasizing usability, user experience, and educational features. These studies collectively underscore the need for user-centric design and assessment in gamified e-learning.

Machine learning approaches, like that of Amirhosseini & Kazemian (2020), offer insights into predicting learner personality types based on textual data. This introduces the potential for dynamic customization of gamified elements based on learners' psychological traits. Such approaches highlight the integration of advanced technologies in enhancing the efficacy of gamification. Incorporating learning style models, studies such as El Bachari et al. (2011) and Hassan et al. (2019) accentuate the importance of aligning gamification elements with learners' cognitive preferences. The researchers further demonstrate that adaptive gamification can enhance student motivation and interaction, suggesting a pathway to keep learners engaged through tailored approaches. Conceptual frameworks, as shown in Alexiou & Schippers (2018) and Piteira, Carlos & Aparicio (2017), offer theoretical perspectives on how game design and personalization influence learning outcomes. While these frameworks provide valuable insights, empirical testing is required to validate their effectiveness in diverse e-learning contexts.

Notably, studies like Çelik (2015) and Ofosu-Asare, Essel & Bonsu (2019) point to the importance of considering pedagogical effectiveness alongside usability. While usability is crucial, it doesn't necessarily guarantee improved learning outcomes. Thus, an integrated approach that aligns usability with learning effectiveness becomes vital. Overall, a comprehensive understanding emerges when these studies are considered collectively. Gamification's potential lies in personalized approaches, user-centered design, incorporation of learning styles, and the integration of advanced technologies. However, further research and empirical validation are required to fully unlock the benefits of gamification in diverse e-learning environments.

The gamification research in e-learning as highlighted above, holds promising potential to enhance the educational experience. Yet it also poses certain considerations and potential downsides. Research findings shed light on various dimensions of this approach, providing insights into its impact on learners and learning outcomes.

Numerous studies highlight the upsides of personalized gamification strategies. Abbasi et al. (2021) and Aljabali et al. (2020) underscore that personalized approaches significantly improve student performance and engagement. By tailoring gamified elements to learners' motivation, personality, and learning styles, these studies reveal how customized interventions can amplify the effectiveness of e-learning. However, it's important to approach gamification with a critical lens. The studies by Alexiou & Schippers (2018) and Freitas (2021) emphasize the need for thorough evaluation and testing. While the former presents a conceptual framework linking game design to user experience and learning outcomes, the latter highlights that the impact of gamification can be influenced by various factors like course design, content quality, and instructor engagement. User-centered design is another facet to consider. Santoso et. al., (2021) show that addressing learners' visual and global preferences through a well-designed e-learning module enhances satisfaction and usability. On the other hand, Hurpur & de Villiers (2015) demonstrate the significance of evaluating usability, user experience, and educational features in mobile learning environments to ensure a seamless learning experience.

While gamification has the potential to motivate learners, certain downsides must also be acknowledged. Kang and Kusuma (2020) note that some personality traits may not align well with gamified elements, potentially hindering engagement. Hassan et al. (2019) reveal that without adequate feedback mechanisms, some learners may not be fully engaged with gamified systems. It is worth noting that not all learners respond equally to gamification. For instance, Zaric, Roepke, Lukarov, and Schroeder, (2021) emphasized that gamification's impact depends on learners' individual tendencies and behaviors. Similarly, the study by Çelik (2015) underscores that usability doesn't guarantee learning effectiveness, implying that a balance between usability and pedagogical principles is vital.

In conclusion, the reviewed literature underscores the crucial importance of addressing the challenge of low motivation and engagement in e-learning through innovative approaches. The current state of research highlights the potential of gamification as a strategy to enhance learner motivation and engagement, while also revealing the complexity of its effects across different contexts and learner profiles. The notion of incorporating gamification in personalized e-learning using personality-based approach has emerged as a promising solution, aligning with the personalized learning paradigm that tailors educational experiences to individual preferences. This aligns seamlessly with the identified research gap – the development of a gamification model that harnesses learners' personalities to create a personalized e-learning experience. By delving deeper into the methodologies, approaches, outcomes, and considerations explored in the literature, this study seeks to contribute to the academic discourse surrounding personalized gamification, ultimately aiming to foster enhanced motivation, engagement, and academic achievements in e-learning platforms. This endeavour represents a significant step toward creating a more rewarding and effective educational journey for learners in the digital age. A literature review table is available in appendix i.

## CHAPTER THREE

## METHODOLOGY

### 3.1. Introduction

This chapter introduces a conceptual framework encompassing key concepts that underpin the research. The subsequent discussions delve into the methods, strategies, and tools employed to achieve the research objectives. This encompasses explaining the process of model development and the creation of the corresponding implementation system. The chapter culminates by examining the evaluation methodologies employed in the study.

### 3.2. Conceptual Framework

The research explored the intersection of multiple components in order to create a personalized e-learning system that enhances the learning experience. The research framework, which is shown in Figure 3.1, is composed of the following components:

1. Learning Styles: This component considers MBTI personality model in detail to identify learning styles that can be used to create novel gamified and personalized e-learning experiences.

2. Game Elements: This component will explore how game elements can be incorporated into the e-learning system to create engaging experiences for learners.

3. ADDIE ID Model: This component uses ADDIE to guide the design and development of the learning system.

4. RUP Methodology: This component will explore how RUP methodology can be used to develop a gamified personalized e-learning system.

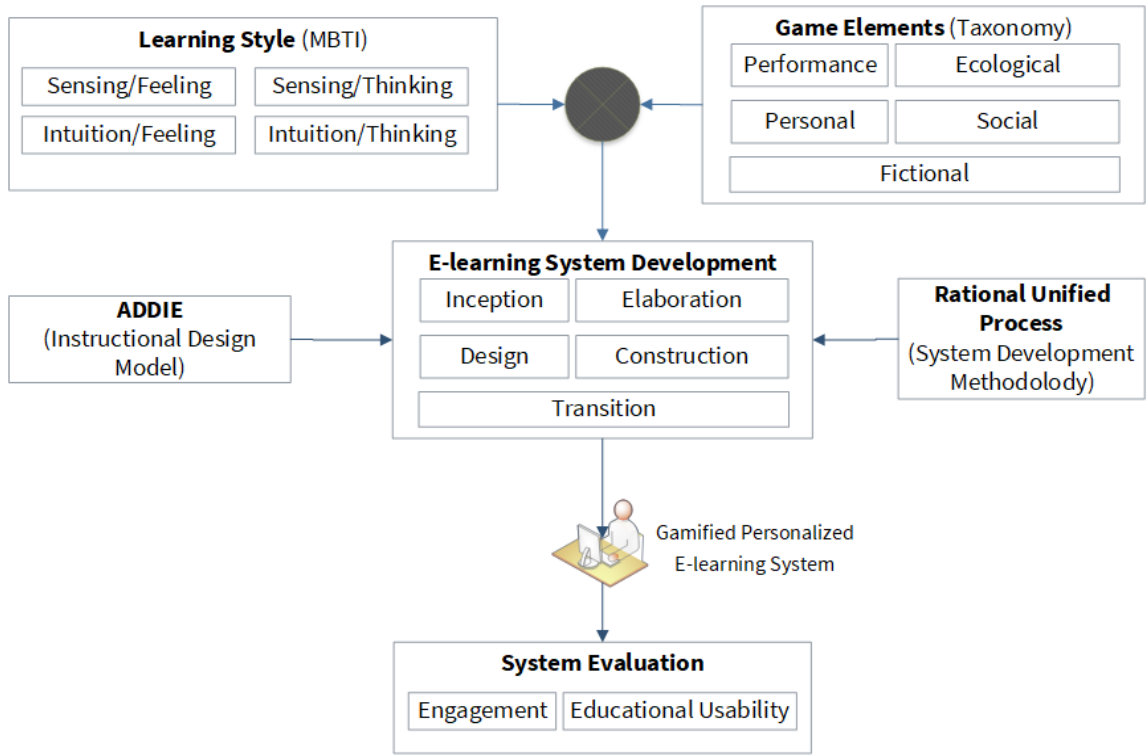

**Figure 3.1** Conceptual Framework

5. Evaluation: This component will evaluate the efficacy of the developed system using two criteria – engagement and educational usability.

The research framework provides a holistic view of the concepts that are necessary to create a personalized e-learning system that is effective and engaging for users. By considering all of these components, e-learning system developers can create systems that are more likely to be used and that will lead to better learning outcomes.

## 3.3. Research Methodology

The research methods involved selecting a means of personalizing the system, identifying gamification elements for e-learning environments, collecting expert opinions from educational psychologists or related disciplines, and designing a gamification model for personalized e-learning. Moreover, the system development was done using a hybrid approach, incorporating aspects of the Rational Unified Process (RUP) system development methodology and the Analysis, Design, Development, Implementation, Evaluation (ADDIE) instructional design model.

## 3.4. Model Development

In order to develop a gamification model for personalized e-learning, a variety of techniques were utilized. This section outlines the processes followed in the model development. First, a review of the literature concerning personalized e-learning is conducted to understand the current state of the field and to identify a suitable personality model to adopt. A summary of works reviewed can be found in the literature review table (Appendix i). Secondly, existing gamification research in e-learning are studied to identify game elements used for e-learning. Thirdly, interviews and surveys are administered to experts in educational psychology to collect opinion on game elements recommended for different learner types based on the learner's personality. Finally, the findings from the literature review, and surveys in addition to the analysis of existing gamification research, are used to create a new model.

### 3.4.1. Personalization requirements

The review of literature in section 2.3.2 highlights a range of frameworks and models used for categorizing learners, including the Felder-Silverman Learning Style Model, Kolb's Learning Style Inventory, the Ocean model, and the Myers-Briggs Type Indicator (MBTI). The MBTI was chosen because it focuses on individual problem-solving approaches, decision-making tendencies, and environmental interaction. The distinctive behavioural patterns within the MBTI's cognitive core according to Pearman and Albritton (1994), namely Intuition and Thinking, Intuition and Feeling, Sensing and Thinking, and Sensing and Feeling, are mainly selected as the foundation for personalization, as they significantly shape individuals' learning style preferences.

### 3.4.2. Identification of game elements

Existing studies on gamification have prescribed various game elements for gamifying activities. However, some game elements are more suitable for other activities such as crowdsourcing or marketing, than for educational purpose. For this reason, a search for a classification or taxonomy of game elements mainly used within learning environment was done to ensure that elements that could be incompatible with learning for the gamification of learning environment are exclude. The five dimensions for gamification taxonomy in educational environments presented by Toda *et. al*. (2019) was adapted and extended to include Avatar / Virtual Identity in the 'Personal' dimension (Figure 3.2). Appendix iv provides a list of the game elements with their descriptions.

### 3.4.3. Expert opinion on game elements and MBTI cognitive core

The purpose of this section was to gather expert opinions on game elements and their suitability for different MBTI cognitive cores in the context of e-learning. The process for obtaining the results in the Expert Opinion on Game Elements and MBTI Cognitive Core study involved the following steps:

1. Survey Design: A Google Forms survey was created to gather expert opinions on game elements and their suitability for different MBTI cognitive cores in the context of e-learning

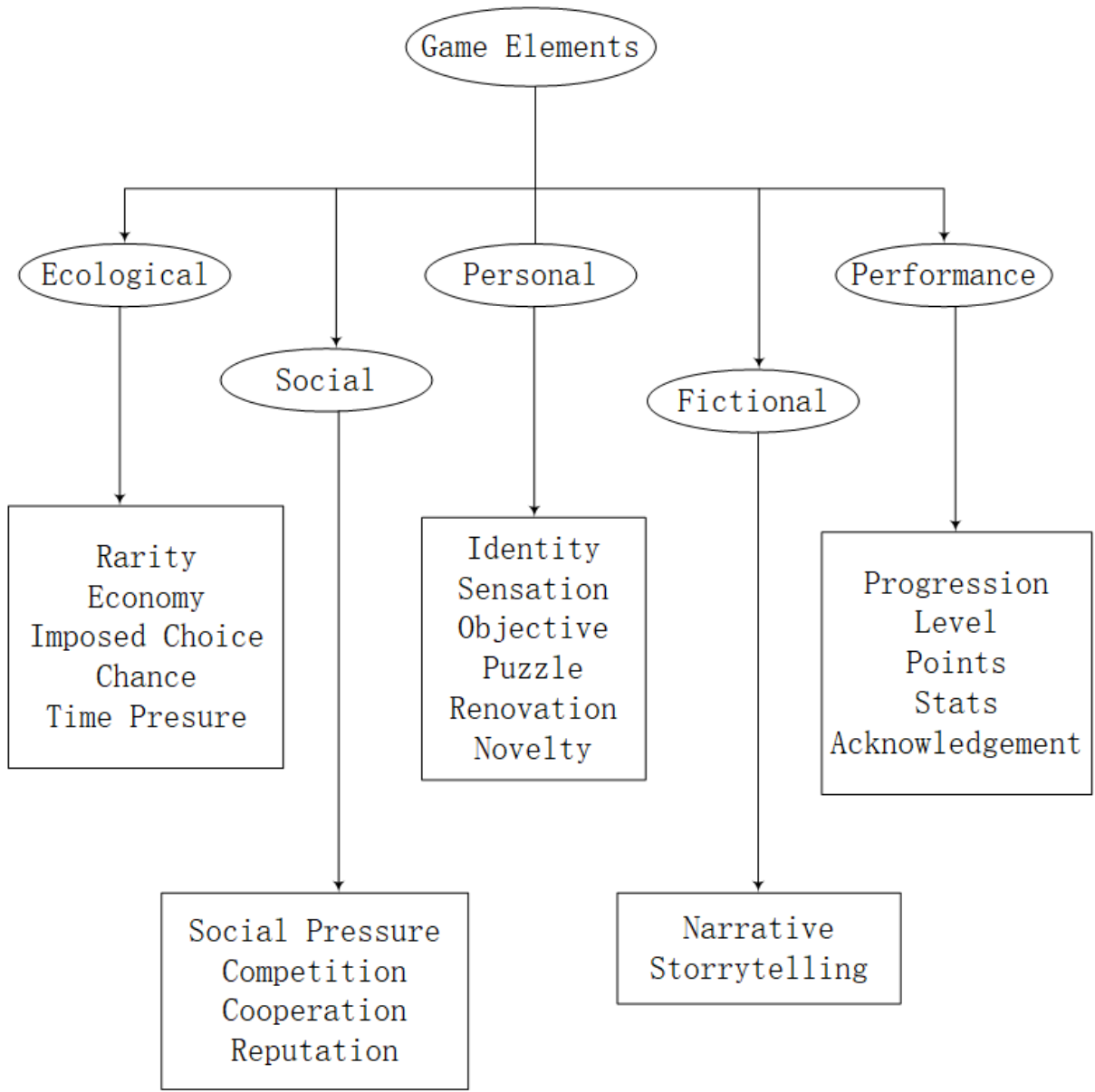

**Figure 3.2** Extended taxonomy of game elements for educational environment

The survey targeted experts with backgrounds in educational psychology, behavioral psychology, personalized learning, e-learning, general psychology, and cognitive psychology and related fields.

2. Questionnaire Development: The survey included game elements classified into five categories: Performance & Measurement, Environment, Personal, Social, and Fictional. Experts were asked to evaluate the suitability of these game elements for learners with different MBTI cognitive cores (Please refer to appendix ii for full questionnaire).

3. Expert Selection: 10 Experts who identified with at least one of the specified backgrounds were invited to participate in the survey. The selection process aimed to ensure a diverse range of expertise and perspectives.

4. Data Collection: The survey was distributed to the selected experts, who provided their opinions by selecting the game elements they deemed most suitable for each MBTI cognitive core within each category.

5. Data Analysis: The responses from the experts were collected, exported to excel spreadsheet and analyzed.

6. Result Presentation: The results were presented in a clear and organized manner, highlighting the selected game elements for each MBTI cognitive core within each category.

### 3.4.4. Model design

The gamification model for personalized e-learning is then designed by mapping the learner types to the game elements recommended by experts from the survey. This process is described by equations below:

$L = \{NT, ST, NF, SF\}$  **Equation 3.1**

This equation defines the set of learner types L, which contains four elements representing the four MBTI Cognitive Core learner types: NT, ST, NF, and SF

Where:

NT is Intuition with Thinking

ST is Sensing with Thinking

NF is Intuition with Feeling

SF is Sensing with Feeling

$$G = \{(A_1, A_2, \ldots, A_n), (B_1, B_2, \ldots, B_n), (C_1, C_2, \ldots, C_n), (D_1, D_2, \ldots, D_n)\}$$ **Equation 3.2**

This equation defines the set of game elements G, which contains four tuples, each consisting of n game elements. These tuples are denoted as $(A_1, A_2, \ldots, A_n)$, $(B_1, B_2, \ldots, B_n)$, $(C_1, C_2, \ldots, C_n)$, and $(D_1, D_2, \ldots, D_n)$.

$$F: L \rightarrow G$$ **Mapping 3.1**

This mapping defines the function F, which maps the set L to the set G. The notation $F: L \rightarrow G$ means that F takes an element of L as input and produces an element of G as output.

$$F(NT) = (A_1, A_2, \ldots, A_n)$$ **Equation 3.3**

This equation specifies that the output of F when the input is NT is the tuple $(A_1, A_2, \ldots, A_n)$.

$$F(ST) = (B_1, B_2, \ldots, B_n)$$ **Equation 3.4**

This equation specifies that the output of F when the input is ST is the tuple $(B_1, B_2, \ldots, B_n)$.

$$F(NF) = (C_1, C_2, \ldots, C_n)$$ **Equation 3.5**

This equation specifies that the output of F when the input is NF is the tuple $(C_1, C_2, \ldots, C_n)$.

$$F(SF) = (D_1, D_2, \ldots, D_n)$$ **Equation 3.6**

This equation specifies that the output of F when the input is SF is the tuple $(D_1, D_2, \ldots, D_n)$.

Together, equations 1-6, as well as the mapping function, define the relationship between the set of MBTI Cognitive Core learner types (L), the set of game elements (G), and the function that maps the learner types to the game elements (F).

The gamification model for personalized e-learning above can best be described in terms of UML Classes. Figure 3.3 illustrates the model using a class diagram. The diagram contains two classes: one for the set L and one for the set G. The Class for the set L has four attributes: NT, ST, NF, and SF. The Class for the set G has four tuples: $(A_1, A_2, \ldots, A_n)$, $(B_1, B_2, \ldots, B_n)$, $(C_1, C_2, \ldots, C_n)$, and $(D_1, D_2, \ldots, D_n)$. The Class for the set G contains a method called F, which is used to map the elements of the set L to the corresponding elements of the set G.

## 3.5.    System Development

The RUP system development methodology and ADDIE instructional design model are two frameworks that can be used to create a personalized gamified e-learning system. The RUP system provides a structured approach to system development, allowing for the efficient and effective design and implementation of the system. The ADDIE model on the other hand provides guidance for designing the system to best meet the individual user's needs, by focusing on the user's goals and objectives, analysing the user's environment, and designing a system to meet their unique needs. Additionally, ADDIE can be used to guide the development of the system, from designing the system architecture to testing the system's effectiveness. By combining these two approaches, a personalized gamified e-learning system can be created that is tailored to each user's individual needs.

### 3.5.1.  Inception (analysis)

The initial phase of the system development process involved conducting a comprehensive analysis of the problem: the creation of a personalized gamified e-learning system. This step encompassed identifying the system's objectives and understanding user needs and preferences.

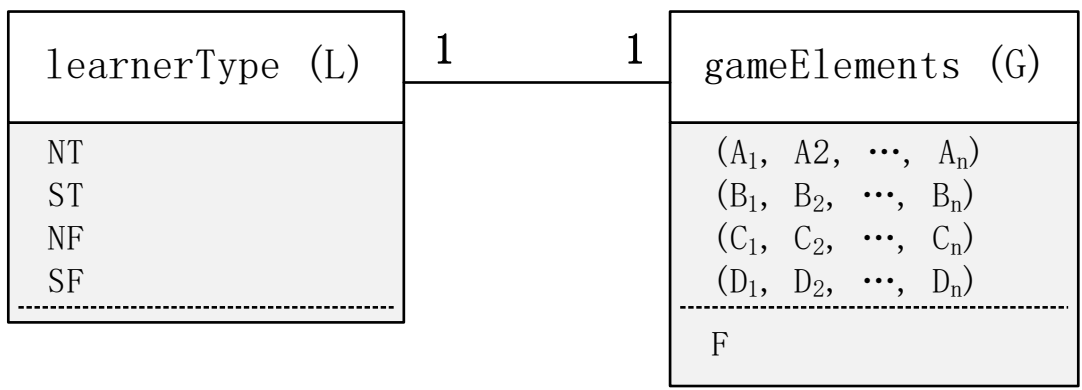

**Figure 3.3** Class diagram of personalized game elements

A depiction of the high-level system architecture, illustrated in Figure 3.4, outlined crucial system requirements. The system was ideated to include a user interface for user access, the ability for users to complete a personality test to determine their preferred learning approach, and the capacity to provide distinct gamified experiences based on individual users' personalities.

### 3.5.2. Elaboration (design)

Following the Inception phase, the Elaboration phase followed suit. During this stage, the focus shifted to developing comprehensive system requirements. This entailed specifying the essential features and functionalities that the system should encompass to meet user expectations.

As illustrated in Figure 3.5, a use case was designed to capture and document these features and requirements, offering insights into the learner journey. The primary users of the system consisted of learners and instructors, both empowered to sign up and log in. Learners could enrol in courses, instructors could create and manage courses, and both groups could engage in forums and access dashboards for interactive interactions.

### a. Functional requirements

The functional requirements of the gamified personalized e-learning system are as follows:

i. User authentication system: A secure user authentication process must be implemented in order to ensure that only authorized users can access the gamified LMS.

ii. User profile management: A user profile management system must be implemented in order to accurately capture user personality, as well as log user activities in the system.

iii. Game mechanics: The game mechanics must be tailored to the user's personality and foster engagement in learning activities. Learning resources: The system must provide the user with access to personalized learning resources.

iv. Feedback: The system must possess means of providing feedback to users based on various events occurring within the system.

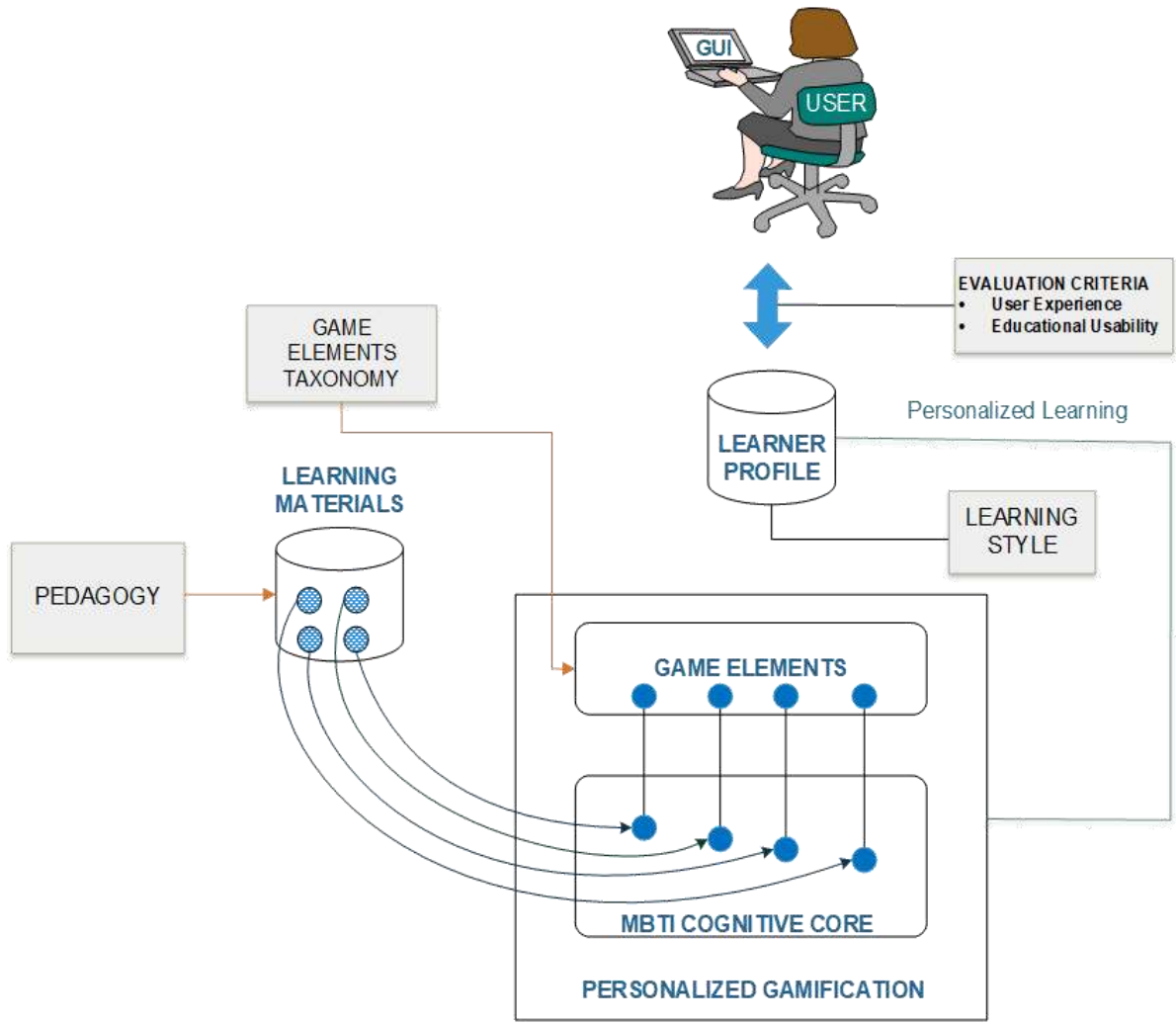

**Figure 3.4** System architecture

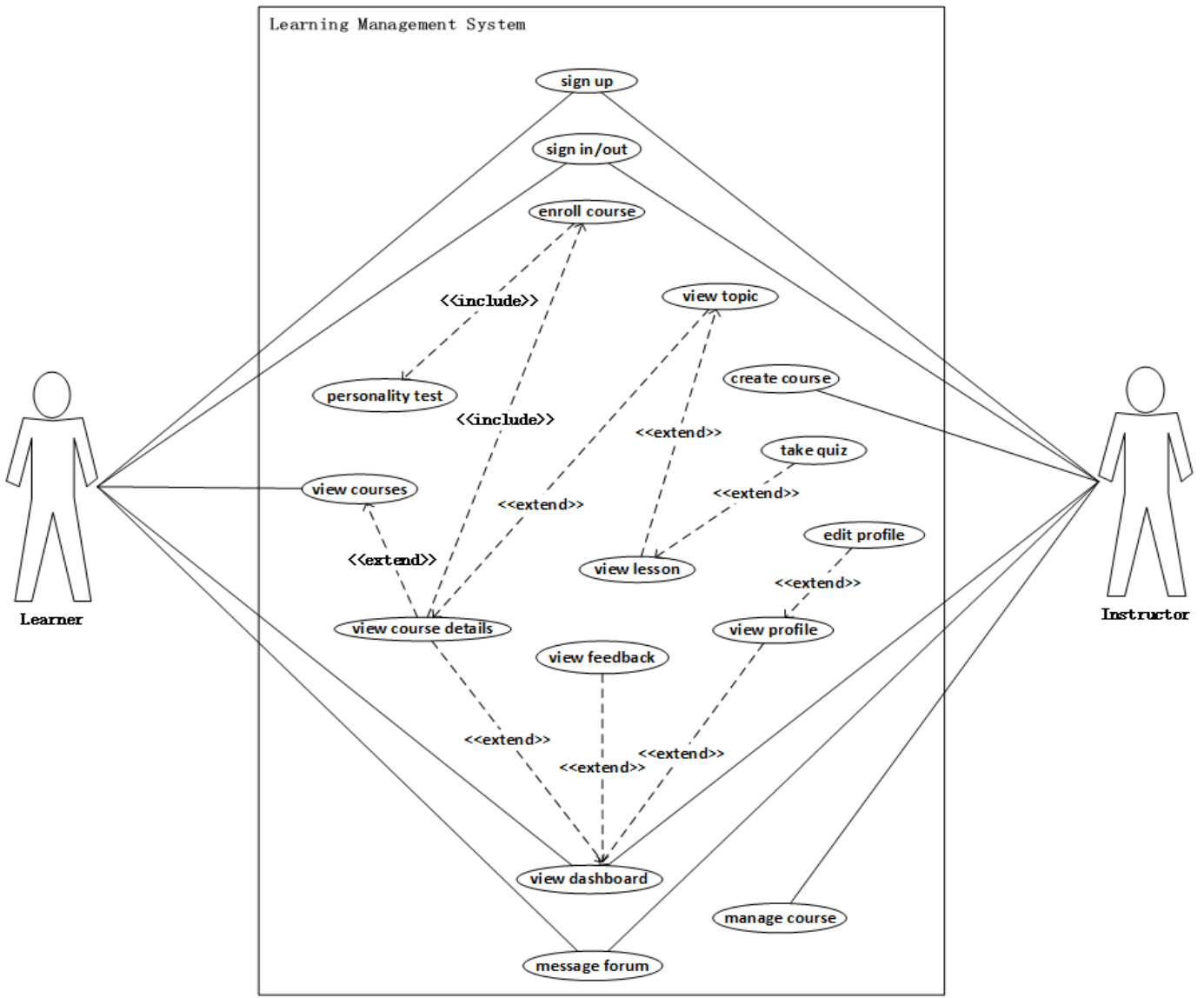

**Figure 3.5** System use case diagram

**b.  Non-functional requirements**

The non-functional requirements of the gamified personalized e-learning system are as follows:

   **i.**    Usability: The system must be easy to use and understand.

   **ii.**   Performance: The system must be able to handle multiple concurrent users.

   **iii.**  Scalability: The system must be able to easily scale to accommodate increasing user numbers.

   **iv.**   Security: The system must adhere to security best practices and comply with industry standards.

   **v.**    Reliability: The gamification model must be able to withstand disruptions and maintain operation.

### 3.5.3.  Design (design and development)

Subsequent to the Elaboration phase, the Design phase was undertaken, primarily focused on creating the system's interaction designs for learners' course progression. In addition to the designed use case diagram, supplementary UML diagrams were employed to model interactions within the system, including the interactions between various game elements. These interactions were elaborated upon in subsequent subsections.:

**a.  Registration and enrolment**

Figure 3.6 illustrates a registration and course enrolment activity by a learner on the system. New learners are required to create an account by registering on the system while existing learners will need to login to the system before proceeding. Upon successful login, the learners can either access their dashboard to view or modify personal details, or advance to the course catalogue to select a course to take. To enrol into the course, a learner must take a personality audit based on the MBTI to identify the learning style of the learner before being granted access to the personalized course.

**b.  Learning scenario**

Once the learner gains access to the course, the learner can consume the available course content, including quizzes and other interactions. As shown in Figures 3.7, a course may contain one or more topics. Each topic may also contain one or more lessons. Upon completion of a lesson, the learner may be provided an activity to partake in such as a quiz. At the end of the quiz, if the learner fails to fulfil the requirements to pass, the learner can retake the quiz again and pass before unlocking more contents. Once all the quizzes associated with a lesson are completed, the next lesson is then unlocked for the learner to continue learning. When the learner has completed the final lesson of a topic, the next topic is unlocked. A learner can access topics, lessons and quizzes and revisit them anytime, granted they have been unlocked. The learner's profile is continuously being updated as the learner progresses in the system. The learner may be awarded points for completing activities such as a quiz, in addition to access to more learning contents. At various stages, the learner may be granted rewards such as completion badges to acknowledge progress. When the entire course has been completed, the learner may receive an achievement badge or a completion certificate for the course.

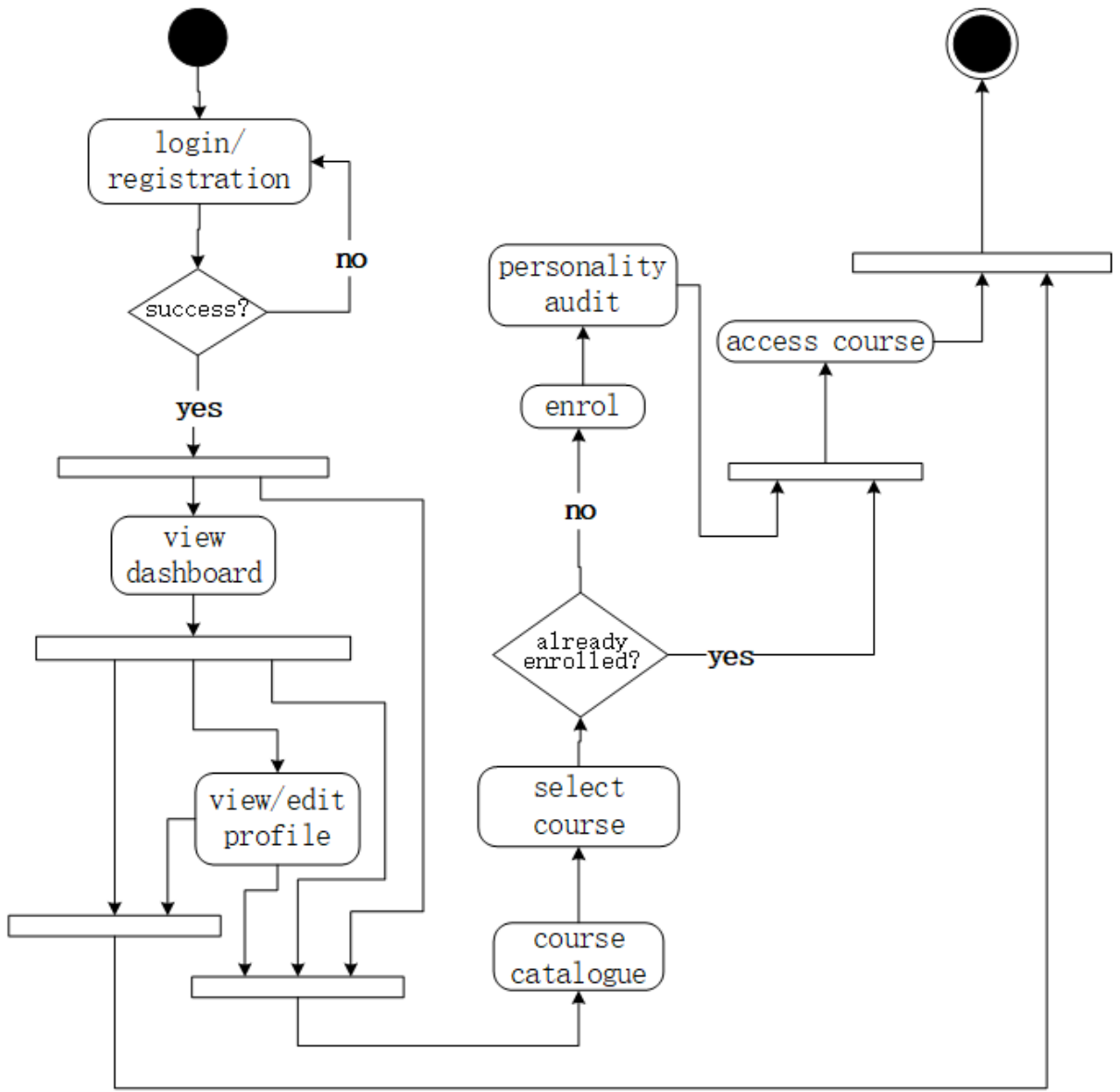

**Figure 3.6** Enrolment activity diagram

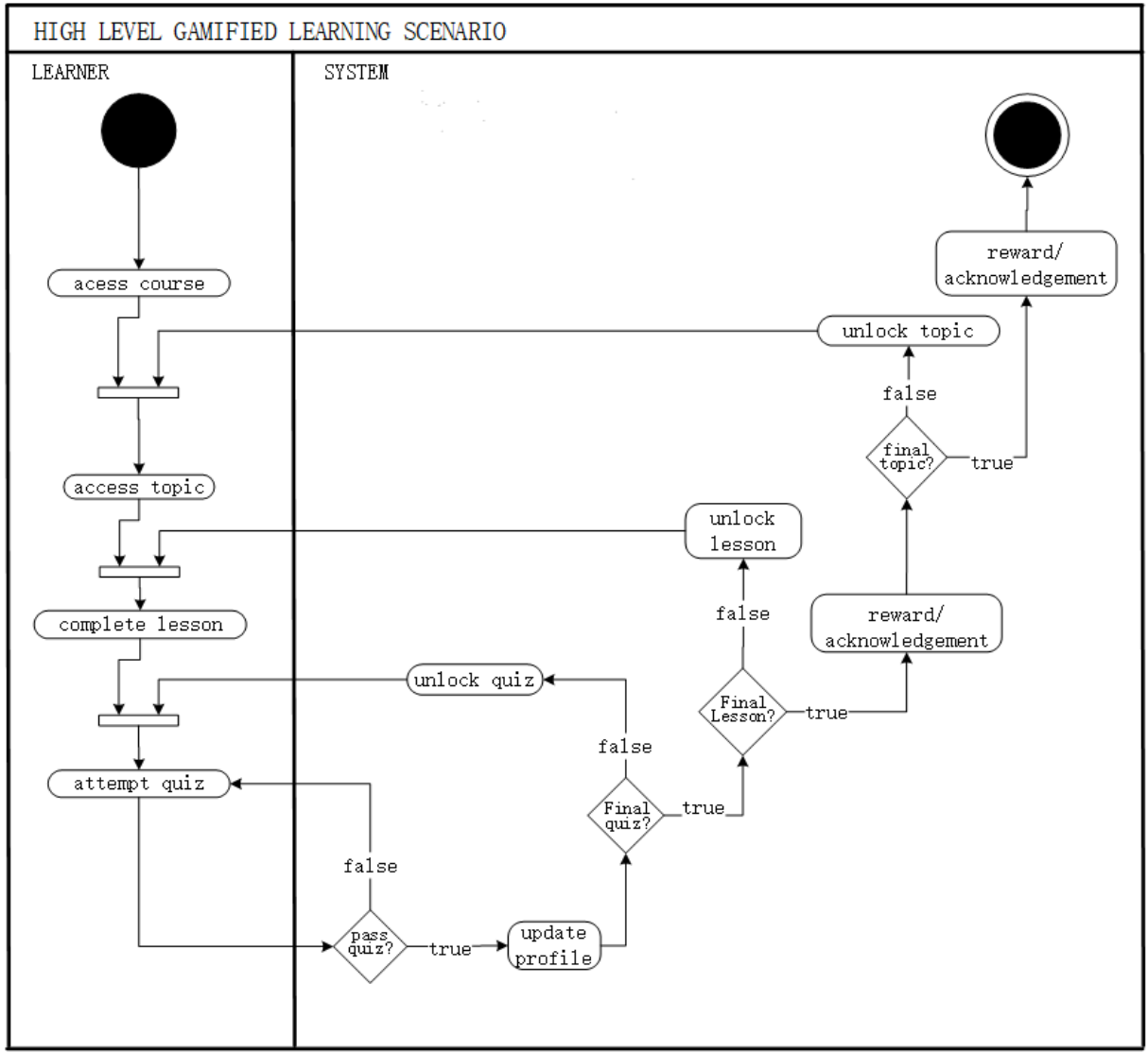

**Figure 3.7** Learning scenario activity diagram

### c.   Course structure

The course is structured by a series of complex interactions among multiple modules. These interactions are modelled in the class diagram shown in Figure 3.8. The class 'user' represents a learner in the system. A user possesses attributes such as 'user_id', 'user_name', and 'user_mail' among other personal attributes. The user class is linked to a class 'course' by an 'enrolls' association. A 'course' class represents courses on the system and has attributes such as 'course_id', 'course_name', and Topics are 'composed of' the class 'lesson' with attributes which includes 'lesson_id', 'lesson_name' and 'last_accessed'. A lesson 'compose of' the 'quiz' class, which has attributes like 'quiz_id', 'last_attempt', and total_score'. The 'user' class 'compose of' a 'feedback' class, which 'depends' on the classes 'course', 'topic', 'lesson', and 'quiz' to provide updates on the learners activities.

### d.   Game elements

As the learner engages the system, it triggers interactions between various game elements in the system. The class diagram in Figure 3.9 shows the relationship between game elements in the system. The 'user' class associates with the 'game elements' class, which is a generalized class for game elements in the system. Each game element possesses attributes such as 'element_id' and 'element_name'. Game elements such as 'level', 'timer' and 'point' can perform their operations solely from feedback from the learners' activities in the system such as completing a lesson or passing a quiz, while others such as 'badge', 'progress bar' and 'scoreboard' have dependency on other game elements. For instance, the 'scoreboard' depends on 'points' accumulated by the learner from an activity, while the 'progress bar' depends on the learners' progression through 'levels' in the system. Figure 3.10 depicts the SQL schema of the system, showing a logical and structural blueprint that defines the organization, structure, relationships, and constraints of the system database.

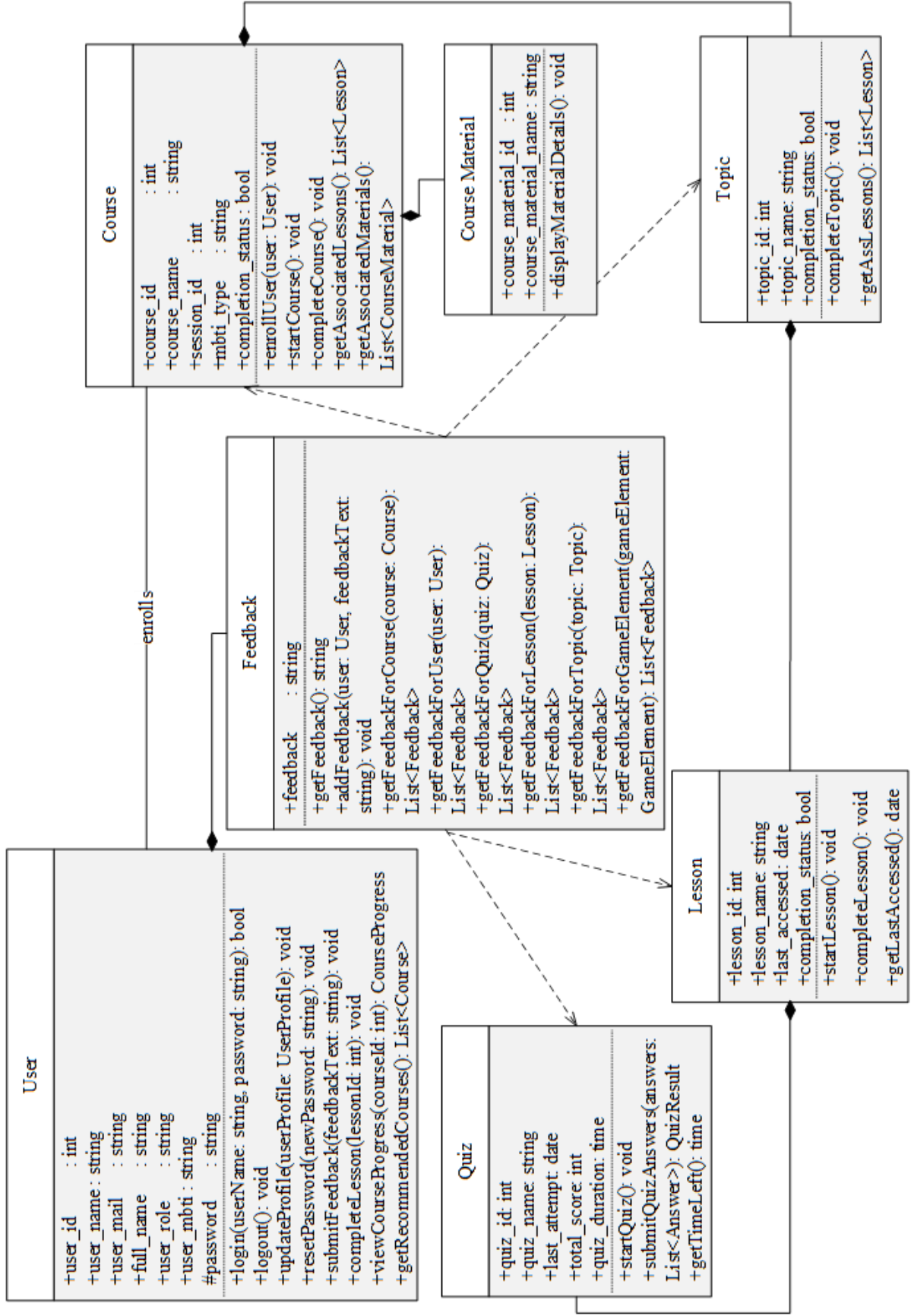

**Figure 3.8** Course class diagram

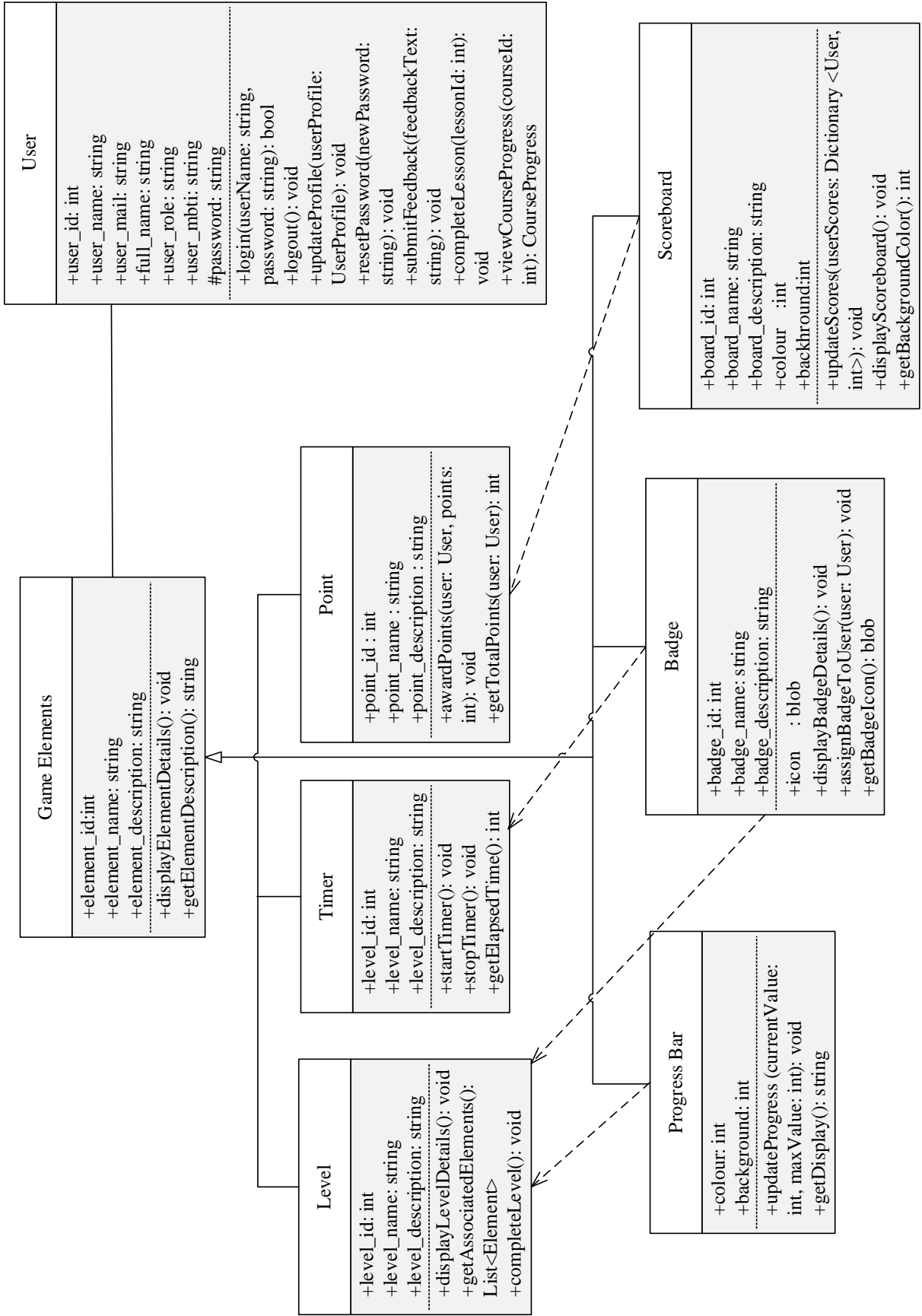

**Figure 3.9** Class diagram of game elements

### 3.5.4. Construction (development)

With the design phase defined, the Construction phase was initiated, dedicated to translating the system's requirements into functional code. Depicted in Figure 3.11, an algorithm guided the system's construction. An abridged MBTI test was devised to identify the learners' cognitive core, serving as pointer to user's learning preferences. A sample of the test questions are provided in appendix iii. User responses were collected and employed to determine each individual's learning path. Consequently, learners were mapped to appropriate course paths tailored to their respective learning preferences, incorporating relevant game elements to enhance the learning experience.

### a. Designing the gamified course

A concise course titled 'Instructional Innovation Course' was developed. This course was composed of six distinct topics, each containing at least one lesson accompanied by quizzes or activities thoughtfully crafted to address the four cognitive cores. It was mandated that learners accomplish the activities and quizzes within a given lesson before unlocking access to subsequent topics. Importantly, learners retained the liberty to revisit completed content in any order of their preference. The course had been meticulously fashioned based on recommended game elements,

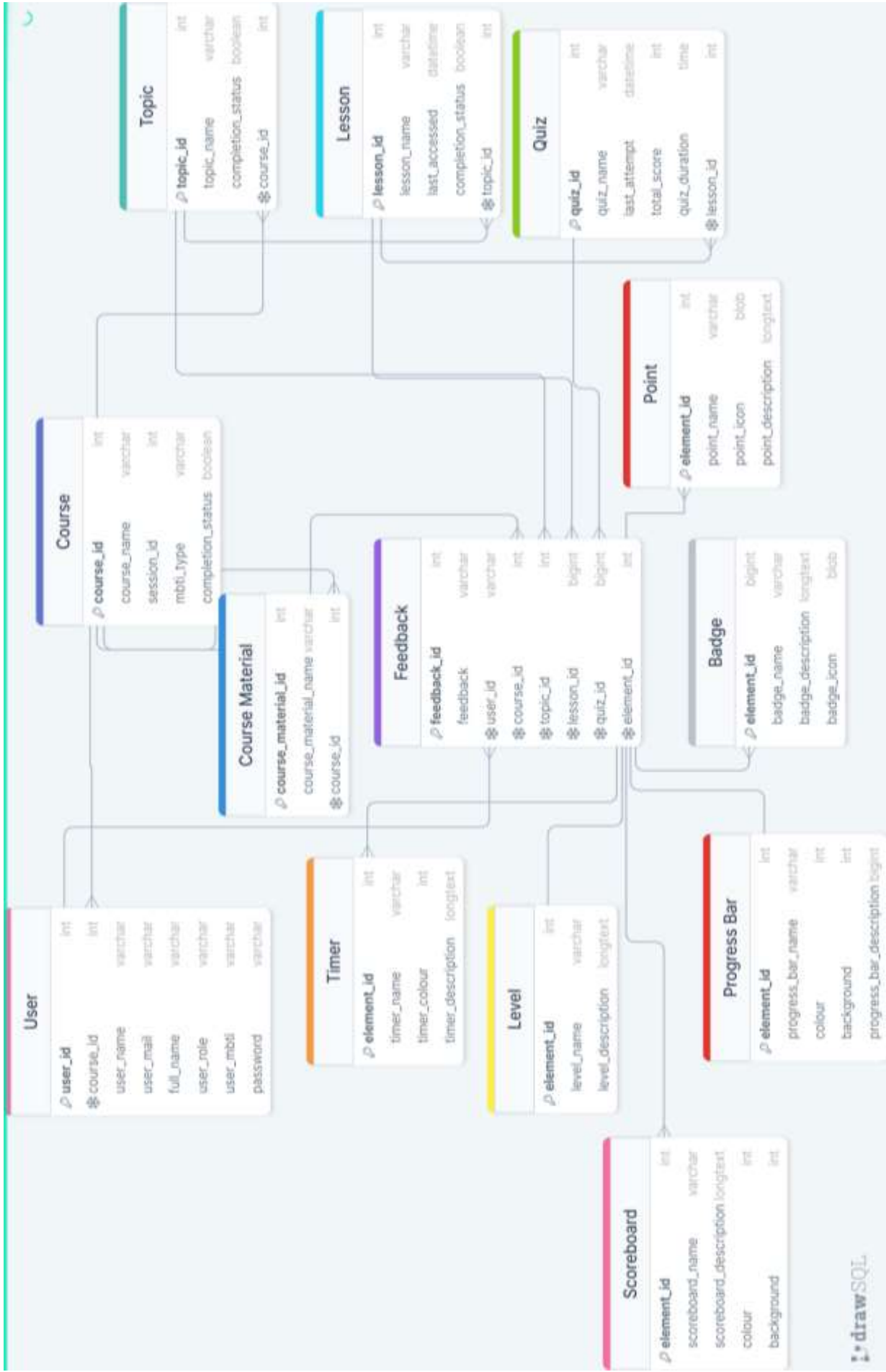

**Figure 3.10 Database SQL Schema**

```
//ALGORITHM
//create the MBTI test
runMBTITest(){
        //ask several questions related to personality
        //collect response
        //calculate and analyze the results
        //store the type of person in a variable
retrieve personType:
        //check what type of person the test taker is
 if(Intuition. Thinking) {
        personType = "NT":
 } else if (Intuition. Feeling) {
        personType = "NF":
 } else if (Sensing. Feeling) {
        personType = "SF":
 } else if (Sensing. Thinking) {
        personType = "ST":
 }
 //return the type of person
 return personType:
}
//trigger MBTI test
let type = runMBTITest():
//store the type of person
console.log(type)
//Mapping type of person to game elements
personType =L:
function F(L):
if L = NT {
        return (A1. A2. .... An):
} else if L = ST {
        return (B1. B2. ..., Bn):
} else if L = NF {
        return (C1. C2 ..... Cn):
} else if L = SF {
        return (D1 . D2 ..... Dn):
}
end function:
//console.log(learningPath):
```

**Figure 3.11** Algorithm for system coding

encompassing factors like time pressure, points, progression, stats, puzzles, choice, and competition. The incorporation of time pressure aimed to motivate learners to finalize activities within stipulated timeframes. Points were bestowed upon learners as a reward for successfully completing activities. Progression was aptly utilized to propel learners through the course, while stats were adeptly displayed, showcasing each learner's advancement relative to their peers.

The introduction of puzzles aimed to present learners with stimulating challenges, whereas well-defined objectives furnished them with clear-cut goals. The provision of feedback was facilitated through diverse channels. Immediate feedback was presented through pop-ups and on-screen displays during activities. Simultaneously, learners' personal dashboards received updates, and email notifications were dispatched, offering a comprehensive overview of their individual progress. In a bid to foster interaction among learners, a dedicated forum was established, encouraging the exchange of ideas, inquiries, and mutual support.

Upon concluding the course, learners were tasked with a final course project. This project necessitated the execution of varied tasks tailored to their respective types. Learners were also expected to express their insights concerning their course experience and undertake an evaluation survey. The evaluation survey utilized a Likert-scale approach to accumulate learners' feedback, which served as a resource for enhancing the course in subsequent iterations.

### b. System specification

The system was developed as an advanced web application, making use of the WordPress Content Management System (Version 5.6.5). Its functionalities encompassed those of a learning management system (LMS) facilitated through LearnDash plugin (Version 3.6). For the user interface, the latest technologies, namely HTML5 and CSS3, were employed. The server-side scripting was achieved via the most recent iteration of PHP (Version 7.4), while data storage relied on the MySQL (Version 5.6) database.

The aim was to create a system that is both modern and user-friendly, providing users with an optimized experience regardless of the device they are using. The use of the latest technologies available allowed the creation of a web application that was fast, reliable, and secure. The system was designed with a responsive layout, meaning it automatically adapts to the size of the device being used. This makes it easier for users to access and use the system, as they don't have to resize or scroll the page to view its content. The layout also ensures that all users get the same experience, regardless of their device. The user interface has been designed to be intuitive and easy to understand. All the features are clearly visible, allowing users to find the information they're looking for quickly and easily. Using the latest technologies, the system is secure and reliable. The system is hosted on a secure server, with an SSL certificate, providing users with a secure connection. The system also has a backup system in place, so that in the event of an emergency, the system can be restored quickly and easily.

### 3.5.5. Transition (implementation and evaluation)

The fifth step in the system development was the transition process or deployment of the system. This step was essential, ensuring that the system was available for use and that it was functioning properly. It was important to ensure that users were able to access the system from their web browsers. To accomplish this, the URL for the system, https://www.skillcomvalley.com/courses/instructional-innovation-course/, was made available to users. Once the URL was available, it was test tested to ensure that it functioned properly. This also included testing the user interface, making sure all the features were working and ensuring that the system was secured and reliable. If any issues were found, they were addressed and fixed before making the system available to the users.

Once the system is tested and running properly, it is time to make the system available to the users. Finally, the system is monitored and changes would be made as required. This includes responding to user feedback and making any necessary improvements to the system to ensure that it is running smoothly. By monitoring the system and making changes as needed, it is possible to ensure that the system is functioning properly and that users are able to get the most out of it.

### 3.5.6. Evaluation

The final step in the development of the learning system is to evaluate its performance. A subset of criteria from Harpur and De Villiers (2015) designed to assess user experience, and education usability were adopted for evaluation in this study. The criteria used included four from the user experience criteria (emotion, user-centricity, appeal and satisfaction) and three education usability criteria (clarity, error recognition and feedback).

The evaluation of a learning system requires assessing the performance and behaviours of learners when using the system to determine if the system's goals were achieved. User experience criteria such as emotion, user-centricity, appeal and satisfaction can help determine how enjoyable and engaging the system was for learners, while education usability criteria like clarity, error recognition and feedback can determine how effective the system was in course delivery and learning. Using these criteria helped to determine if the system objectives were accomplished and to what extent.

In order to evaluate the system, learners were required to complete a survey at the end of the course. This survey contained statements pertaining to the criteria being evaluated, and the responses were collected using a 5-point Likert-scale instrument. This allowed for more accurate assessment of the systems performance. Table 3.1 presents the statements used in the survey.

**Table 3.1:** Questionnaire for system evaluation

| S/N | CRITERIA | STATEMENTS |
|---|---|---|
| 1 | User-centricity | The activities increased my interaction with the course |
| 2 | | The activities encouraged me to revisit the course materials |
| 3 | | I tried to beat my completion time for activities |
| 4 | Emotion | The activities were worth time spent |
| 5 | | The activities/quizzes added fun to the course |
| 6 | | Achievement points or badges earned were motivating |
| 7 | Appeal | The course interface is easy to navigate and use |
| 8 | | The course seemed to be designed for my learning style |
| 9 | Satisfaction | I enjoyed the entire course experience |
| 10 | | I enjoyed the hands-on nature of the course project |
| 11 | Clarity | The goals of each activity/quiz were clear |
| 12 | | The final course project was difficult to complete |
| 13 | Error recognition / correction | The course activities/quizzes gave me the chance to learn from mistakes |
| 14 | | Activities are graded with grades providing instant feedback and correction |
| 15 | Feedback | Prompt feedback was provided |
| 16 | | The progress bar helped me keep track of my advancement in the course |
| 17 | | My advancement on the progress bar also encouraged me to continue the course |

*Rows 1–10 are grouped under the vertical label "Engagement"; rows 11–17 are grouped under the vertical label "Educational Usability".*

<div align="center">

**CHAPTER FOUR**

**RESULTS AND DISCUSSION**

</div>

## 4.1.  Introduction

This chapter presents and discusses findings from the study. The findings are in accordance with the stated research objectives as highlighted in chapter 1 which includes a gamification model for personalized e-learning, an implemented system as well as results obtained from evaluating the system.

## 4.2.  Model Development

The development of the personalized e-learning gamification model began by identifying behavioural tendencies through the MBTI personality model. The subsequent stages encompassed a literature review to identify pertinent game elements, followed by a survey administered to experts. This section comprehensively showcases the outcomes of these phases, structured under the subsequent headings:

### 4.2.1.  Behavioural tendencies of MBTI cognitive core

The following are the behavioural tendencies documented for each MBTI cognitive core functions

a. **Intuition with Thinking (NT):** Psychologically-minded individuals who possess intuition and analytical thinking skills enjoy autonomy and have a keen sense of objectivity. They are often clever and ingenious, demonstrating a focused attentiveness to theory and models. Additionally, these individuals take pride in having well-defined interests.

b. **Sensing with Thinking (ST):** Calculated, rational, pragmatic, concrete, able to evaluate evidence (check, measure), organized, disciplined, motivated by power and influence, self-assured, consistent.

c. **Sensing with Feeling (SF):** People with this style tend to be fact-focused and gentle, seeking to meet the needs of others. They are often seen as being compassionate, responsible, conscientious, and patient when it comes to dealing with the details at hand. They primarily focus on the concrete things that affect the well-being of those around them.

d. **Intuition with Feeling (NF):** Soft-hearted, excitable, perceptive, eager to take on new challenges, fond of complexity, malleable, able to recognize patterns and connections, sensitive to beauty and creativity, imaginative, independent in thinking, and unconstrained by rules and conventions.

### 4.2.2. Game elements for e-learning

The following game elements were identified for use in educational environments: Chance (randomness, luck, fortune or probability), Alternate paths / Imposed Choice, Economy (transactions, market, exchange), Rarity (limited items, collectibles, exclusivity) and Time Pressure (timed events and clock counts), Competition (conflict, leaderboards, scoreboards, player vs player contests), Cooperation (teamwork, co-op, groups), Reputation (classification or status), and Social Pressure (peer pressure or guild missions). Other game elements such as Sensation (haptic, visual or audio stimulation), Objectives (missions, side-quests, milestones), Puzzle (challenges, cognitive tasks, actual puzzles), Novelty (updates and changes), and Renovation (boosts, extra life, renewal), storylines (where learners follow a predefined script of events and activities) and narratives (where learners actions and decisions have an impact on future events), in addition to Point (scores, experience points, skill points), Progression (progress bars, steps, maps), Levels, Performance Statistics, and Acknowledgement (badges, medals, trophies and achievement cards). Appendix iv presents these game elements with a description for each of them.

### 4.2.3. Expert opinion on game elements and MBTI cognitive core

The purpose of this section was to gather expert opinions on game elements and their suitability for different MBTI cognitive cores in the context of e-learning. The results were presented in a clear and organized manner, highlighting the selected game elements for each MBTI cognitive core within each category. The findings were visually represented through figures as interpreted in the subsequent subsections.

### a. Performance and measurement

In the Performance and Measurement category, the suitability of game elements for different MBTI cognitive cores was examined. The results, summarized in Figure 4.1, shed light on the preferences of learners with distinct cognitive styles regarding various game elements. For the case of learners with Sensing-Thinking cognitive style, the most preferred game element was Stats, which received five selections. Progression and Acknowledgement were chosen three times each. Points were selected twice, while Level did not receive any preferences. On the other hand, learners with Sensing-Feeling cognitive style showed a different pattern. Acknowledgement was the most favoured game element, selected five times. Points, Progression, and Level were each chosen three times. Stats received only one selection. For learners with Intuition-Feeling cognitive style, Progression emerged as the most preferred game element, with four selections. Points followed closely behind with three selections. Level, Stats, and Acknowledgement were each selected twice. Similarly, learners with Intuition-Thinking cognitive style displayed similar preferences. Points and Progression were equally favoured, selected three times each. Level, Stats, and Acknowledgement were each chosen twice.

**Implication:** Based on the survey results in the Performance and Measurement dimension, the implications for each cognitive style are as follows: Sensing-Thinking cognitive style: The most preferred game element for learners with this cognitive style is Stats. This suggests that they value concrete and measurable data and enjoy tracking their performance and progress. Progression and Acknowledgement were also chosen multiple times, indicating that learners with this cognitive style appreciate clear milestones and recognition for their achievements. Points received mild preference, implying that they might find the concept of earning points as a measure of success appealing. Sensing-Feeling cognitive style: The most favoured game element for learners with this cognitive style is Acknowledgement. This indicates that they value external recognition, feedback, and positive reinforcement. Points, Progression, and Level were also chosen multiple times, suggesting that learners with this cognitive style appreciate a sense of advancement, clear goals, and a structured progression system. Stats received minimal preference, indicating that they may not be as concerned with detailed performance metrics and data. Intuition-Feeling cognitive style: Progression emerged as the most preferred game element for learners with this cognitive style. This suggests that they value a sense of growth, advancement, and exploration in a game. Points received moderate preference, indicating that they find the concept of earning points appealing, possibly as a measure of their accomplishments. Level, Stats, and Acknowledgement were chosen moderately, implying that learners with this cognitive style also appreciate a sense of structure, performance metrics, and social recognition. Intuition-Thinking cognitive style: Points and Progression were equally favoured by learners with this cognitive style. This suggests that they appreciate both the tangible rewards and a sense of advancement and growth in a game. Level, Stats, and Acknowledgement were chosen moderately, indicating that learners with this cognitive style value a structured framework, performance metrics, and social recognition to some extent.

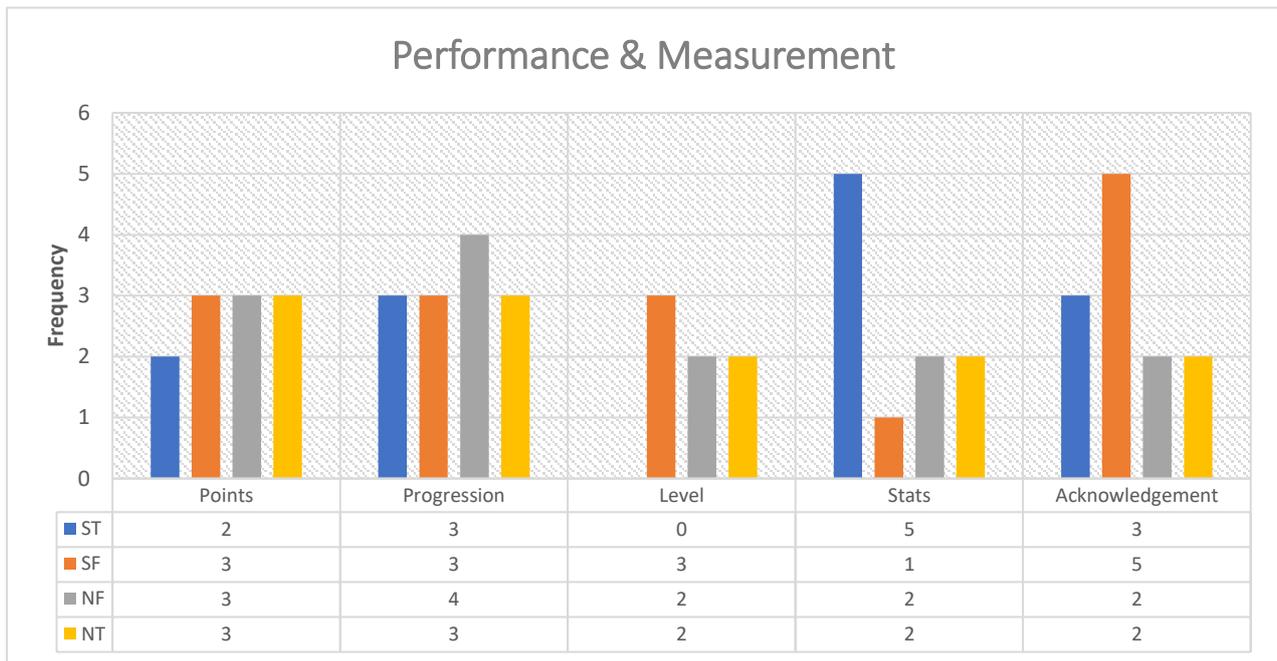

**Figure 4.1** Expert opinion results (a)

### b. Environment

In the category of environment, the suitability of game elements for different MBTI cognitive cores was examined. The summarized results, presented in Figure 4.2, provide insights into the favoured environmental factors for each cognitive style. Sensing-Thinking learners displayed a preference for Economy, with four selections. Time Pressure also received four selections, indicating its significance. Choice was chosen three times, while Chance received one selection. Rarity did not receive any preferences from this group. Sensing-Feeling learners, on the other hand, showed a different pattern. Choice emerged as the most favoured environmental factor, with five selections. Economy received three selections, highlighting its importance. Rarity and Time Pressure were each chosen twice, indicating their relevance. Chance received one selection. Intuition-Feeling learners displayed a preference for Choice, which received four selections. Economy received three selections, indicating its significance. Chance, Rarity, and Time Pressure each received two selections, suggesting their relevance within this cognitive style. Intuition-Thinking learners exhibited distinct preferences. Time Pressure was the most preferred environmental factor, with four selections. Rarity followed closely behind with three selections. Economy and Chance were each chosen twice, while Choice received only one selection. These results demonstrate that learners with different cognitive styles will have varying preferences for environmental gamification.

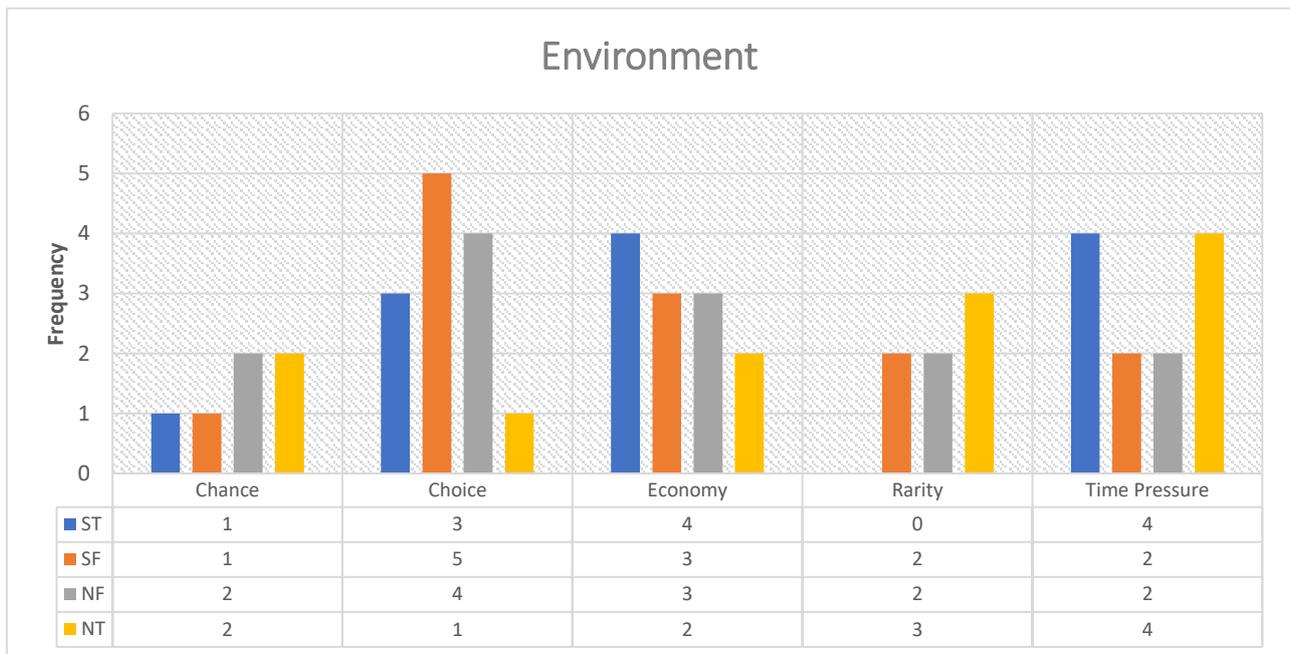

**Figure 4.2** Expert opinion results (b)

**Implication:** The implications of the results obtained for the Environment category highlight the varying preferences of learners with different cognitive styles for game elements related to the environment. Here is a breakdown of the implications for each cognitive style: Sensing-Thinking learners showed a preference for Economy and Time Pressure. This suggests that incorporating elements related to resource management, efficiency, and time constraints can engage and motivate these learners. Designing learning environments that require strategic decision-making, optimizing resources, and handling time-sensitive tasks can enhance their learning experience. Sensing-Feeling learners displayed a preference for Choice. This indicates that providing opportunities for them to make decisions, exercise autonomy, and have control over the game environment can enhance their engagement and satisfaction. Offering choices that allow for personal expression, values alignment, and emotional connections can create a more meaningful learning experience for this group. Intuition-Feeling learners exhibited a preference for Choice and Economy. This suggests that allowing them to explore different possibilities, make meaningful decisions, and consider the consequences of their choices can enhance their engagement. Incorporating elements that highlight economic considerations, such as resource allocation or cost-benefit analysis, can also resonate with their cognitive style. Intuition-Thinking learners showed a preference for Time Pressure and Rarity. This indicates that designing game elements that involve time-sensitive challenges, urgency, and limited availability of resources or opportunities can engage and stimulate their cognitive processes. Creating environments that require quick thinking, problem-solving under pressure, and the ability to identify rare or unique opportunities can enhance their learning experience.

### c. Social

For the social category results of each MBTI cognitive styles, learners with Sensing-Thinking cognitive style showed a strong preference for Reputation, which received six selections. Cooperation was also favoured, chosen four times. Competition and Social Pressure each received two selections. Learners with Sensing-Feeling cognitive style exhibited different preferences. Cooperation emerged as the most favoured game element, selected four times. Competition, Reputation, and Social Pressure received equal attention, with each receiving three selections. Learners with Intuition-Feeling cognitive style demonstrated a significant preference for Cooperation, which received six selections. Competition closely followed with five selections. Reputation and Social Pressure received one selection each. Learners with Intuition-Thinking cognitive style showed a strong inclination towards Competition, which received five selections. Cooperation was also valued, chosen four times. Reputation received two selections, while Social Pressure received one selection. The result summary is presented in figure 4.3.

**Implication:** The implications of these results for each cognitive style in e-learning can be examined based on the results obtained for learners within the social category. For learners with Sensing-Thinking cognitive style, the strong preference for Reputation suggests that incorporating social recognition and status-based elements into e-learning environments can be effective in engaging and motivating these learners. Recognizing and showcasing their achievements and accomplishments through badges, leaderboards, or public acknowledgments can foster a sense of accomplishment and drive their participation. Cooperation being favoured by learners with Sensing-Feeling cognitive style indicates that creating collaborative and team-based activities in e-learning can be beneficial for this group. Promoting opportunities for learners to work together, share ideas, and engage in group discussions can enhance their learning experience. Facilitating a supportive and cooperative learning environment can also foster a sense of community and emotional connection among learners. For learners with Intuition-Feeling cognitive style, the strong preference for Cooperation suggests that e-learning platforms should emphasize collaborative and interactive components. Encouraging group work, peer-to-peer learning, and opportunities for social interaction can enhance their engagement and learning outcomes. Providing avenues for learners to connect with others, such as through online forums or collaborative projects, can help satisfy their need for social interaction and a sense of belonging. The inclination towards Competition for learners with Intuition-Thinking cognitive style indicates that incorporating elements of challenge, gamification, and opportunities for goal-oriented tasks can be effective in engaging this group. Designing activities that allow learners to compete against others, track their progress, and achieve measurable goals can enhance their motivation and satisfaction. Implementing leaderboards, timed quizzes, or performance-based assessments can cater to their desire for intellectual challenge and achievement.

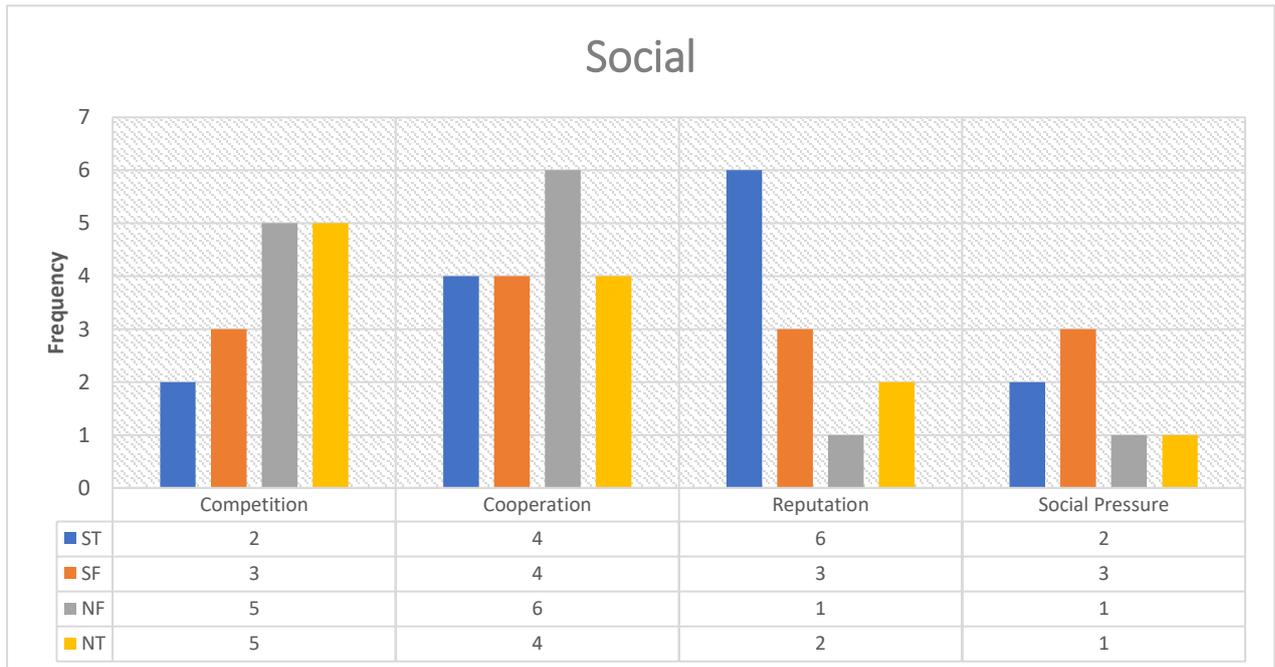

**Figure 4.3** Expert opinion results (c)

### d. Personal

The results for the personal category in the study focused on examining the preferences of learners with different MBTI cognitive styles for game elements. The summarized results, as presented in Figure 4.4, provide insights into the preferences of learners with distinct cognitive styles regarding various game elements within the personal category: Learners with Sensing-Thinking cognitive style displayed a strong preference for Puzzle, receiving seven selections. Sensation followed with four selections, and Objective received three selections. Novelty received one selection, while Renovation and Avatar did not receive any preferences. Learners with Sensing-Feeling cognitive style showed a more diverse pattern. Novelty and Sensation were equally favoured, each receiving four selections. Puzzle followed with three selections. Renovation received two selections, while Avatar and Objective each received one selection. Learners with Intuition-Feeling cognitive style demonstrated a preference for Novelty and Objective, each receiving four selections. Puzzle received two selections. Renovation, Sensation, and Avatar each received one selection. Learners with Intuition-Thinking cognitive style indicated a preference for Puzzle, which received five selections. Novelty, Objective, and Sensation followed with two selections each. Renovation and Avatar each received one selection.

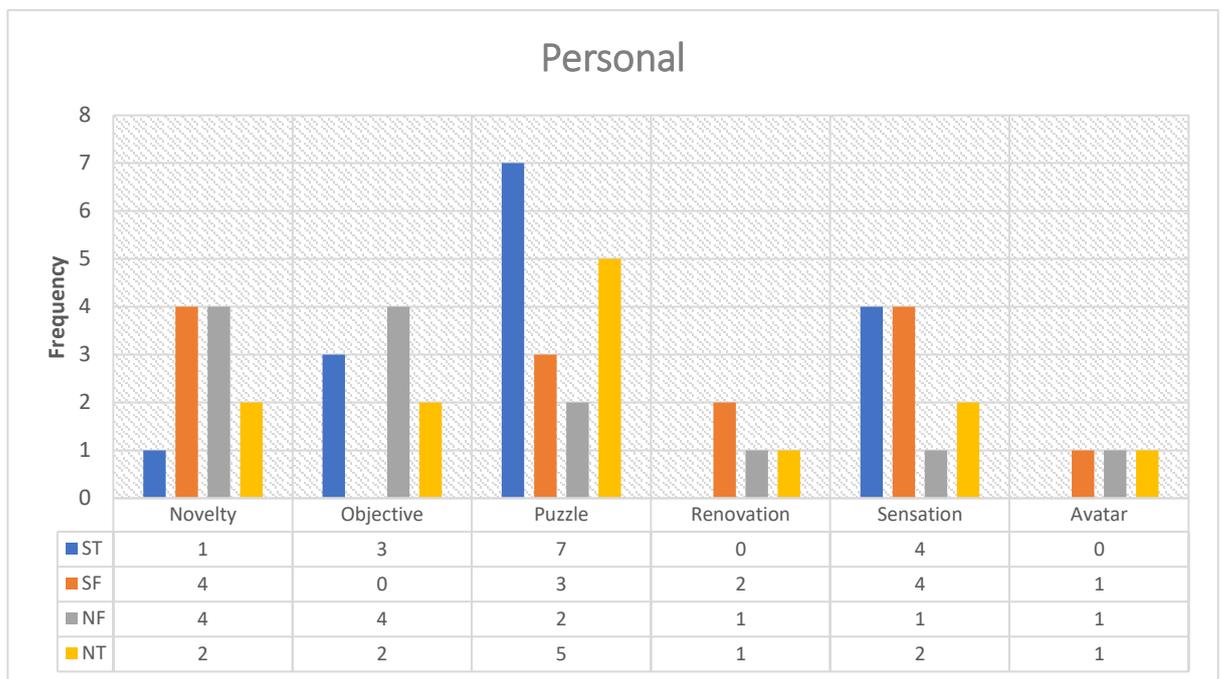

**Figure 4.4** Expert opinion results (d)

**Implication:** These results have implications for the design of personalized learning experiences based on different cognitive styles within the personal category: For learners with Sensing-Thinking cognitive style, the preference for Puzzle suggests that incorporating activities that require problem-solving, logical reasoning, and analysis can be effective in engaging them. Providing opportunities for them to engage with hands-on tasks, data analysis, or organizing information can enhance their learning experience. Learners with Sensing-Feeling cognitive style showed preferences for both Novelty and Sensation. Designing learning activities that offer new and unique experiences, as well as incorporating sensory elements like visuals or multimedia, can capture their attention and create an emotionally engaging learning environment. For learners with Intuition-Feeling cognitive style, a balance between Novelty and Objective is important. Designing activities that promote creative thinking, exploration of new ideas, and goal-oriented tasks can enhance their motivation and engagement. Learners with Intuition-Thinking cognitive style exhibited a preference for Puzzle elements. Incorporating complex problem-solving tasks, critical thinking exercises, and opportunities for creative solutions can cater to their inclination for intellectual challenges and analytical thinking.

### e. Fictional

The results for the fictional category in the study focused on examining the preferences of learners with different MBTI cognitive styles for game elements. The summarized results, illustrated in Figure 4.5, provide insights into the preferences of learners with distinct cognitive styles regarding narrative and storytelling elements: Learners with Sensing-Thinking cognitive style showed a preference for both Narrative and Storytelling, with each receiving five selections. This suggests that these learners value immersive and engaging storytelling experiences that allow them to follow a structured narrative. Learners with Sensing-Feeling cognitive style exhibited a preference for Narrative, receiving five selections. Storytelling also received significant attention, with four selections. This indicates that these learners appreciate emotionally engaging narratives and the power of storytelling in creating meaningful learning experiences. Learners with Intuition-Feeling cognitive style displayed a strong preference for Narrative, receiving seven selections. Storytelling also received four selections. This suggests that these learners value rich, imaginative narratives that stimulate their emotions and allow for deeper connections with the content. Learners with Intuition-Thinking cognitive style showed a preference for Narrative, receiving six selections. Storytelling followed closely with four selections. This indicates that these learners appreciate narratives that stimulate their intellectual curiosity and offer opportunities for critical thinking and analysis.

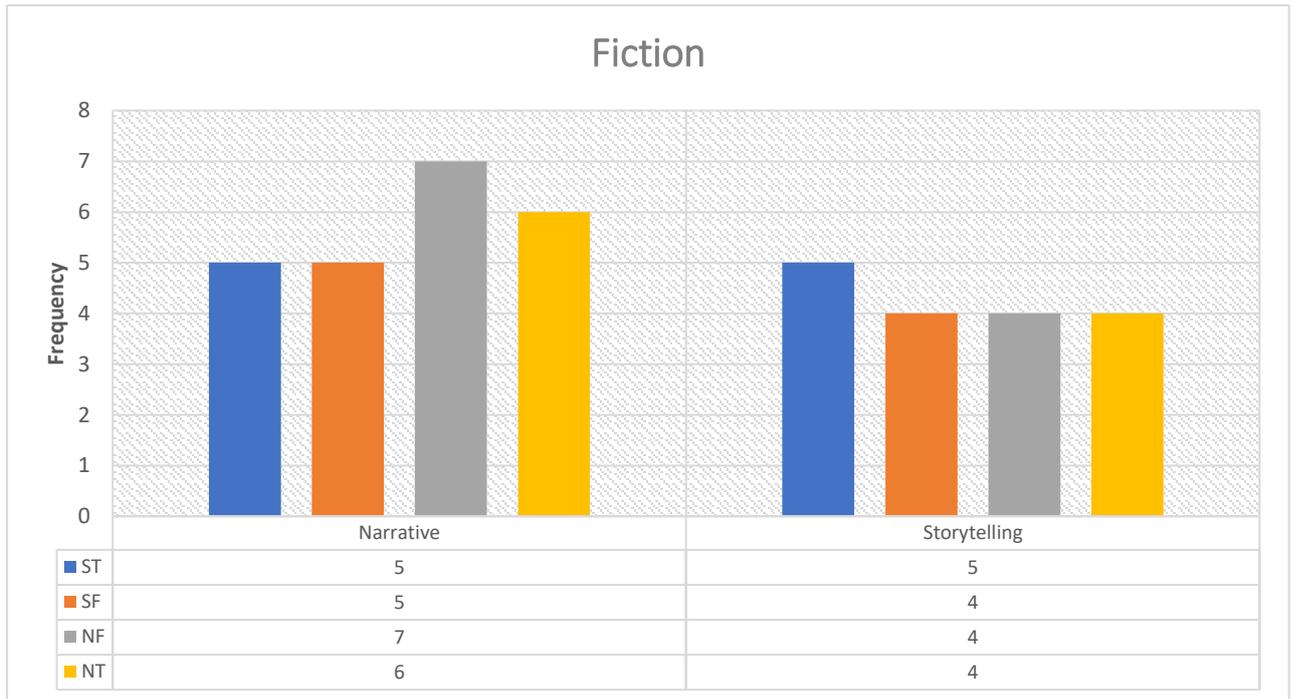

**Figure 4.5** Expert opinion results (e)

**Implication:** These results have implications for the design and integration of game elements in the Fictional category for different cognitive styles: For learners with Sensing-Thinking cognitive style, incorporating well-structured narratives and storytelling elements can enhance their engagement. Providing clear plotlines, character development, and logical story progression can create immersive experiences that resonate with their preference for organized and structured thinking. Learners with Sensing-Feeling cognitive style value narratives that evoke emotions and provide opportunities for personal connections. Incorporating storytelling techniques, such as compelling characters, dramatic arcs, and emotionally resonant themes, can enhance their learning experiences. For learners with Intuition-Feeling cognitive style, rich and imaginative narratives that allow for exploration of ideas and emotions are crucial. Designing game elements that provide deep and immersive storytelling experiences, incorporating symbolism, metaphor, and open-ended narratives, can cater to their preference for introspection and creative expression. Learners with Intuition-Thinking cognitive style appreciate narratives that stimulate their intellectual curiosity and critical thinking. Designing game elements that offer complex narratives, non-linear storytelling, and opportunities for analysis and problem-solving can enhance their engagement and satisfaction.

### 4.2.4. Developed model

As described in chapter 3, the results from the expert opinion received was used to develop the gamification model for personalized e-learning. Inputting the game elements into the equations and mapping functions from chapter 3, the following were obtained:

L = {NT, ST, NF, SF}                                                **Equation 4.1**

This equation defines the set of learner types L, which contains four elements representing the four MBTI Cognitive Core learner types: NT, ST, NF, and SF

Where:

NT is Intuition with Thinking

ST is Sensing with Thinking

NF is Intuition with Feeling

SF is Sensing with Feeling

G = (Time, Competition, Puzzle), (Economy, Stats, Puzzle), (Progression, Choice, Competition), (Acknowledgement, Choice, Sensation)}                                     **Equation 4.2**

This equation defines the set of game elements G, which contains four tuples, each consisting of n game elements.

F: L → G                                                                                     **Mapping 4.1**

### 4.2.5. Model implementation

Using the recommended game elements from the preceding section, a gamified course titled 'Instructional innovation course' was designed and integrated into the Skills and Competence Factory's open LMS as presented in the following sub-sections.

**Course**

The course was designed to be asynchronous, which influenced the selection of game elements used. Participants' profiles were mapped to personalized versions of the course based on an abridged MBTI test taken upon enrolment to the course.

**Course design and gamification elements**

The 'Instructional Innovation Course' aimed to promote participant engagement through gamification. Game elements were selected based on different personality types. The following elements were used to gamify the course for each personality category:

**ST (Sensing-Thinking)**

  i.  Economy: Participants could unluck contents having earned sufficient course points from previous activities.

 ii.  Stats: Participants could track and analyse their performance statistics.

iii.  Puzzle: Engaging puzzles were incorporated to challenge participants' problem-solving abilities.

**SF (Sensing-Feeling)**

  i.  Acknowledgement: Participants received acknowledgment and positive reinforcement for their accomplishments.

 ii.  Choice: Participants had opportunities to make choices, providing a sense of autonomy.

iii.  Sensation: Sensory elements to enhance the course experience for SF participants.

**NF (Intuition-Feeling)**

  i.  Progression: Participants progressed through different levels, representing growth and achievement.

 ii.  Choice: Participants were given choices that aligned with their preferences and values.

iii.  Competition: Competitive elements were introduced to motivate NF participants.

**NT (Intuition-Thinking)**

  i.  Time: Time constraints were imposed, encouraging efficient decision-making.

 ii.  Competition: Competitive elements stimulated NT participants to perform their best.

iii.  Puzzle: Engaging puzzles were incorporated to challenge participants' problem-solving abilities.

**Feedback mechanisms**

Apart from gamification elements, the 'Instructional Innovation Course' incorporated various feedback mechanisms to enhance the learning experience. Feedback was provided through email notifications, on-screen popups, dashboard notices, and reminders. Additionally, the course page featured a course drip, allowing learners to view the content outline and track their completion levels.

### Course content and interactive quizzes

The course included 10 video materials with a combined playback time of approximately 1 hour and 40 minutes (01:39:19) at normal speed. These videos served as the primary instructional content, conveying essential concepts and information. Interactive quizzes were implemented with gamification to assess participants' understanding and reinforce learning outcomes.

### Participant recruitment

Participants for the 'Instructional Innovation Course' were recruited using the snowballing technique. Initial participants recruited others, who were also given the opportunity to recruit additional participants. This snowball sampling method aimed to create a diverse participant pool with varying backgrounds and perspectives.

## 4.3.   Results

The subsequent sections provide a detailed presentation of results obtained at the end of the study experiment, as well as visual illustration presenting findings.

### Data collection methods

Two modes of data collection were employed to gather information about participants' experiences and the effectiveness of the course. The first mode involved a questionnaire that participants completed, providing qualitative and quantitative data on their perceptions, satisfaction, engagement, and learning outcomes. The second mode involved analysing system logs to collect objective data about participants and their interactions with the course platform. System log analysis encompassed metrics such as module completion rates, time spent, and other relevant data.

**System logs**

Analysing the system logs revealed that out of the total number of participants (n=37) who enrolled within the period of 3 months (March to May, 2022), the gender distribution of participants were 17 females (46.0%), and 20 males (54.0%). The personality distribution of participants by MBTI cognitive core was NF (n=6, 16.2%), NT (n=8, 21.6%), SF (n=8, 21.6%), and ST (n=15, 40.6%). The system logs provided crucial insights into various aspects of user engagement. These included the frequency of activities like quizzes, instances of retaking quizzes, performance trends across trials, and the time spent on each activity, all illustrated in Figure 4.6. Furthermore, Figure 4.7 presents a snapshot capturing a learner's interaction with the system. For further visual references depicting system usage, refer to Appendix v.

**Questionnaire**

A questionnaire was designed and integrated into the course and placed as the final activity of the course. The purpose of this questionnaire was to gather as much information as possible about the participants perceived experience with the gamified course. The questionnaire intended to collect data on how the research method affects the learner's engagement (using appeal, emotion, user-centricity and satisfaction as criteria) and the educational usability (based on clarity, error recognition and feedback) of the method. The questionnaire had 17 questions, each with a Likert-scale ranging from 1 to 5, where 1 indicated a negative experience and 5 indicated a positive experience. By the time the course had closed after 3 months, 23 participants (62.2%) had completed the main course activities and had gained access to the questionnaire. However, only 21 participants (91.3%) submitted their response. The resulting data for each MBTI cognitive core from the questionnaire presented in subsequent sections.

| user_id | quiz_id | score | total | date | points | points_total | percentage | time_spent | passed | course_id |
|---------|---------|-------|-------|------|--------|--------------|------------|------------|--------|-----------|
| 195 | 30381 | 0 | 1 | 13-04-22 | 0 | 10 | 0 | 1m 45s | NO | 31285 |
| 199 | 30381 | 1 | 1 | 20-04-22 | 10 | 10 | 100 | 1m 23s | YES | 31285 |
| 200 | 30381 | 0 | 1 | 20-04-22 | 0 | 10 | 0 | 41s | NO | 31285 |
| 201 | 30381 | 0 | 1 | 20-04-22 | 0 | 10 | 0 | 35s | NO | 31285 |
| 202 | 30381 | 1 | 1 | 20-04-22 | 10 | 10 | 100 | 40s | YES | 31285 |
| 110 | 30386 | 2 | 2 | 11-03-22 | 10 | 10 | 100 | 32s | YES | 30142 |
| 141 | 30386 | 1 | 2 | 09-03-22 | 5 | 10 | 50 | 55s | NO | 30142 |
| 141 | 30386 | 1 | 2 | 09-03-22 | 5 | 10 | 50 | 47s | NO | 30142 |
| 141 | 30386 | 1 | 2 | 09-03-22 | 5 | 10 | 50 | 41s | NO | 30142 |
| 141 | 30386 | 2 | 2 | 09-03-22 | 10 | 10 | 100 | 26s | YES | 30142 |
| 145 | 30386 | 2 | 2 | 20-03-22 | 10 | 10 | 100 | 55s | YES | 30142 |
| 101 | 30386 | 1 | 2 | 09-05-22 | 5 | 10 | 50 | 45s | NO | 31285 |
| 101 | 30386 | 0 | 2 | 09-05-22 | 0 | 10 | 0 | 18s | NO | 31285 |
| 101 | 30386 | 1 | 2 | 09-05-22 | 5 | 10 | 50 | 17s | NO | 31285 |
| 101 | 30386 | 2 | 2 | 09-05-22 | 10 | 10 | 100 | 16s | YES | 31285 |
| 102 | 30386 | 2 | 2 | 04-03-22 | 10 | 10 | 100 | 2m 10s | YES | 31285 |
| 104 | 30386 | 1 | 2 | 28-02-22 | 5 | 10 | 50 | 36s | NO | 31285 |
| 104 | 30386 | 2 | 2 | 28-02-22 | 10 | 10 | 100 | 10s | YES | 31285 |
| 106 | 30386 | 1 | 2 | 22-04-22 | 5 | 10 | 50 | 28s | NO | 31285 |
| 106 | 30386 | 2 | 2 | 22-04-22 | 10 | 10 | 100 | 12s | YES | 31285 |
| 107 | 30386 | 2 | 2 | 21-04-22 | 10 | 10 | 100 | 39s | YES | 31285 |
| 123 | 30386 | 2 | 2 | 02-03-22 | 10 | 10 | 100 | 19s | YES | 31285 |
| 124 | 30386 | 2 | 2 | 23-03-22 | 10 | 10 | 100 | 37s | YES | 31285 |
| 128 | 30386 | 2 | 2 | 08-04-22 | 10 | 10 | 100 | 25s | YES | 31285 |
| 140 | 30386 | 2 | 2 | 09-03-22 | 10 | 10 | 100 | 36s | YES | 31285 |
| 144 | 30386 | 2 | 2 | 15-03-22 | 10 | 10 | 100 | 1m 9s | YES | 31285 |
| 171 | 30386 | 2 | 2 | 28-03-22 | 10 | 10 | 100 | 7s | YES | 31285 |
| 171 | 30386 | 1 | 2 | 28-03-22 | 5 | 10 | 50 | 26s | NO | 31285 |

**Figure 4.6** Screenshot of system log

**Figure 4.7** Screenshot of learner viewing course report

### 4.3.1. Educational usability

The system was assessed by learners categorized under the four different MBTI cognitive core groups, and their mean ratings for various criteria were obtained. The results are as follows: Intuition with Thinking group gave a rating of 4.0 for clarity, 4.6 for error correction, and 4.8 for feedback. Intuition with Feeling group provided a rating of 3.7 for clarity, 4.6 for error correction, and 4.8 for feedback. Sensing with Thinking group gave a rating of 3.5 for clarity, 4.5 for error correction, and 4.4 for feedback. Sensing with Feeling group gave a rating of 4.1 for clarity, 4.8 for error correction, and 5.0 for feedback. These ratings were visualized in Figure 4.8, illustrating the differences in mean ratings across the different MBTI groups for clarity, error correction, and feedback criteria during the system assessment.

### 4.3.2. Engagement

Learners provided mean ratings to assess their engagement with the system, and the results for different MBTI groups are as follows: Intuition with Thinking group rated the system 4.3 for appeal, 4.4 for emotion, 4.3 for user-centricity, and 4.4 for satisfaction. Intuition with Feeling group gave a rating of 4.2 for appeal, 4.3 for emotion, 4.4 for user-centricity, and 4.2 for satisfaction. Sensing with Thinking group rated the system 3.9 for appeal, 4.2 for emotion, 4.4 for user-centricity, and 4.0 for satisfaction. Sensing with Feeling group provided a rating of 4.8 for appeal, 4.7 for emotion, 4.8 for user-centricity, and 4.6 for satisfaction. These results are visualized in Figure 4.9, which presents a graphical representation of the mean ratings for appeal, emotion, user-centricity, and satisfaction across the different MBTI groups. Additionally, Table 4.1 provides a statistical summary of the results, encompassing both educational usability and engagement factors.

### 4.4. Evaluation of Results

The results obtained from the learners review of the system are evaluated in the subsequent sections as follows:

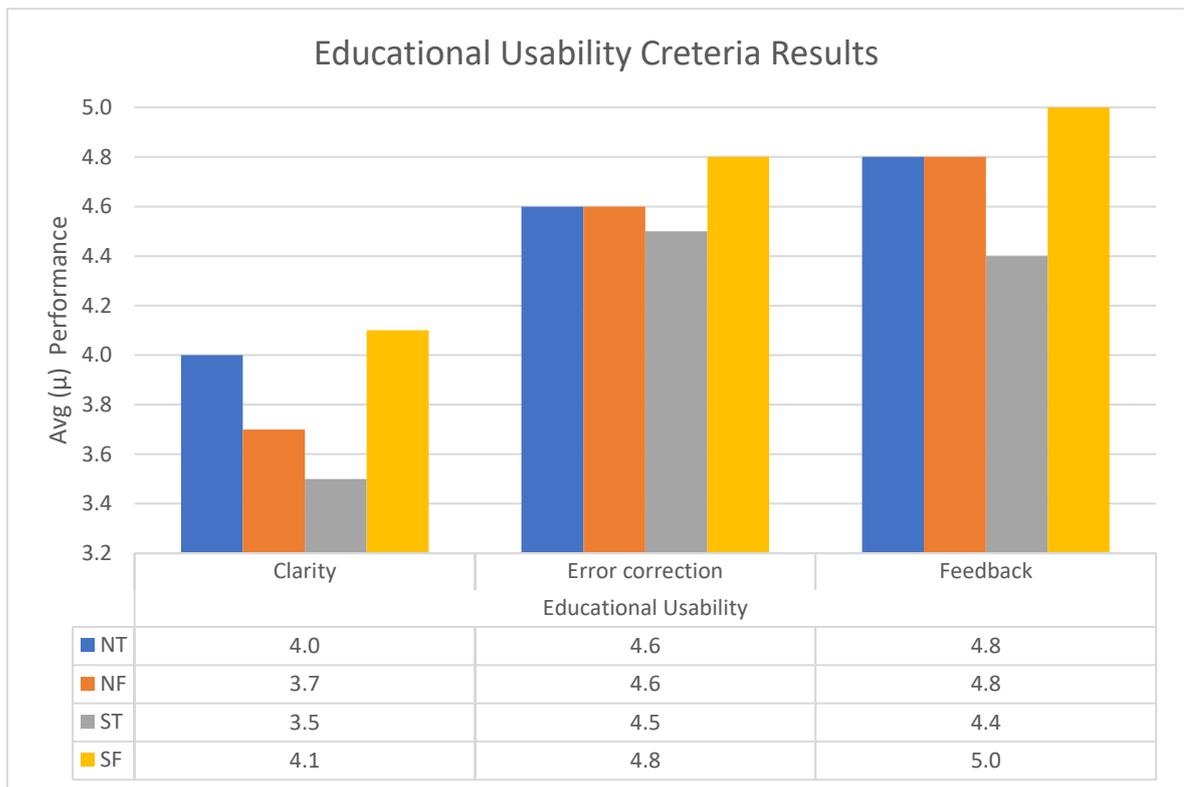

**Figure 4.8** Results from student survey questionnaire (a)

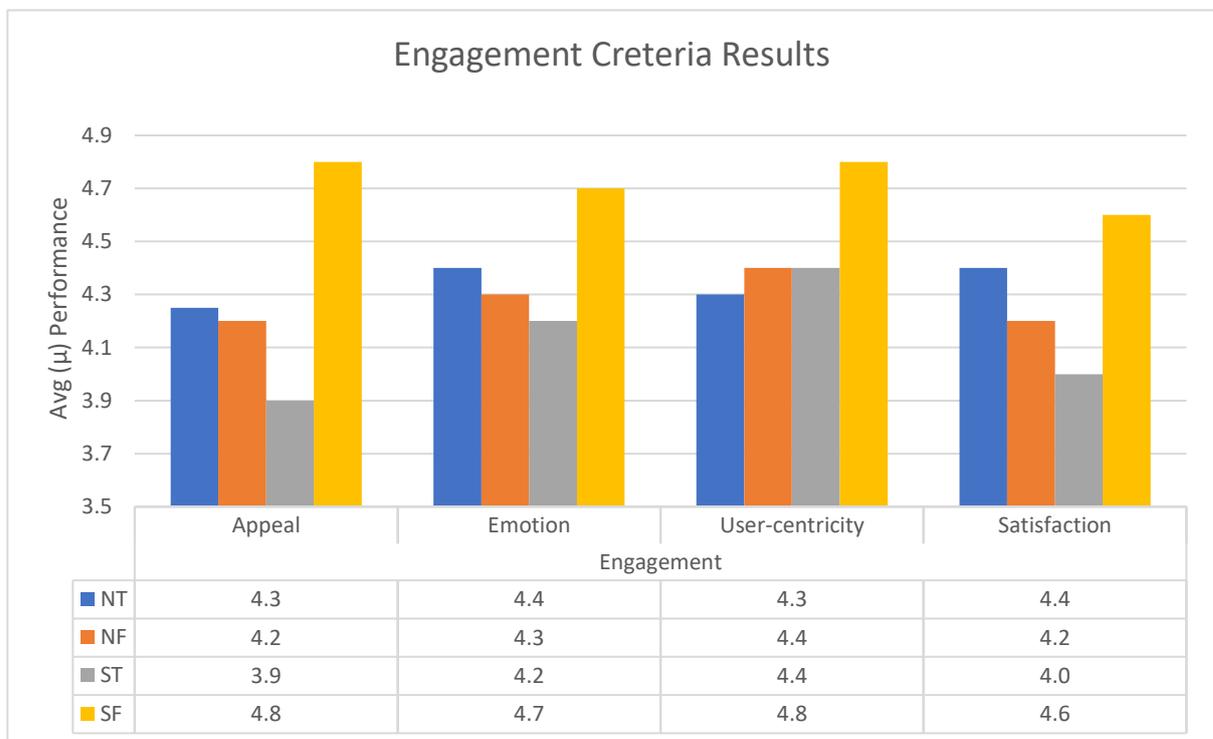

**Figure 4.9** Results from student survey questionnaire (b)

### 4.4.1. Engagement

When analysing the results of the Likert scale ratings for learners' engagement and educational usability, statistical ranges of negative (1 to 2.6), neutral (2.6 to 3.4), and positive (3.4 to 5) were used. The score for each criterion was calculated by taking the mean ($\mu$) of the statement scores for that criterion. All four criteria in the learners' engagement category were rated positively. The criterion in the engagement category that received the highest overall mean rating as shown in Table 4.1 was emotion ($\mu = 4.5$) with a standard deviation (s.d) = 0.83. Both user-centricity and satisfaction received the same mean rating ($\mu = 4.4$) with s.d = 0.86 and 0.92 respectively while appeal received the least mean rating ($\mu = 4.3$), with a s.d = 1.03. The positive results for the emotion category indicate that the participants found the gamified course very interesting and generally engaging. The result was also evaluated based on the four learner types as visualized in Figure 4.10. Participants identified as sensing with feeling gave the system highest mean rating ($\mu = 4.7$), followed by intuition with thinking ($\mu = 4.4$), sensing with thinking scored ($\mu = 4.3$) while intuition with feeling gave a rating of ($\mu = 4.0$).

### 4.4.2.    Educational usability

For the educational usability criteria, feedback received the highest mean rating ($\mu = 4.8$) with s.d = 0.51. Error recognition / correction was rated ($\mu = 4.0$) with s.d = 0.71 while clarity received the least mean rating ($\mu = 3.9$) with s.d = 1.18. The results, when viewed by personality rating as in Figure 4.10, shows that learners who were sensing with feeling gave the highest rating ($\mu = 4.7$), both intuition with thinking and sensing with thinking gave a mean rating of ($\mu = 4.5$) while intuition with feeling gave the system a rating of ($\mu = 4.4$) for educational usability. Overall, feedback was the criterion with the highest mean rating ($\mu = 4.8$) while clarity received the least rating ($\mu = 3.9$), which is still considered a positive rating.

### 4.5. Discussion of Findings

The reported results from the research on gamification of personalized e-learning using MBTI cognitive core (function pairs) have revealed distinct patterns in learners' rankings, with Sensing-

Table 4.1: Summarized Results for System Engagement and Educational Usability

| S/N | Criteria | Statements | AVG. (μ) | SD (σ²) | AVG. (μ) | SD (σ²) | Min. | Max. |
|-----|----------|-----------|----------|---------|----------|--------|------|------|
| 1 | User-centricity | The activities increased my interaction with the course | 4.6 | 0.66 | 4.4 | 0.86 | 3 | 5 |
| 2 | | The activities encouraged me to revisit the course materials | 4.5 | 0.66 | | | 3 | 5 |
| 3 | | I tried to beat my completion time for activities | 4.0 | 1.09 | | | 2 | 5 |
| 4 | Emotion | The activities were worth time spent | 4.5 | 1.05 | 4.5 | 0.83 | 1 | 5 |
| 5 | | The activities/quizzes added fun to the course | 4.4 | 0.73 | | | 3 | 5 |
| 6 | | Achievement points or badges earned were motivating | 4.4 | 0.66 | | | 3 | 5 |
| 7 | Appeal | The course interface is easy to navigate and use | 4.3 | 1.03 | 4.3 | 1.03 | 2 | 5 |
| 8 | | The course seemed to be designed for my learning style | 4.3 | 1.03 | | | 1 | 5 |
| 9 | Satisfaction | I enjoyed the entire course experience | 4.5 | 0.79 | 4.4 | 0.92 | 3 | 5 |
| 10 | | I enjoyed the hands-on nature of the course project | 4.2 | 1.02 | | | 1 | 5 |
| 11 | Clarity | The goals of each activity/quiz were clear | 4.6 | 0.66 | 3.9 | 1.18 | 3 | 5 |
| 12 | | The final course project was difficult to complete | 3.2 | 1.18 | | | 1 | 5 |
| 13 | Error recognition / correction | The course activities/quizzes gave me the chance to learn from mistakes | 4.5 | 0.91 | 4.7 | 0.71 | 1 | 5 |
| 14 | | Activities are graded with grades providing instant feedback and correction | 4.9 | 0.35 | | | 4 | 5 |
| 15 | Feedback | Prompt feedback was provided | 4.8 | 0.43 | 4.8 | 0.51 | 4 | 5 |
| 16 | | The progress bar helped me keep track of my advancement in the course | 4.9 | 0.47 | | | 3 | 5 |
| 17 | | My advancement on the progress bar also encouraged me to continue the course | 4.8 | 0.61 | | | 3 | 5 |

(Engagement spans rows 1–10; Educational Usability spans rows 11–17)

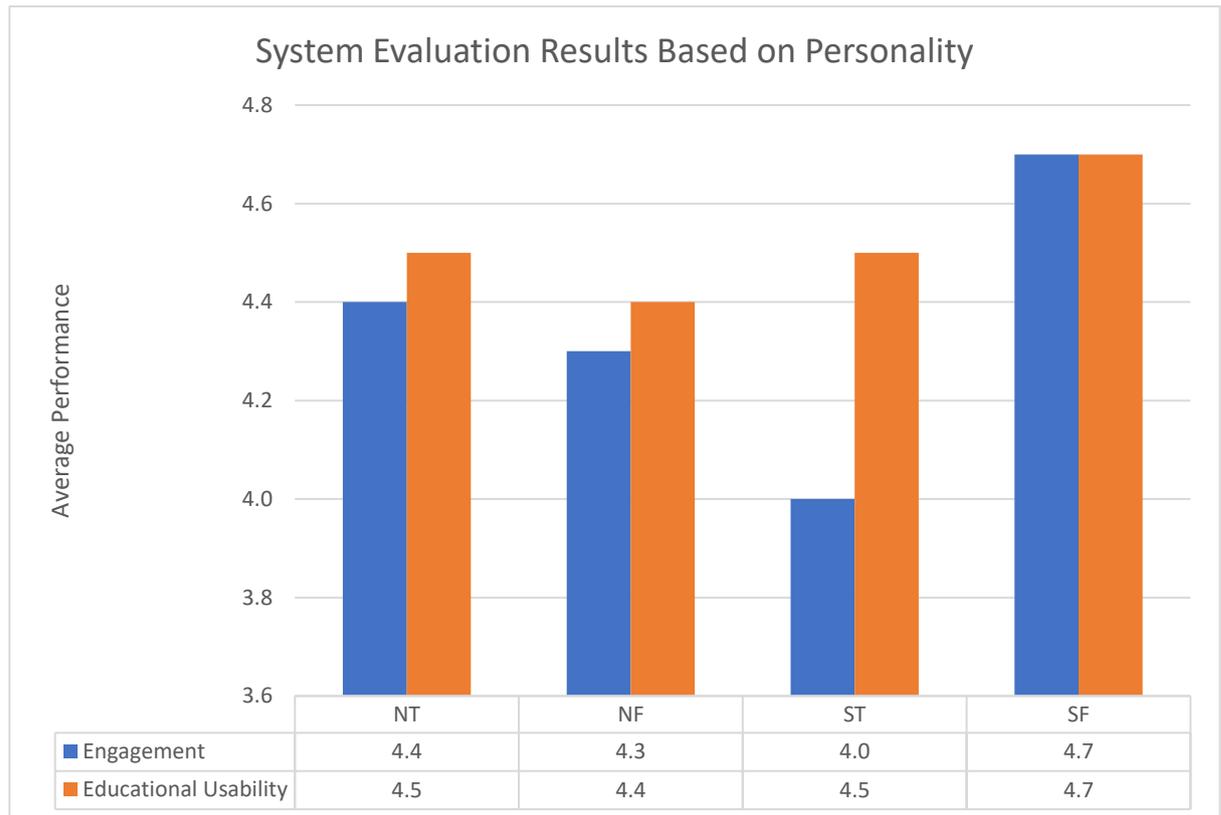

**Figure 4.10** Overall system results

Feeling (SF) individuals consistently reporting better results than other cognitive function pairs. These patterns can be discussed and interpreted through several potential reasons within the context of engagement and educational usability.

**Engagement patterns:**

The observed ranking of engagement criteria suggests that participants across all cognitive function pairs generally found the gamified course engaging and interesting. Notably, the "emotion" criterion received the highest mean rating, indicating that participants perceived the course as emotionally stimulating and captivating. The positive results suggest that the gamification approach successfully piqued participants' interest and maintained their engagement throughout the learning process. When categorizing the results by personality type, participants identified as Sensing with Feeling gave the highest mean rating for engagement. This could be attributed to the personalized gamification strategy aligning well with their cognitive preferences. Sensing individuals tend to focus on practical, tangible experiences and value interactions that evoke emotions. The personalized approach could have resonated strongly with their learning style, leading to higher levels of engagement and interest.

**Educational usability patterns:**

Moving on to the educational usability criteria, the findings reveal that feedback received the highest mean rating, indicating that participants perceived the system as providing effective feedback on their progress. The relatively lower ratings for "clarity" and "error recognition / correction" suggest potential areas for improvement in terms of user understanding and error-handling mechanisms. Nevertheless, these ratings remain positive overall, indicating that participants found the course usable and conducive to their learning experience. When analysed by personality type, a similar trend emerges. Sensing with Feeling individuals gave the highest mean rating for educational usability. This could be attributed to their preference for practical and experiential learning, aligning with the system's feedback-oriented approach. Their emphasis on emotions might have made them more receptive to feedback, enhancing their perception of educational usability.

**Interpretation and implications:**

The consistent trend of SF individuals reporting better results might be attributed to the alignment between the personalized gamification approach and their cognitive preferences. Sensing individuals tend to focus on concrete experiences, while Feeling individuals value emotional resonance. The gamified activities might have been designed to cater to these preferences, resulting in enhanced engagement and usability.

These findings underscore the importance of catering to diverse learning styles when implementing gamification strategies. Designing personalized experiences that resonate with learners' cognitive functions can lead to improved engagement and positive educational outcomes. Future research could delve deeper into the specific elements of gamification that align with different cognitive function pairs, providing insights into tailoring educational experiences to individual learners.

To further understand the results of the study and to analyse them in relation to existing research, it is necessary to explain that a version of the proposed model was put into practice with some of the proposed game elements that fit the structure of the course. This is because not all the recommended game elements were deemed feasible due to the size and asynchronous nature of the course that was implemented. The findings of the study indicate that learners identified as sensing with feeling generally benefited more from the approach compared to the other personalities, as evidenced by the higher ratings they gave the system in Figure 4.10. However, learners with other personalities also rated the system favourably, suggesting that personalized gamification was useful for all types of learners. Educational usability was rated slightly higher than engagement as visualized in Figure 4.11, which could imply that learners found the system to be helpful in their learning process. The evaluated criteria are further discussed in this section. The findings suggest that learners found the tailored gamification of their learning experience to be suitable for their individual learning styles. This is in line with literature that demonstrates users find personalized gamified versions of learning systems to be impactful in enhancing their learning outcomes (Aljabali, et al., 2020). Furthermore, research has demonstrated that users are more likely to have a better learning experience and improved learning results when the material presented in online courses is in accordance with their learning style (El-Bishouty, et al., 2019). Additionally, the results suggest that users found the personalized gamification experience to be emotionally appealing. This is inconformity with literature which indicates that games can evoke a wide range of emotions, such as joy, excitement, frustration, and anger (Lazzaro, 2004; Werbach & Hunter, 2012). Moreover, literature suggests that games can minimize the perceived stakes of a task (Lee & Hammer, 2011) and redefine failure as an opportunity for learning (Tzouvara & Zaharias, 2013). This suggests that gamification can be an effective tool for engaging users and motivating them to complete tasks. The research has further strengthened the idea that in order to motivate system users a, it is important to focus on the individual preferences of the users and design a suitable gamification element (Burgers, Eden, van-Engelenburg, & Buningh, 2015; Hamari & Koivisto, 2015; Kamunya, et. al., 2020). The positive perception of the learners towards the system's appeal that has been revealed through this research further supports the notion that gamification can be an effective tool to engage users and make them more interested in the system. By incorporating gamification elements into the system, it can help to create a more enjoyable and engaging experience for the users, thus increasing their motivation and engagement with the system.

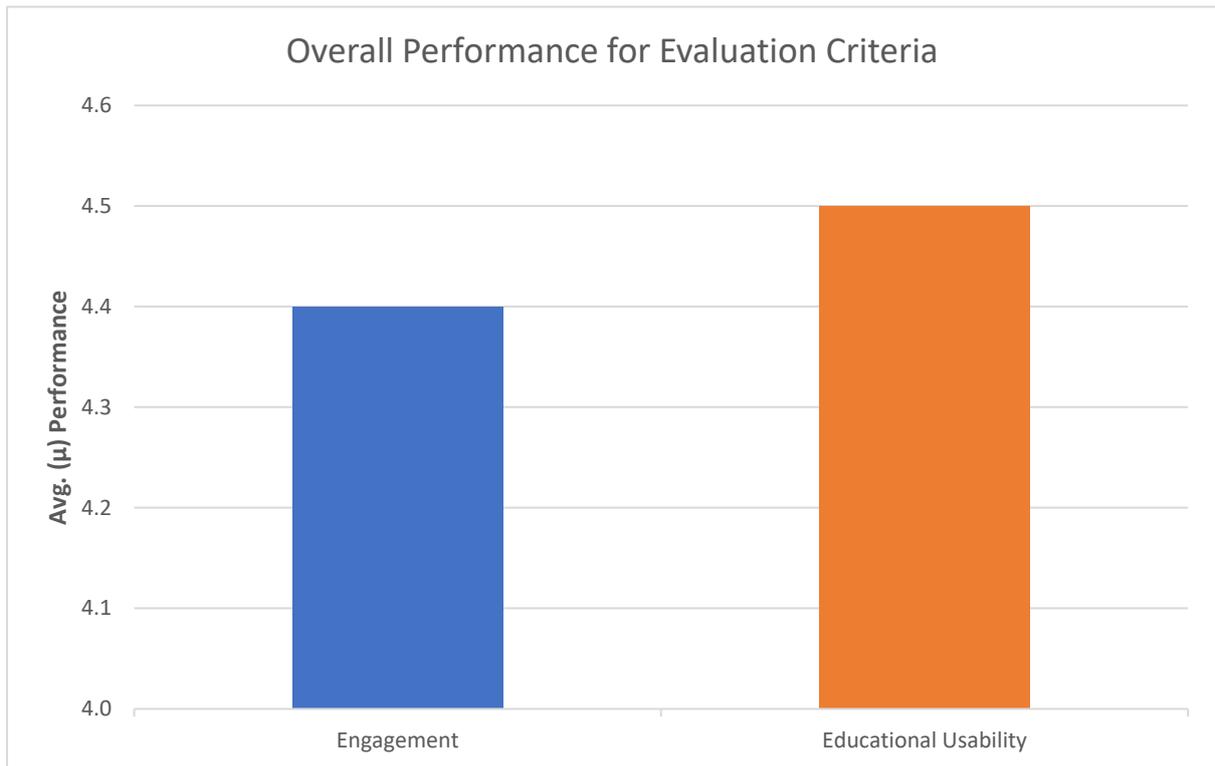

**Figure 4.11** Overall engagement and educational usability

In the area of perceived satisfaction, the results of the study demonstrate that learners are more likely to remain engaged with personalized experiences when they feel that their needs are being met. Personality and games have been identified as key components in user satisfaction in previous studies (Johnson & Gardner, 2010; Tondello, et al., 2016). These studies suggest that providing personalized experiences that are tailored to the individual's needs and interests can help to increase user engagement and satisfaction. Furthermore, incorporating elements of gaming into the learning experience can help to make the experience more enjoyable and engaging. By providing personalized experiences that meet the needs of the learner, educators can ensure that learners remain engaged and motivated to continue learning.

# CHAPTER FIVE

## SUMMARY, CONCLUSION AND RECOMMENDATION

### 5.1.    Summary

Drawing to a close, this research presents an innovative gamification model for personalized e-learning, anchored in the foundational principles of the MBTI cognitive core. The research has interwoven individual profiles with gamification strategies to achieve heightened engagement and improved learning outcomes. This breakthrough approach signals a pivotal advancement in the realm of personalized education. By uniting personality-based customization and game elements in learning, this research shapes a distinctive trajectory for the evolution of e-learning methodologies, promising enduring implications for educational landscapes.

### 5.2.    Conclusion

The research underscores the resounding success of the gamification model hinged on the MBTI cognitive core. Findings highlight not only elevated engagement but also improved knowledge acquisition. The alignment of learners' cognitive preferences with game elements serves as a beacon for future personalized e-learning endeavours. The conclusive insights validate the effectiveness of this innovative approach, bolstered by the seamless union of personality traits and gamification strategies.

### 5.3.    Contribution to Knowledge

The research has made a substantial contribution to the field of Information Systems by introducing a pioneering model that gamifies e-learning through the utilization of the MBTI cognitive core. This innovative approach not only broadens the horizons of personalized education but also provides valuable insights into the integration of cognitive preferences and engaging educational strategies within the Information Systems framework. This model stands as a foundational milestone, offering a robust platform for further investigations and advancements in the dynamic intersection of cognitive profiling and interactive learning methodologies.

## 5.4. Recommendation

Based on the findings of this study, it is recommended that e-learning platforms incorporate personalized gamification strategies that align with learners' cognitive function pairs, particularly emphasizing the preferences of Sensing-Feeling individuals. This approach can enhance engagement and educational usability, catering to diverse learning styles. Future research could explore the specific gamification elements that resonate with different cognitive preferences, further optimizing personalized learning experiences. Additionally, investigating how other personality traits intersect with gamification could provide a more comprehensive understanding of effective strategies for fostering engagement and motivation in e-learning.

**APPENDICES**

**i. Literature review table**

| SN | Author/Year | Title | Methodology | Result | Remark |
|---|---|---|---|---|---|
| 1 | Abbasi *et. al.* (2021) | Personalized Gamification in E-Learning with a Focus on Learners' Motivation and Personality | A pretest-posttest experimental design was used to assess effect of gamification using 5 groups: 3 personalized gamification groups (PG) (motivation and personality, personality, motivation), non-personalized gamification, and a control group | Students in the PG groups significantly outperformed the students in the other groups on the posttest. The students in the PG groups also spent more time engaged in the learning process than the students in the other groups. | Provides valuable insights into the potential of personalized gamification for improving the effectiveness of e-learning. |
| 2 | Adewale, Agbonifo & Osajiuba (2019) | Development of a Myers-Briggs Type Indicator Based Personalised Elearning System | The study proposes a personalized teaching strategy for e-learning systems based on the MBTI dominant cognitive functions. | 66% of the students ranked the system user friendly, 30% ranked it not user friendly while 4% were not sure. 52% rated it very good, 44% rated good. 96% reported easy navigation. | Freedom to fail enables learners revisit unclear learning materials for better understanding |
| 3 | Alexiou & Schippers (2018) | Digital game elements, user experience and learning: A conceptual framework | The authors conducted a literature review to identify relevant articles on game design, user experience, and learning, which served as the basis for the proposed framework. The conceptual framework consists of three main components: game design elements, user experience, and learning outcomes. | The results of the study demonstrated that game design elements play a significant role in shaping user experience and learning outcomes for digital games. The authors emphasized the importance of considering the relationship between game design elements, user experience, and learning outcomes in the development of digital games for education. | Further research is required to test the effectiveness of the proposed framework |
| 4 | Aljabali *et. al.*, (2020) | An experimental study: personalized gamifed learning based on learning style | Applied FSLS model as personalization parameter along with 10 game elements to propose a personalized gamified learning model and examine the proposed model in improving the students' scores in Data Flow Diagram | The experimental group performed better than the control group with a mean score of 31.89 against 29.78 out of 50 in a test, while reporting a mean rating of 4.22 against 4.19 out of 5 for | The evaluation process was undertaken based on the experiment in a 1-hour class. It remains to be seen how such gamification play out |

| | | | (DFD) lesson during class learning process. | perceived usefulness of the gamified course. | in asynchronous setting. |
|---|---|---|---|---|---|
| 5 | Amirhosseini & kazemian (2020) | Machine Learning Approach to Personality Type Prediction Based on the Myers-Briggs Type Indicator | The study used a supervised learning approach, with the MBTI personality types as the target variables and the textual contents of social media posts as the independent variables. The dataset was pre-processed to remove stop words, punctuation and converted to lowercase. | The authors compared the performance of several machine learning algorithms, including Random Forest, Naive Bayes, and Support Vector Machines. The results showed that the Random Forest algorithm achieved the best performance in predicting personality types, with an accuracy of 83.6% on the test set. | The study provided insights into the effectiveness of machine learning algorithms in predicting personality types based on social media data. |
| 6 | Bachari, *et. al.* (2011) | E-learning personalization based on dynamic learners' preference | Dominant MBTI cognitive functions were used as learning styles. Each group was designed a distinct learning strategy and different media formats which would be revised with time and changed if and when needed | From an experiment conducted during the research, 24 students each formed the control and experimental groups. The experimental group had a mean score of 14.52 of 20 while the control group had a mean score of 12.02 out of 20. | The study provides a foundation for personalizing e-learning using MBTI, but does not feature gamification |
| 7 | Böckle *et. al.* (2018) | A Design Framework for Adaptive Gamification Applications | Used design science method to design a versatile gamification framework purpose of adaptivity, adaptivity criteria adaptive interventions, as well as adaptive game mechanics and dynamics. | Implemented model increased the overall activity engagement on the system | The framework could serve as basis for integrating gamification to systems in different contexts |
| 8 | Buckley *et. al.* (2018) | A Gamification – Motivation Design Framework for Educational Software Developers | Used deductive process to map game elements to components of SDT | In a survey presented to participants, the average level of agreement was 70.9% for competence, 60.8% for Autonomy, and 57.2% for Relatedness | Framework included game elements such as combat and boss fight, which may not be applicable in educational software systems |
| 9 | Butler (2014) | A framework for evaluating the effectiveness of gamification techniques by personality type | Behavioral tendencies of each personality trait in the MBTI model was grouped with its accompanying game elements | The framework consisted of a number of tendencies and accompanying mechanic that can potentially improve gamification attempts. | Implementing and evaluation will help validate the |

| | | | | effectiveness of the framework |
|---|---|---|---|---|
| 10 | Çelik (2015) | Development of Usability Criteria for E-learning Content Development Software | Used a mixed-method approach to develop usability criteria for e-learning content development software. First phase involved a literature review to identify existing usability criteria for e-learning software. Next, a survey of pre-service instructional designers to identify the most important usability criteria for e-learning software. The third phase of the study involved a user study to evaluate the usability of 15 leading e-learning content development software. | The overall findings of the suggested 31 usability criteria classified under three categories as technical (14), media (8), and assessment (9) competence levels for the e-learning content development tools. The study also found that the usability of the software was affected by the user's experience level. Experienced users were able to use the software more easily than inexperienced users. | The study only focused on the usability of the software. The study did not consider the pedagogical effectiveness of the software. It is possible that a software that is easy to use is not necessarily effective for learning |
| 11 | El-Bishouty *et. al.* (2018) | Use of Felder and Silverman learning style model for online course design | Study investigated impact of designing online courses based on Felder and Silverman learning style model. Combination of learning style questionnaires and academic performance data used to examine relationship between learning style prefs. and course performance. | Students who were taught according to their learning style preferences had a significantly higher average grade than those who were not. The research also found that students who had access to course materials designed according to their learning style had a higher level of satisfaction with the course. | The results of the study establish a foundation for designing more student-centric e-learning courses. |
| 12 | Freitas, (2021) | Gamification for MOOC online courses | The study used a systematic literature review to investigate the effects of gamification on Massive Open Online Courses (MOOCs). The study searched for papers published in peer-reviewed journals and conference proceedings from 2014 to 2020 | The results of the study showed that gamification can have a positive effect on student engagement, motivation, and learning outcomes in MOOCs. The study found that the most commonly used gamification elements in MOOCs were points, badges, and leaderboards. The study also found that gamification can be effective in MOOCs with a variety of student populations, including traditional-age students, working adults, and lifelong learners. | Study did not consider the impact of other factors, such as the design of the MOOC, the quality of the content, and the instructor's teaching style, on the effects of gamification. |

| 13 | Hassan *et. al.* (2019) | Adaptive gamification in e-learning based on students' learning styles | Gamification elements were matched to the learning dimensions of the FSLS model | The results indicated a 25.49% increase in student motivation level and 22% more student interaction with the system | The system does not provide feedback on uncompleted activity. Feedback could potentially increase engagement levels with the system. |
|----|----|----|----|----|----|
| 14 | Hurpur & de Villiers (2015) | MUUX-E, a framework of criteria for evaluating the usability, user experience and educational features of m-learning environments | The study used a design-based research approach to develop a framework for evaluating the usability, user experience, and educational features of mobile learning environments. The researchers evaluated the MUUX-E framework using a case study of an m-learning environment for teaching mathematics to primary school students. | The researchers found that MUUX-E was a useful tool for evaluating the m-learning environment. The researchers were able to identify strengths and weaknesses of the m-learning environment, and make recommendations for improvement. | The framework could be used to guide the design and development of m-learning environments effectively. |
| 15 | Issa & Jusoh (2019) | Usability Evaluation on Gamified E-Learning Platforms | The study used a mixed-method approach, which involved both quantitative and qualitative data collection methods. Quantitative data was collected through a survey of 10 undergrad students from programming class students. Qualitative data was collected through interviews with all the participants. | The students found that the gamified elements of the platform helped to increase their motivation and engagement in the learning process. The students also found that the gamified platform was easy to use and navigate. | Impact of factors such as the students' learning style, on the effectiveness of the gamified e-learning platform were not considered. |
| 16 | Kamunya *et. al.* (2020) | An Adaptive Gamification Model for E-Learning | Used DSRM that includes four phases to develop and analyze a gamified intervention. The design, development, and research of this intervention is conducted through iterative cycles that involve collecting feedback, observations, and analyzing the impact of the model on student engagement. | The study developed an adaptive gamification model to guide and implement adaptivity within e-learning platforms. The model was developed through a literature review and expert consultations. The researchers also tested the model through a prototype developed for a digital logic gates platform. | Not Implemented. Provides valuable evidence that adaptive gamification can be an effective way to improve student engagement and performance in e-learning. |

| 17 | Kang and Kusuma (2020) | The effectiveness of personality-based gamification model for foreign vocabulary online learning | An online system was developed by applying the Big Five, game elements, FSLS, and the ARCS Motivation to help increase students' academic achievement and motivation. | The mean rank for the test scores were 41.64 for the hybrid model and 25.36 for the non-hybrid model. As for the motivation scores, the hybrid model had 43.86 mean rank while the non-hybrid model had 23.14 mean rank. | Extraversion & Neuroticism were considered non-gamifiable |
|----|----|----|----|----|----|
| 18 | Ofosu-Asare, Essel & Bonsu (2019) | E-learning Graphical User Interface Development using the Addie Instruction Design Model and Developmental Research: The Need to Establish Validity and Reliability | The study used a mixed-method approach, which involved both quantitative and qualitative data collection methods. The quantitative data was collected through a survey of students who used an e-learning course with a GUI that was developed using the ADDIE instruction design model. The qualitative data was collected through interviews with the students who used the e-learning course | The results of the study showed that the developed GUI was valid and reliable, as evidenced by the positive user acceptance and satisfaction scores. The study also identified some design features that needed improvement to enhance the usability and user experience of the GUI. | It is unclear the impact of the developed GUI on student learning performance. |
| 19 | Piteira, Carlos & Aparicio (2017) | A Conceptual Framework to Implement Gamification on Online Courses of Computer Programming Learning: Implementation | The framework was designed based on target, goals and learning outcomes, contents, educational learning principles, game mechanics, cognitive absorption, flow and personality | The gamifice framework provided overall increased experience with the course | Lack of personalization led to some learners getting demotivated by certain game elements |
| 20 | Santoso, Puntra & Hendra (2021) | Development & Evaluation of E-Learning Module Based on Visual and Global Preferences Using a User-Centered Design Approach | The study used a user-centered design approach to develop a prototype e-learning module using Figma that catered to both visual and global learning styles of FSLS. A content analysis was conducted to identify the essential components of the e-learning module. A usability test and a survey were then conducted to evaluate the effectiveness of the module. | The study found that the e-learning module was effective in catering to the visual and global learning preferences of the participants. The usability test showed that the module had a high level of usability, while the survey revealed that the participants were satisfied with the module's design and its ability to cater to their learning styles. | The study was limited to the use of design prototypes. |
| 21 | Scholtz, Raga, & Baxter (2016) | Design and Evaluation of a "Gamified" System for | Designed a gamified computer science career information system based on an extended MDA framework | 4 out of 5 the gamified tasks in the system recorded 100% success rate while 1 task recorded 83% success rate. | The study suggests users are more comfortable with |
| | | Improving Career Knowledge in Computing Sciences | | The 3 evaluation criteria: usability, ux and educational usability all received favourable ratings by the participants | textual content than audio contents. |
| 22 | Sincharoenkul & Tongtep (2022) | A Conceptual Framework of the Assessment of the Effectiveness of Gamification in Learning Process | Used MBTI and ILS to determine personalized learning styles while Flow theory to assess the effectiveness of gamification of learning based on board games. | A framework for assessing effectiveness of gamification strategy in learning was developed. | Personality traits impact effectiveness of game characteristics in learning process. |
| 23 | Zaric *et. al.* (2021) | Gamified Learning Theory: The Moderating role of learners' learning tendencies | FSLS model was used with gamification to boost students' engagement by triggering the intuitive behavior in sensing students in a non-linear course setup with Moodle LMS. | Results from an experiment set up to assess online activeness, online engagement, emotional engagement, effort invested, effort regulation, accomplishment, challenge, and playfulness indicates experimental group performed better and enjoyed the system more than the control group. | Gamification can produce change only if the targeted behavior can itself be changed. |

**ii.    Questionnaire for expert opinion**

Title: Personalized Gamification for E-learning

Dear Sir/Ma

Thank you for being part of this survey being conducted by Ibisu Afvensu Enoch of Department of Computer Science & Engineering in Obafemi Awolowo University, Ile-Ife, Nigeria on 'Developing a Gamification Model for Personalized E-learning'. The survey aims to document expert opinions on preferred game elements by different personalities for E-learning.

All information provided will be treated with the strictest confidence.

*Remember, there is no right or wrong answer. You could expect this survey to take 5 to 10 minutes.

**Name*:** ___________________________________________

Please select the option(s) that BEST describes your specialty*:

☐    General psychology                    ☐    Behavioural psychology
☐    Human Computer Interaction            ☐    E-learning
☐    Education psychology                  ☐    Game-based learning
☐    Software / System Development         ☐    Gamification
☐    Cognitive psychology                  ☐    Personalized learning
☐    Other: _____________________

Qualification*
☐ Practitioner / developer with 2 or more years of experience
☐ Post graduate / Researcher
☐ Other:

Please state your location (Country)*
_________________________________

Personality & Gamification (Group 1)
This section presents a group of Personality traits. Please indicate which game elements you consider MOST useful for improving motivation and engagement in e-learning for this personality

Group 1
They are objective and analytical, and like to focus on realities and practical applications in their work. They are mainly interested in facts, that can be collected and verified directly by the sense – by seeing, hearing, touching, etc. They make decisions on these facts by impersonal analysis, with step-by-step process of reasoning from cause to effect, from premise to conclusion.

Performance & Measurement*
Which of these game elements would provide the best feedback to this personality type?
☐ Performance Statistics
☐ Point (scores, experience points, skill points, etc)
☐ Acknowledgement (badges, medals, trophies and achievement cards, etc)

☐ Progression (progress bars, steps, maps)

☐ Levels

☐ Other: _______________________

Environment*

Which of this game elements would best give context for this personality type?

☐ Rarity: Limited items, collection, exclusivity

☐ Chance (Randomness, luck, fortune or probability)

☐ Choice or Path selection

☐ Economy: Transactions, market, exchange

☐ Time Pressure: Countdown timers or clocks

☐ Other: _______________________

Social*

Which of these game elements would provide the best social interaction to this personality type?

☐ Social Pressure (Peer pressure or guild missions)

☐ Competition (Conflict, leader boards, scoreboards, player vs player)

☐ Cooperation (Teamwork, co-op, groups)

☐ Reputation (Classification, status)

☐ Other: _______________________

Personal*

Which of these game elements would motivate this personality to engage more with the learning system?

☐ Objectives: Missions, side-quests, milestones

☐ Puzzle: Challenges, cognitive tasks, actual puzzles

☐ Sensation: Visual or sound stimulation

☐ Novelty: Updates and changes

☐ Renovation: Boosts, extra life, renewal

☐ Avatar, Virtual Identity

☐ Other: _______________________

Fiction*

Which of these game elements would motivate this personality to engage more with the learning system?

☐ Storytelling: Audio, graphical or text queues

☐ Narrative: (Where previous decisions impact future experiences)

☐ Other: _______________________

Please provide any additional comments regarding any of the questions above

_____________________________________________________________

Personality & Gamification (Group 2)

This section presents a group of Personality traits. Please indicate which game elements you consider MOST useful for improving motivation and engagement in e-learning for this personality

Group 2

They tend to approach life and work in a warm people-oriented manner, liking to focus on realities and hands-on activities. They are also interested in facts, but make their decisions with personal warmth, because they base their decision making on feeling, with its power to weigh how much things matter to themselves and others.

Performance & Measurement*

Which of these game elements would provide the best feedback to this personality type?

☐ Performance Statistics

☐ Point (scores, experience points, skill points, etc)

☐ Acknowledgement (badges, medals, trophies and achievement cards, etc)

☐ Progression (progress bars, steps, maps)

☐ Levels

☐ Other: _______________________

Environment*

Which of this game elements would best give context for this personality type?

☐ Rarity: Limited items, collection, exclusivity

☐ Chance (Randomness, luck, fortune or probability)

☐ Choice or Path selection

☐ Economy: Transactions, market, exchange

☐ Time Pressure: Countdown timers or clocks

☐ Other: _______________________

Social*

Which of these game elements would provide the best social interaction to this personality type?

☐ Social Pressure (Peer pressure or guild missions)

☐ Competition (Conflict, leader boards, scoreboards, player vs player)

☐ Cooperation (Teamwork, co-op, groups)

☐ Reputation (Classification, status)

☐ Other: _______________________

Personal*

Which of these game elements would motivate this personality to engage more with the learning system?

☐ Objectives: Missions, side-quests, milestones

☐ Puzzle: Challenges, cognitive tasks, actual puzzles

☐ Sensation: Visual or sound stimulation

☐ Novelty: Updates and changes

☐ Renovation: Boosts, extra life, renewal

☐ Avatar, Virtual Identity

☐ Other: _______________________

Fiction*

Which of these game elements would motivate this personality to engage more with the learning system?

☐ Storytelling: Audio, graphical or text queues

☐ Narrative: (Where previous decisions impact future experiences)

☐ Other: _______________________

Please provide any additional comments regarding any of the questions above

________________________________________________________

Personality & Gamification (Group 3)

This section presents a group of Personality traits. Please indicate which game elements you consider MOST useful for improving motivation and engagement in e-learning for this personality

Group 3

They tend to approach life and work in a warm and enthusiastic manner, and like to focus on. They prefer Intuition, their interest is not in facts but in possibilities, such as new projects, things that have not happened yet might be made to happen, new truths that are not yet known but might be found out, or above all, new possibilities for people.

Performance & Measurement*
Which of these game elements would provide the best feedback to this personality type?
☐ Performance Statistics
☐ Point (scores, experience points, skill points, etc)
☐ Acknowledgement (badges, medals, trophies and achievement cards, etc)
☐ Progression (progress bars, steps, maps)
☐ Levels
☐ Other: _______________________

Environment*
Which of this game elements would best give context for this personality type?
☐ Rarity: Limited items, collection, exclusivity
☐ Chance (Randomness, luck, fortune or probability)
☐ Choice or Path selection
☐ Economy: Transactions, market, exchange
☐ Time Pressure: Countdown timers or clocks
☐ Other: _______________________
Social*
Which of these game elements would provide the best social interaction to this personality type?
☐ Social Pressure (Peer pressure or guild missions)
☐ Competition (Conflict, leader boards, scoreboards, player vs player)
☐ Cooperation (Teamwork, co-op, groups)
☐ Reputation (Classification, status)
☐ Other: _______________________
Personal*
Which of these game elements would motivate this personality to engage more with the learning system?
☐ Objectives: Missions, side-quests, milestones
☐ Puzzle: Challenges, cognitive tasks, actual puzzles
☐ Sensation: Visual or sound stimulation
☐ Novelty: Updates and changes
☐ Renovation: Boosts, extra life, renewal
☐ Avatar, Virtual Identity
☐ Other: _______________________
Fiction*
Which of these game elements would motivate this personality to engage more with the learning system?
☐ Storytelling: Audio, graphical or text queues
☐ Narrative: (Where previous decisions impact future experiences)
☐ Other: _______________________

Please provide any additional comments regarding any of the questions above
_________________________________________________________________

Personality & Gamification (Group 4)
This section presents a group of Personality traits. Please indicate which game elements you consider MOST useful for improving motivation and engagement in e-learning for this personality

Group 4
They tend to approach life and work in a logical and objective manner, and like to make use of their ingenuity to focus on possibilities, particularly possibilities that have a technical application with the human element sometimes ignored.

Performance & Measurement*
Which of these game elements would provide the best feedback to this personality type?
☐ Performance Statistics
☐ Point (scores, experience points, skill points, etc)
☐ Acknowledgement (badges, medals, trophies and achievement cards, etc)
☐ Progression (progress bars, steps, maps)
☐ Levels
☐ Other: _______________________

Environment*
Which of this game elements would best give context for this personality type?
☐ Rarity: Limited items, collection, exclusivity
☐ Chance (Randomness, luck, fortune or probability)
☐ Choice or Path selection
☐ Economy: Transactions, market, exchange
☐ Time Pressure: Countdown timers or clocks
☐ Other: _______________________

Social*
Which of these game elements would provide the best social interaction to this personality type?
☐ Social Pressure (Peer pressure or guild missions)
☐ Competition (Conflict, leader boards, scoreboards, player vs player)
☐ Cooperation (Teamwork, co-op, groups)
☐ Reputation (Classification, status)
☐ Other: _______________________

Personal*
Which of these game elements would motivate this personality to engage more with the learning system?
☐ Objectives: Missions, side-quests, milestones
☐ Puzzle: Challenges, cognitive tasks, actual puzzles
☐ Sensation: Visual or sound stimulation
☐ Novelty: Updates and changes
☐ Renovation: Boosts, extra life, renewal
☐ Avatar, Virtual Identity
☐ Other: _______________________

Fiction*
Which of these game elements would motivate this personality to engage more with the learning system?
☐ Storytelling: Audio, graphical or text queues
☐ Narrative: (Where previous decisions impact future experiences)

☐ Other: _______________________

Please provide any additional comments regarding any of the questions above
________________________________________________________________

**iii.    Personality evaluation questions for MBTI cognitive core**

| | Sensing or Intuitive |
|---|---|
| | **Sensing or Intuitive** |
| 01 | If I were a teacher, I would you rather teach<br>   a.  facts-based courses<br>   b.  courses involving opinion or theory |
| 02 | I would rather be considered<br>   a.  a practical person<br>   b.  an out-of-the-box-thinking person |
| 03 | I would rather have a friend<br>   a.  who is consistent<br>   b.  who always comes up with something new |
| 04 | I usually get along better with<br>   a.  realistic people<br>   b.  imaginative people |
| 05 | When doing things, I would rather<br>   a.  do them in the accepted way<br>   b.  try out my own way of doing it |
| 06 | I prefer things to be<br>   a.  plain with no hidden meanings<br>   b.  figurative with deeper meanings |
| 07 | I prefer<br>   a.  facts<br>   b.  Ideas |
| | **Feeling or Thinking** |
| 08 | When relating with people or things, I am often<br>   a.  sympathetic towards them<br>   b.  analytic towards them |
| 09 | Strangers would easily consider me<br>   a.  lively<br>   b.  calm |
| 10 | I consider myself more<br>   a.  warm-hearted<br>   b.  firm-minded |
| 11 | When resolving conflicts, I will likely come to<br>   a.  a human-centered conclusion<br>   b.  a logical conclusion |
| 12 | Which of these describes me more?<br>   a.  devoted<br>   b.  determined |
| 13 | It is a higher compliment to be called<br>   a.  a person of real feeling<br>   b.  a consistently reasonable person |
| 14 | I value<br>   a.  feelings more than logic<br>   b.  logic more than feelings |

**iv.    Description of game elements**

| Category | Game Element | Description |
|---|---|---|
| **Ecological** | *These elements mirror real-world dynamics, introducing surprise, scarcity, and resource management to enhance learners' engagement and decision-making.* | |
| | Rarity | Implementing limited collectibles or exclusive rewards motivates learners to actively engage and strive for accomplishments that stand out. |
| | Economy | Integrating transactions, markets, or exchanges of virtual resources like points creates a sense of value and strategy, enhancing learner engagement. |
| | Imposed Choice | Offering learners alternate paths or imposing choices in their learning journey encourages decision-making. This dynamic shapes their experience and fosters critical thinking. |
| | Chance | Incorporating randomness or probability-based events adds an element of surprise to the learning journey. Learners may encounter lucky opportunities or unforeseen challenges, enhancing engagement. |
| | Time Pressure | Timed events and clock counts create urgency, mirroring real-world scenarios and promoting quick decision-making in learning tasks. |
| **Social** | *These dynamics encourage learners to engage with peers, fostering teamwork, healthy rivalry, and a sense of accomplishment within a social context.* | |
| | Social Pressure | Introducing peer pressure or guild missions encourages learners to participate and perform well to meet collective goals, enhancing engagement. |
| | Competition | Leaderboards, scoreboards, or player-vs-player contests foster |

| | | healthy competition among learners, boosting motivation and engagement. |
|---|---|---|
| | Cooperation | Encouraging teamwork and collaboration through group activities or cooperative tasks strengthens a sense of community and shared learning objectives. |
| | Reputation | Classifying learners' status or achievements within the gamified environment motivates them to maintain a positive reputation and strive for excellence. |
| **Personal** | *The "Personal" category tailors the gamified e-learning experience to individual learners. It encompasses elements that enable learners to set goals, engage in cognitive challenges, and interact with updates that match their preferences. This category empowers learners to take charge of their learning journey and offers a dynamic and personalized educational environment.* | |
| | Identity/Avatar | Allowing learners to create and personalize avatars or digital personas enhances their sense of immersion and ownership within the e-learning environment. |
| | Sensation | Incorporating haptic, visual, or audio stimuli enriches the learning experience, making it interactive, immersive, and memorable. |
| | Objective | Structuring missions, side-quests, or milestones provides learners with clear goals and a sense of purpose, enhancing their motivation and progress. |
| | Puzzle | Challenging learners with cognitive tasks, challenges, or puzzles stimulates critical thinking and problem-solving skills, enhancing engagement. |
| | Renovation | Providing boosts, extra lives, or renewal elements offers learners second chances and boosts their engagement by promoting perseverance. |

| | Novelty | Regular updates and changes maintain learner interest by offering new challenges or content, preventing monotony. |
|---|---|---|
| **Fictional** | *Fictional elements introduce imaginative and narrative-driven aspects, such as storylines and narratives. These elements immerse learners in fictional worlds where their actions and decisions shape outcomes, fostering an enriched and interactive learning experience.* | |
| | Narrative | Allowing learners' actions and decisions to shape future events in the narrative enhances their agency and immersion in the learning experience. |
| | Storytelling | Guiding learners through predefined events and activities as they follow a scripted narrative adds context and engagement to the learning process. |
| **Performance** | *Performance elements encompass the measurement and acknowledgment of learners' accomplishments. Point systems, progress indicators, and performance statistics provide a clear overview of learners' progress, motivating them to excel and offering a tangible sense of achievement.* | |
| | Progression | Displaying progress through progress bars, maps, or steps offers learners a visual representation of their advancement, motivating them to reach higher levels. |
| | Level | A leveling system acknowledges learners' accomplishments and advancement, providing a sense of achievement and motivating sustained engagement. |
| | Point | Accumulating scores, experience points, or skill points reinforces positive behavior and provides learners with tangible markers of progress. |
| | Stats | Visualizing learners' statistics and progress fosters self-awareness and encourages continuous improvement. |
| | Acknowledgment | Rewarding learners with badges, medals, trophies, or achievement cards acknowledges their achievements, boosting their self-esteem and motivation. |

**v.    Screenshots from developed system**

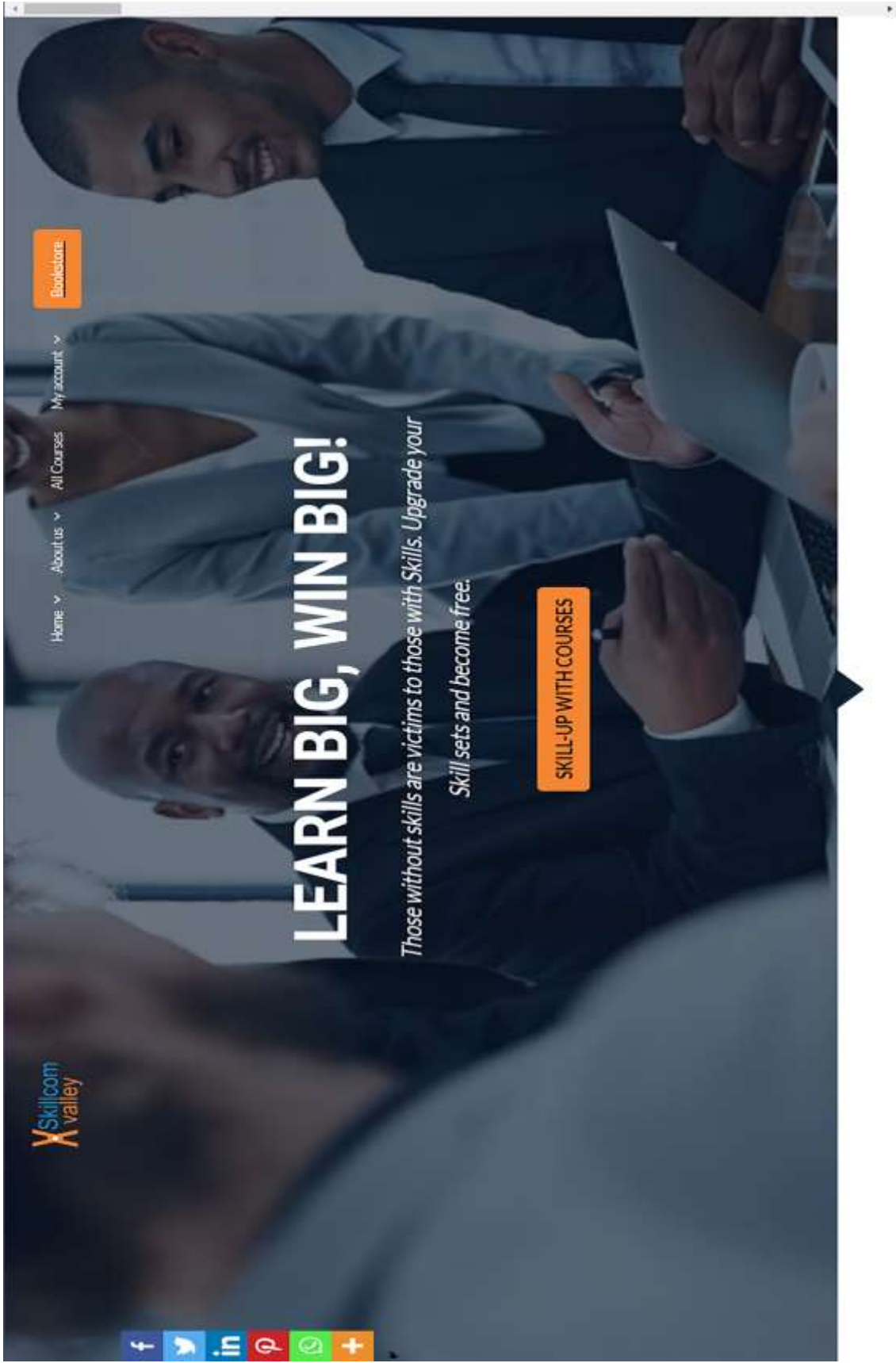

Skillcom Valley landing page

Login / Registration page

Lesson in session

Statistical performance feedback

Feedback from quiz